# CELLULAR AUTOMATA THEORY AND PHYSICS

A New Paradigm For The Unification of Physics

**Tom Ostoma and Mike Trushyk**

48 O'HARA PLACE, Brampton, Ontario, L6Y 3R8
emqg@rogerswave.ca

July 7, 1999

## ACKNOWLEGMENTS

We wish to thank R. Mongrain (P. Eng.) for his encouragement, constructive criticisms, and for the lengthy conversations on the nature of space, time, light, matter, and CA theory.

# ABSTRACT

*A new paradigm for the unification of physics is described. It is called Cellular Automata (CA) theory (ref. 2,4 & 6), which is the most massively parallel computer model currently known to science. We maintain that at the tiniest distance and time scales the universe is completely deterministic, and utterly simple. Our universe is a Cellular Automaton consisting of a huge array of cells capable of storing numeric information. These cells form a vast, 3D 'geometric' CA, where each cell has 26 surrounding neighboring cells that influence the state of a given cell. One set of common mathematical rules exists for each given cell, and the rules are applied to the inputs of the numeric state of the cell's immediate 26 neighbors and on the numeric state of the cell itself. The result of this 'computation' is then stored back in the cell on the next 'clock cycle'. In this way, all the cells are updated simultaneously, everywhere (not in the context of our measure of time) This process repeats itself for all cells, and for each and every 'clock cycle'. The CA structure automatically has a built-in, low-level space and time in the form of cells and 'clock cycles'. This can be thought of as a special kind of absolute and quantized 3D space, and separate, quantized absolute time. Low-level space and time units are totally <u>inaccessible</u> to direct measurement, and are independent of the usual 4D relativistic space-time units (which can be affected by motion and by gravitational fields). Here we reconcile the known relative nature of Einstein's 4D space-time of special and general relativity with the absolute nature of the quantized 3D CA space and quantized, time of the cellular automaton.*

*CA theory directly implies that all the laws of physics must result from interactions that are strictly local, therefore forbidding any form of action at a distance. CA theory suggests that space, time, matter, energy, and motion are all the same thing: the end result of information changing state in the CA. Matter in motion is an illusion resulting from the 'shifting' of stable particle-like information patterns from cell to cell in low-level 3D CA space. The CA model automatically contains an inherent <u>maximum speed limit</u> for which information can be moved from place to place. At the CA level we can represent CA low-level space as a rectangular array of integers or cells; $C_{i,j,k}$ with respect to an arbitrary cell that acts like the origin. The entire array of numbers is updated at every new CA 'clock cycle' $\Delta t$, where $N\Delta t$ is the measure of the number of CA low-level time periods elapsed between two events. The 26 inputs surrounding a given cell $C_{i,j,k}$ act to modify the state of the cell at the next 'clock cycle' according to a fixed (possibly integer) function or algorithm F acting on the 27 inputs (and on the state of the cell itself) as follows: $C_{i,j,k} = F\ C_{i+x,j+y,k+z}$ ; where x,y,z take on all combinations of -1,0, and 1. CA low-level space or time are not affected by any physical interactions, and are inaccessible by direct measurement.*

*We propose that light (photon) motion is the fixed, simple shifting of a photon information pattern from cell to adjacent cell at every 'clock cycle'. Thus photons 'travel' only at one fixed speed, which is unaffected by any possible source motion. By adopting absolute CA space and time coordinates for the description of a pair of observers in inertial reference frames with a relative velocity 'v', then the Lorentz transformation follows mathematically. We argue that 4D flat space-time is a direct consequence of the behavior of the universal CA, as seen by inertial observers who are not aware of, and cannot access the absolute low-level CA units of space and time. We also provide some speculative arguments that may reconcile the apparent non-local character of entangled quantum particle interactions in quantum theory, and the strictly locality that is required for particle 'motion' in relativity and in CA theory.*

*We establish a deep connection between the CA model and the existence of virtual particles in the quantum vacuum. We find that the particle exchange paradigm of quantum field theory is also manifestly compatible with the CA model. However, classical Newtonian inertia is not compatible with CA, since it is a global law regarding the property of the total mass to resist being accelerated. We briefly present a new theory of inertia based on a proposal by R. Haisch, A. Rueda, and H. Puthoff (ref. 5) which we modified and called Quantum Inertia (QI). QI is also manifestly compatible with Cellular Automata theory. According to QI, inertia is due to the strictly local electrical force interactions of matter consisting of quantum particles with the surrounding electrically charged virtual*

*particles of the quantum vacuum. The sum of all the tiny electrical forces (from photon exchanges) originating from each electrically charged particle in the mass with the surrounding electrically charged virtual particles of the quantum vacuum is the source of the total inertial force opposing accelerated motion in Newton's law 'F = MA'. QI resolves the problems and paradoxes of accelerated motion introduced in Mach's principle by suggesting that the state of acceleration of the electrically charged virtual particles of the quantum vacuum with respect to a mass serves as Newton's absolute space, for accelerated motion only.*

*We have developed a new quantum theory of gravity called Electro-Magnetic Quantum Gravity (EMQG) (ref. 1) for the purpose of unifying Quantum Inertia and gravity into one common quantum framework. EMQG theory is designed to be manifestly compatible with Cellular Automata theory. What is unique about EMQG as a quantum gravity theory is that it is based on two boson force exchange particles; the graviton and the photon. Furthermore, the photon and graviton are physically nearly identical, with the same quantum numbers, but varying greatly in the strength of the force coupling. We invoked Einstein's principle of equivalence of inertial and gravitational mass to understand the origin of gravitational mass from the perspective of Quantum Inertia. We concluded that gravity also involves the same 'inertial' electrical force component that exists for inertial mass. We proposed that the general relativistic Weak Equivalence Principle (WEP) is a physical phenomenon, originating from common 'lower level' quantum processes occurring in both gravitational mass and inertial mass. The magnitude of the gravitational mass results from the electrical force interactions of the electrically charged mass particles interacting with the surrounding electrical charged virtual particles of the quantum vacuum. However, now the quantum vacuum particles are falling due to graviton exchanges between the earth and the surrounding quantum vacuum particles, which posses mass. Thus, both the electrical force (photon exchanges) and the pure gravitational force (graviton exchanges) are involved in gravity. However inertial mass is strictly the result of only the electrical force process specified in quantum inertia (with negligible graviton processes present). For a gravitational test mass near the earth, the graviton exchange process occurring between the earth, the test mass, and the surrounding vacuum particles upsets perfect equivalence of inertial and gravitational mass, with the gravitational mass being slightly larger than inertial mass. Thus, for a large and a tiny test mass dropped simultaneously on the earth, the larger mass falls faster by a very small amount. This tiny deviation from perfect equivalence might be detected experimentally.*

*Since the virtual matter particles (specifically virtual fermions) of the quantum vacuum posses mass, they exchange gravitons with the earth and are therefore falling. This can be visualized as a downward directed accelerated 'flow' of electrically charged vacuum particles. The reversal in the relative accelerated state of the quantum vacuum particles with respect to a test mass in an accelerated or in a gravitational frame still leads to the same measure of mass. This is the reason why there is mass equivalence. 4D curved space-time is now understood as a consequence of the behavior of matter (particles) and energy (photons) under the influence of this (statistical average) downward accelerated 'flow' of electrically charged virtual particles of the quantum vacuum. This coordinated downward 'accelerated flow' can be thought of as a 'Fizeau fluid' that 'flows' through all matter subjected to a gravitational field. Like in the Fizeau experiment (which was performed with a constant velocity water flow), the behavior of light and measuring instruments is now affected by the accelerated flow of the 'Fizeau fluid' (virtual particle fluid). This is the root cause of 4D space-time curvature. This process also occurs in accelerated motion, where the quantum vacuum acceleration is an 'apparent' acceleration actually caused by the acceleration of the observer. In EMQG, space-time measurements based on instruments composed of matter are now affected by the action of this accelerated 'Fizeau-like' quantum vacuum fluid to give an 'apparent' 4D space-time curvature. Because our measuring instruments are incapable of measuring the true absolute and quantized CA 3D space and quantized CA time units, 4D space- time appears to be curved in accelerated frames and inside gravitational fields.*

# **TABLE OF CONTENTS**

# 1. INTRODUCTION

*" It always bothers me that, according to the laws as we understand them today, it takes a computing machine an infinite number of logical operations to figure out what goes on in no matter how tiny a region of space and no matter how tiny a region of time. .... why should it take an infinite amount of logic to figure out what one tiny piece of space-time is going to do?"*

*- Richard Feynman*

In spite of the phenomenal success of modern physics in explaining all aspects of reality, the current state of fundamental physics is both messy and incomplete. The two most fundamental theories we have are quantum theory and relativity. Relativity is primarily about the nature of space and time, and quantum theory is mostly about matter at the smallest distance scales. The different disciplines of fundamental physics are based on postulates, or fundamental principles that are specific to their particular regime, and in some sense seem unrelated to each other.

For example, in special relativity the founding postulates are the constancy of light velocity and the principle of relativity. In general relativity we have the principle of equivalence and the principle of general covariance. In quantum mechanics we have the DeBroglie wave hypothesis and associated Schrodinger's wave function, the Bohr wave-particle duality, the Heisenberg uncertainty principle, the principle of indistinguishability of particles in the same state, the Pauli's exclusion principle, etc. When these different postulates are carefully compared against each other, we find a general lack of consistency in the use of certain basic notions such as space, time, and matter.

The concept of mass is not handled the same way in relativity as in quantum theory. In special and general relativity, mass is treated as a continuous mass (mass-energy) distribution, and *not* as a collection of elementary particles that are held together by forces due to exchanged particles (as required by the principles of quantum field theory). In another example, general relativity requires the existence of some form of a global 4D space-time structure in order to understand gravity and the overall cosmological evolution of the universe. Yet special relativity is quite different in that 4D space-time is purely dependent on the *local reference frame chosen* by an observer, where observers in different states of relative acceleration experience different 4D space-time curvatures. Quantum mechanics seems to require the quantization of all physical quantities on the small scale, yet space and time are still treated as a classical space-time continuum, where there are an infinite number of space points between any two given locations, no matter how close.

It is for these reasons that most physicists agree that the current set of fundamental physical laws is incomplete. What is really lacking in fundamental physics is a single, unified model of the universe that allows us to understand the inner workings of nature. It has been known for quite some time that the laws of physics are mathematical in nature, and that this somehow implies that the universe must function in some sort of mathematical way. Is there hidden *machinery* that causes the universe to function the way it does? Is it possible for the humans to comprehend the complete inner workings of our universe? We have reason to believe that the answer to these questions is yes.

We believe that Cellular Automata (CA) theory is currently our *best* model we have for understanding *how* the universe functions. A Cellular Automaton computer is the most massively parallel computer model known to computer science. We believe that our universe is indeed a massive numeric computation that runs on some form of a CA 'machine'. We refer to this CA 'machine' as the universal CA. We propose that our universe is the 'software' (or changing information patterns) running on this universal CA 'hardware'. The 'hardware' of the CA machine exists outside our universe by definition, and it's functions is therefore not necessarily limited by our laws of physics, or by any future technologies.

A Cellular Automaton consists of a huge array of 'cells' or memory locations that are capable of storing numerical data. The numeric state of *all the cells,* everywhere, changes at a regular synchronized interval called a 'clock cycle'. The change in numeric state depends on the numeric values of a cell's neighbors and on the same set of fixed rules or algorithm, which are located in each and every cell of the cellular automaton. *It is important to note that the 'clock cycle' marks events related to the CA operation, and is not the same thing as events in our measure of time.*

Unlike the ordinary desktop computer, the cellular automata computer updates *all* it's memory locations in a single 'clock cycle', thus making the CA computer vastly more faster and powerful than any of the common computers systems in use today. It should be noted that there already exists computational machines modeled on the CA architecture, which are being used by physicists for various physical modeling projects. These CA machines are used to model nature, such as problems in fluid flow and turbulence, rather then for computing things like income tax or other business applications. However, it is interesting to note how well the CA computer model is suited to modeling various particle interactions, such as for elementary particle physics problems, fluid mechanics, and turbulence problems. The CA structures used for these type of simulations are known as 'Lattice Gases'. Reference 38 and 39 contain a detailed discussion on the uses of lattice gases for modeling of physical problems.

The cellular automaton computer was discovered theoretically by Konrad Zuse and Stanislav Ulam in the late 1940's, and later was put to use by John von Neumann to model the real world behavior of complex spatially extended structures. The best known example of a CA is the 'game of life' originated by John Horton Conway and published in the famous mathematical games department of Scientific American magazine in the 1970's. A direct consequence of CA theory is that our universe must be *utterly simple* on the smallest scales. Furthermore, the universe must be completely deterministic, or at least in principle. The apparent randomness of nature set forth by quantum theory must reflect our general ignorance of the exact numeric state of a system on the CA. If we know the structure of the universal CA, the exact mathematical rules that govern the cells of the CA, and the exact numeric state of all the cells in the universe at a given instant, then we have all the information needed to predict the exact future state of the system. Of course this is completely impossible to do in practice, for a variety of reasons. Even if we can somehow gather all this information, we would need a computer just as powerful as the universal CA to process the data in a reasonable period of time!

The purpose of this work is to survey fundamental physics using the CA model as our guiding principle in order to select the postulates, physical concepts, and ideology that most closely fits this CA paradigm. Furthermore we introduce an important new quantum gravity theory that was designed to be manifestly compatible with the CA paradigm called ElectroMagnetic Quantum Gravity (EMQG) theory. EMQG incorporates CA theory, quantum theory, and a new theory of quantum inertia and quantum gravity into a unified framework. According to EMQG, the general relativistic, curved, 4D space-time on the earth turns out to be the end result of the activities of electrically charged quantum particles inside a mass, which interact with the vast sea of surrounding electrically charged virtual particles of the quantum vacuum, that are falling during their brief existence. Quantum particles are stable numeric information patterns that move in a 'low-level' and quantized 3D CA space, and separate quantized 'low level' CA time.

According to the basic premises of CA theory, elementary particles can be thought of being stable 'numeric information patterns', which shift around from cell to cell in CA cell space. The CA cell space is more closely analogous to an absolute 3D space, where numeric information patterns are free to shift from place to place. Furthermore, CA theory suggests that *all* the laws of physics should be the result of numeric interactions that are strictly local in nature, which therefore forbids any kind of action at a distance. If CA theory is correct, then the laws of physics which are formulated in a global form (like the Newton's law of inertia F=MA) cannot be fundamental. All physical laws should somehow be the result of the local actions of particles (particles interacting with particles through exchange particles), which exist directly as information patterns in the cells of the CA.

We will see that special relativity is already manifestly compatible with the cellular automata model, with an extra assumption about the nature of light motion (photons) on the CA. We believe that light is a messenger from the shadowy CA world, which is isolated from us by tremendously small distance scales of the CA. Photons 'move' in the *simplest possible way* on the CA, and it is this simple motion that is responsible for the rather odd behavior of light from our perspective. The photon patterns always shift over one cell at every 'clock cycle', regardless of the motion of the source. As Einstein pointed out, light is fundamental to our perception of space and time. It turns out that the Lorentz transformation of special relativity can actually be *derived mathematically* from these simple assumptions about light.

Much of the current difficulties with understanding the nature of space and time in modern physics comes from the assumption that space and time are actually fundamental entities in our universe. In CA theory this is definitely not the case. The space and time that we can access and measure turn out to be the end result of complex particle interactions. This action involves another important player in physical theory, the virtual particles of the quantum vacuum and their associated interactions with the observable real matter particles that make up a mass.

We argue that everything is the result of a *vast amounts* of numerical information being processed on the universal CA computer. The dynamics of information inside this computer is indeed our *whole* universe, including ourselves! All elements of reality are due to numerical information being processed on the 'cells' at incredibly 'high speeds' with respect to our measurement of our time (where more than $10^{43}$ 'clock cycles' occur every second!) on the universal computer. However, the computer 'hardware' is inaccessible to us, because we ourselves are also highly evolved numeric information patterns residing on the cells.

How can such mathematical simplicity lead to the incredible complex phenomena that we observe in nature around us? The newly developed mathematical fields of chaos theory and complexity theory gives a hint on how this is possible. Recent developments in chaos theory have taught us that simple rules do indeed lead to complex and apparently unpredictable behavior. For example, in the famous Lorenz system, a simple toy weather system was modeled with a simple set of differential equations. Although the system is predictable given sufficient computing resources to model the mathematical system, the output generated from the system looks random and unpredictable. In fact, unless the initial conditions of the system are known precisely, the future behavior of this system remains unpredictable and apparently random. In another example, the Mandelbrot set, named after the inventor, is an infinitely complex mathematical object. You can display the set on a computer screen and zoom as much as you want and still uncover patterns not seen before. Yet the Mandelbrot set is specified mathematically by a very simple iteration given by: $z^2 \Rightarrow z^2 + c$, where z is a complex variable, and c is a constant. How can so simple an equation lead to literally an infinite complexity?

The laws of physics that govern the 'hardware' functioning of the universal computer do not even necessarily have to be the same as our own physical laws. In fact, the computer 'hardware' that governs our reality can be considered by definition of the word 'universe' to be outside our own universe, since this hardware is totally inaccessible to us. We can only infer the existence of the CA through observation of the behavior of any stable numeric information patterns interacting with other numeric information patterns. In other words, to probe the ultimate nature of matter, one must utilize matter particles to probe matter particles. Ultimately what we learn in this way is how particles interact with other particles. We believe that this is the reason for many of the problems that arise in the interpretation of quantum mechanics.

Our universal CA is so powerful that it is capable of updating it's entire memory, no matter how big, in a ***single CA 'clock cycle'***! Contrast this to the familiar ordinary desktop computer, which is a version of the serial Von Neumann computer architecture, named after the inventor. The desktop computer takes millions of 'clock cycles' to update it's entire memory, as compared to one 'clock cycle' for the CA. This is because memory cells or locations must be accessed sequentially by a central processor. Furthermore, the more memory the desktop computer has, the more 'clock cycles' that are required to update it. However, a cellular automaton architecture is quite different in this respect. CA's are inherently symmetrical, because one set of rules is programmed and repeated for each and every 'memory' cell. If you know the mathematical rules or logic behind one cell, then you know the entire logic of all the cells. We believe that this accounts for the high degree of spatial symmetry found in our universe. In other words, the laws of physics are the same no matter where you are, or how you are oriented in space. Therefore, the perfect symmetry of CA 'cell space' accounts for the perfect symmetry of our actual space.

Not only is the CA a parallel computer, it is also the ***fastest** known parallel computer*. However, the high degree of parallelism of the CA computer comes at a high price. The CA model places severe limitations on just how numeric data can be processed. We will show that one of these limitations results in the CA having an *absolute maximum speed* for which information can be shifted from a certain location in 3D cell space to another. In fact, this limitation can be observed in our universe, which in fact does posses a maximum speed for which elementary particles can achieve. It is well known that this maximum speed limit is the speed of light. Before we can investigate this, we must review some of the basic concepts of CA theory.

## 2.  CELLULAR AUTOMATA THEORY

*"All (the universe) is numbers"*

*- Pythagoras*

The ancient Greek mathematician and philosopher Pythagoras once proclaimed that the universe is really numbers, as in the quotation above. We believe he was very close to the truth. The missing link that Pythagoras lacked to connect his idea to reality is the mathematical concept of a computer, which was invented much later. We have reason to believe the Cellular Automaton computer model is the perfect choice for constructing a universe. In this paper we assume that the universe is a CA. We then examine the consequences of this reality for physics, and for beings such as ourselves that live in such a universe.

## 2.1 INTRODUCTION TO CA THEORY AND PHYSICS

*"Digital Mechanics is a discrete and deterministic modeling system which we propose to use (instead of differential equations) for modeling phenomena in physics.... We hypothesize that there will be found a single cellular automaton rule that models all of microscopic physics; and models it exactly. We call this field Digital Mechanics."*

*- Edward Fredkin*

On remotely small scales of distance and time (on the order of the Plank scale, i.e. $10^{-35}$ meters and $10^{-43}$ seconds) our 'view' of the universe is completely unlike anything we know about physical reality from our senses. Figure 7 gives a 'schematic view' of what 3D cell space might look like to an observer outside the CA computer (the lines radiating from the central cell represent schematically the information links from the adjacent cells). Space, time, and matter no longer exist as real and separate physical entities. Elementary particles of matter reveal themselves as oscillating numeric information patterns. Motion turns out to be an illusion, which results from the 'shifting' of particle-like information patterns from cell to cell in the 3D CA space. However there is no real movement as we know it! It would be more correct to say that numeric data that makes up a particle shifts (or transforms) from place to place incrementally, rather than a smooth motion.

Forces are now recognized as resulting from the exchange of huge numbers of discrete particles, or information patterns called vector bosons, which are exchanged between two or more particle information patterns. The absorption of a vector boson information pattern changes the internal oscillation state of a particle, and causes an impulse of motion (an acceleration) to occur along a particular direction, usually towards or away from the source particle. This turns out to be the quantum origin of all forces. Therefore, forces can be thought of as being 'digital' rather than 'analog' in nature, using the language of modern computer theory.

What appear as smoothly applied forces are the result of a countless numbers of tiny force impulses transmitted by the exchange particles. This force particle exchange paradigm was originated in Quantum ElectroDynamics (QED) in a highly successful way (ref. 40). This model for forces is found to be manifestly compatible with the basic ideology of Cellular Automata theory. Recall that CA theory strictly forbids forces as being caused by any sort of action at a distance or from some form of a field. Any coherent disturbance in any number pattern must propagate sequentially from cell to cell at some rate that is limited in speed.

The quantization of space that has long been sought by physicists is simply the collection of cells or storage locations for the numbers in the universal CA computer. The quantization of time simply represents the number of regular 'clock cycles' elapsed between two CA events. According to CA theory, the 'clock cycle' is required in order to coordinate or synchronize the change of numeric state of each of the cells. Therefore space at the lowest levels is not 'nothing', it really is something. It is memory locations or cells. Particle information patterns (numbers) residing in the cells are dynamic, and shifting (but *not actual* moving)! They simply change state as the computer simulation evolves.

Therefore at the lowest levels *our universe is utterly simple, deterministic, and mathematical in nature*. All changes that occur, must occur as *localized* changes and must be predictable in nature (at least in principle), given the detailed structure and numeric state of the CA. Since the distance and time scales are so remotely small, this level of reality is not directly accessible to us. Furthermore, we cannot 'read out' the numeric state of a given cell even if this scale were accessible to us. We can only infer this type of information by observing larger scale patterns, or by proposing a detailed CA model and checking the results with reality. Matter is composed of elementary particles, which on the CA are oscillating or dynamic number patterns. This includes clocks, rulers, and biological entities like ourselves! Therefore we are very limited in what can be determined about the basic nature of matter on the CA by using our senses or measuring instruments.

Quantum theory states that all elementary particles have wave functions. However, when a particle is detected it appears to be a point-like interaction, which is usually measured as a change in the energy level of an electron in some atom (or the release of an outer electron from the atom). When the particle is not disturbed, it appears as an oscillating wave with a specific wavelength called the 'De Broglie' wavelength. Nobody has *directly* 'observed' this wave function, although it's existence reveals itself indirectly in quantum interference experiments. Only the particle-like interactions between particles with wave functions have been directly observed in the laboratory. The amplitude of the wave function is interpreted as the probability distribution of finding a point-like interaction at a certain place and time. This interpretation of the wave function was first discovered by Max Born in the 1920's. This probabilistic nature of matter has led some physicists like Einstein to question the validity and the completeness of quantum theory.

In accordance with CA theory, we take the position that everything is deterministic, or this is so at least in principle. The CA model also tells us why we cannot achieve precise deterministic results in quantum experiments. Indeterminacy results because we simply cannot know the exact numeric state of a particle, and at this time we do not even know the exact rules of the CA either. In fact if the universe were truly random at the lowest scales, then it can be shown that a computing device would require an infinite number of operations or steps to reproduce this random behavior. It is well known that very difficult and computationally expense to generate a perfectly random sequence of numbers.

Therefore quantum theory *cannot be complete* as an ultimate description of physical reality, as Einstein suspected. The probabilities that are generated from experiments represent our ignorance of the exact numerical state of the CA. We are forced to observe the behavior of the CA with measuring instruments composed of matter, which are incapable of measuring the exact state of any quantum particle. Therefore the wave function is incomplete as the full description of the exact state of an elementary particle.

Any computer has certain limitations that restrict the performance that it can achieve. The cellular automaton is no exception! Most of these restrictions come from the finite speed of operation and the limited capacity of storage locations or memory of the computer. Therefore if the universe is really a computer, should we not be able to see some of these limitations?

In fact, the cellular automata structure contains a definite and inherent maximum speed limit that any particle information pattern is able to achieve. This is due to the following two reasons. First, there is fixed and constant rate in which cells can change from one state to another (the 'clock cycle'). Secondly for a change in state of a particular cell to propagate to some distant cell, the CA model requires that the state change be transferred sequentially from cell to adjacent cell. The fastest rate that this can occur is *only one cell per 'clock cycle'*. This is simply a limitation of the structure of any cellular automata computer model. This limitation naturally leads to the law of strict locality on the CA, which is a concept that is manifestly compatible with special relativity. If this speed limit is interpreted as the speed light, then we can say that the first postulate of special relativity follows from CA theory.

Since all the cells of the CA are 'programmed' with the same rules, and connected in the same way, there exists a simple spatial symmetry of the cells in the CA model. This is the origin of the spatial symmetry in our own universe. Since the 'motion' of particle information patterns cannot be gauged with our measuring instruments in terms of absolute cell locations, some distinct and recognizable information pattern residing on the CA must be used to determine the *relative* velocity of another stable numeric particle information pattern. This situation corresponds directly to the principle of relative motion in special relativity. In relativity one can only gauge motion by using a material reference frame. Therefore, we conclude that the second postulate of special relativity follows from the cellular automaton model.

Since the first and second postulates of special relativity follow from the structure of the CA, it is easy to see that special relativity is manifestly compatible with CA theory. We will explore these implications in detail in the special relativity (section 7). According to special relativity 3D space and time must be united into 4D space-time, which is relative. This results in observers, who are in different inertial frames of reference, **to disagree on** measurements of space and time. This would be difficult to understand from the basic structure of the CA, where 3D CA space and time are absolute, and independent off the numeric activity going on.

In special relativity time depends on the state of motion of the observer, and not on the absolute 'clock cycles' of the CA. In some sense, relativistic 4D space-time is an observer dependent formulation of space-time. What is the connection between absolute 3D CA space and time and the relativistic 4D space-time that is very much observer dependent. This problem is fully addressed in the section 7 on special relativity. It turns out that light plays a very important role in this, and leads to an observer dependent formulation of space and time defined by readings of ordinary clocks and rulers. In fact, we believe that light is a messenger from the CA cell space. We will show that the bizarre nature of relativistic 4D space-time arises from the special type of motion that light (photons) takes on the CA. This subject is covered in detail in the special relativity (section 7).

General Relativity is *not* compatible with the Cellular Automata model for a number of important reasons. First, general relativity is a field theory, where the gravitational field is represented by a 4D tensor field. This does not fit well with a CA. Secondly, 4D space-time is curved when an observer is subjected to accelerated motion, or to a gravitational field. 4D space-time has a global property near a large mass, and takes on a specific value given by the $G_{\alpha\beta}$ (the Einstein tensor) which depends on the amount of *absolute* mass-energy present. $G_{\alpha\beta}$ is the mathematical statement of space-time curvature that is generally covariant. $T_{\alpha\beta}$ is the stress-energy tensor, which represents the quantity of mass-energy of a large mass. Einstein's field equations relate $G_{\alpha\beta}$ and $T_{\alpha\beta}$ as follows:

$$G_{\alpha\beta} = \frac{8\pi G}{c^2} T_{\alpha\beta} \quad \text{.... Einstein's Gravitational Field Equations} \quad (2.1)$$

G is Newton's gravitational constant, and c the velocity of light. The constant G is a remnant from Newton's law of gravity. The constant $8\pi G/c^2$ was chosen to adjust the strength of coupling between matter and 4D space-time curvature, so that in the limit the motion of matter under the influence of curved space-time corresponds to the motion of matter according to Newton's law of gravity.

The 4D space-time depends on the value of the large mass in an *absolute* way. However, it also still retains the relative (and observer dependent) nature of special relativity, which modifies space-time depending on the state of motion of the observer. Therefore the global 4D space-time curvature near a gravitational field can be canceled out by adopting an accelerated free fall reference frame. It is totally unclear as to how a 4D space-time structure that depends so strongly on the motion of the observer can be constructed from a CA. We will investigate this in section 18.

It is obvious that the principle of equivalence cannot be a fundamental principle in the CA model. Inertia and gravity are not fundamental constructs on the CA. What is also lacking is a causal mechanism that couples the amount of 4D space-time near a mass to the quantity of mass present nearby. In the section on general relativity (section 17) we present such a mechanism, based on the particle exchange paradigm, which also provides a strictly local process that accounts for the inertia, gravity, and the equivalence principle.

In general relativity (and special relativity) inertia is treated in exactly the same as in Newton's Inertial force law F=MA, with no deeper understanding of inertia given. Inertia does not seem to be a local process, as would be required by CA theory. Therefore, there must be some hidden processes that accounts for inertia in a strictly local fashion. In order to resolve this problem a new theory of inertia has been developed, called Quantum Inertia. This is described in section 8. Quantum Inertia involves another important aspect of physical reality called the virtual particles of the quantum vacuum, described in section 3. Quantum Inertia is a theory where the inertia results from electrical interactions between charged particles that make up a mass and the charged virtual particles that make up the quantum vacuum. Thus, inertia involves a previously unknown electrical force component. Quantum Inertia is described in section 8.

In order to resolve the problems with general relativity, and also to incorporate this new Quantum Inertia theory on a common framework with gravity, we have developed a new quantum theory of gravity called 'ElectroMagnetic Quantum Gravity' or EMQG. EMQG is the quantization of general relativity, that is manifestly compatible with the Cellular Automaton model. EMQG also involves the same electrical force component in inertia for the description of gravitational interactions. Figure 8 gives a block diagram showing the relationship of CA theory, Quantum Inertia, and EMQG with the rest of physics. EMQG is discussed in detail in section 17.

A basic prediction of quantum field theory is the existence of the virtual particles of the quantum vacuum (ref. 23). It is known that even the most perfect vacuum is teeming with activity, where all types of quantum virtual particles from the various quantum fields appear and disappear spontaneously, everywhere. We have seen this type of behavior in 2D cellular automata. In the simple 2D CA called the 'game of life', most simple initial states (or 'seed' patterns as they are called) evolve into a complex soup of activity, everywhere, after a sufficiently large number of clock cycles have elapsed. This activity is very reminiscent of the quantum vacuum. We will find that quantum vacuum plays a very important role in understanding the origin of inertia, the principle of equivalence, and gravity. The quantum vacuum is discussed in detail in section 3.

We will focus on the relationship between Newton's laws, Special Relativity, General Relativity with the theory of Cellular Automata, and attempt to resolve the (apparently) different views that these theories imply for the fundamental nature of space, time, and matter. Before we can do this, we will need to take a brief look at some of the mathematical results of CA computer theory.

2.2     MATHEMATICS OF CELLULAR AUTOMATA

*"Truth is much to complicated to allow anything but approximations."*
- ***John Von Neumann***

We now give a brief presentation of some of the important mathematical concepts behind Cellular Automata theory and the connections. References 2, 4, and 6 give a much more complete account of the growing body of knowledge of CA theory. We also include some *new* material here, that we developed in regards to the CA space dimensionality. As we said, the cellular automaton computer was discovered theoretically by Konrad Zuse and Stanislav Ulam in the late 1940's, and later put to use by John von Neumann to model the real world behavior. The best known example of a CA is the 'game of life' originated by John Horton Conway in the 1970's. The game of life is one of the most studied two dimensional CA, and was very popular in the 1970's and 1980's. In the game of life, one can see analogies of some of the basic behaviors of elementary particles.

To our knowledge, Edward Fredkin (ref. 2) has been first credited with introducing the notion that our universe is a huge cellular automaton computer simulation. The cellular automaton (CA) model is a special mathematical computing device that consists of a large collection of cells or storage locations for numbers. Each cell contains some initial number (say 0 or 1, if the binary number system is chosen), and each cell contains the same set of rules which are applied after every 'clock cycle' for *each* and every cell. The rules specify how these numbers are to be changed at the next computer 'clock' interval. Mathematically, a 'clock' is required in order to synchronize the next state of all the cells. The logical rules of a cell specifies the new state of that cell on the next 'clock' period, based on the cells current state, and on that of all the cell's immediate neighbors. Each cell in the CA has the same fixed quantity of neighboring cells. The number of neighbors that influence a given cell is what we call the 'connectivity' of the cellular automaton. In other words, the number of neighbors that connect (or influence) a given cell is called the CA connectivity. The connectivity can be any positive integer number.

We define a 'geometric' cellular automata as a CA configuration where each cell has only the correct number of neighbors such that the CA connectivity allows a simple cubic geometric arrangement of cells that is required to build Cartesian spaces of any dimension (a cube in 3D space, for example). This arrangement can be visualized as the stacking of cells into squares, cubes, hypercubes, etc, where there are connections with the immediate surrounding cells. This structure is well suited for constructing mathematical 3D spaces. The mathematical spatial dimension required to contain this geometric CA is defined as the 'dimensionality' of the geometric cellular automata.

For example, a 2D geometric CA each cell has 8 surrounding neighbors, which can be thought of as forming a simple 2D space. One set of rules exists for every given cell, based on the input from its immediate 8 neighbors and on the state of the cell itself. The result of this 'computation' is then stored back in the cell on the next 'clock' period. In this way, all the cells are updated simultaneously in every cell, and the process repeats on each and every clock cycle. In a 1D geometric CA, each cell has 2 neighbors, a 2D geometric CA has 8 neighbors, a 3D geometric CA has 26 neighbors, a 4D has 80 neighbors, and a 5D has 242, and so on. In general, if $C_D$ is the number of neighbors of an Nth dimensional geometric CA, and $C_{D-1}$ is the number of neighbors of the next lower N-1th dimensional geometric CA, then: $C_D = C_{D-1} + C_{D-1} + C_{D-1} + 2$, or $C_D = 3 C_{D-1} + 2$, which is the number of neighbors of an Nth dimensional geometric CA expressed in terms of the next lowest space.

The geometric CA model that is explored here for our universe is a simple geometric 3D CA, where each cell has 26 neighboring cells (figure 7). Your first impression might be that the correct geometric CA model would be a 4D geometric CA, so that it is directly compatible with relativistic space-time. There are several problems with this approach. First 3D space and time have to be united in this CA model, which is not easy to do. More difficult still, is that 4D space-time is relative, and in some cases it is curved. This means that on the earth a falling reference frame observers flat 4D space-time, while a stationary observer exhibits curved 4D space-time. Furthermore, the 4D space-time curvature is directional near the earth. Curvature varies along the radius vectors of the earth, but does not vary parallel to the earth's surface (over small distances).

We will see that there are actually two space-time structures in our universe. First, there is the relativistic 4D space-time that is measured with instruments, and influenced by motion or gravity. Secondly, there is an absolute 'low level' 3D space, and separate 'time' that is not directly accessible to us, which exists at the Cellular Automaton level. The 3D space comes in the form of cells, which because of the structure of the CA are automatically quantized. Time is also automatically quantized in the CA model, and is extremely simple in nature. Time is the unidirectional evolution of the numeric state of the CA based on the local rules. The state of the CA changes on each and every 'clock' pulse, which can be thought of as a primitive form of time. Thus, our *low-level space* and separate *low-level time* are simple concepts in a 3D geometric CA. This model restores utter simplicity to the structure of our universe on the smallest scales. Everything is deterministic, and the future state of the universe can be determined if the exact state of the CA is known now.

What is the scale of this 3D geometric CA computer with respect to our distance and time scales? Assuming that the quantization scale corresponds to the Plank Scale (ref. 11) the number of cells per cubic meters of space is astronomically large: roughly $10^{105}$ cells. (Later, we will see that the quantization scale is actually much finer then the Plank Scale of distance and time, according to EMQG theory.). Remember that all the cells in the universe are all updated in one single 'clock' cycle! This is a massive computation indeed! The number of CA 'clock' pulses that occur in one of our seconds is a phenomenal $10^{43}$ clock cycles per second (based on the assumption that the Plank distance of $1.6 \times 10^{-35}$ meters is the quantization scale of space, and the Plank time of $5.4 \times 10^{-44}$ seconds is the quantization scale of time). Because of the remotely small distance and time scales, we as observers of the universe are very far removed from the low-level workings of the CA.

Why do humans exist at such a large scale compared to the tiny CA scales, as to be completely removed from the CA cell level? The simple answer to this is that life is necessarily complex! Even an atom is remarkably complex object according QED. A lot of storage locations (cells) are required to support the structure of an atom, especially in the light of the complex QED processes going on (ref. 40). It is not possible to assemble anything as complex as life forms without using tremendous numbers of atoms and molecules.

Chaos and complexity theory teaches us that simple rules can lead to enormous levels of complexity. We can see this in a simple example of a 2D geometric CA called the game of life. Being a 2D geometric CA, there are 8 neighbors for each cell, which forms a primitive geometric 2D space. This primitive 2D space can be viewed on a computer screen. Here the rules of life are very simple, and are given below:

**Rules for the famous 2D Geometric CA - Conway's Game of Life:**

(1) If a given cell is in the one state, on the next clock pulse it will stay one when it is surrounded by precisely two or three ones among it's eight neighbors on a square lattice. If the cell is surrounded by more than three neighbors, it will change to zero; if fewer than two neighbors have a one, it changes to zero.

(2) If a given cell is in the zero state, on the next clock pulse: it will change to a one when surrounded by precisely three ones, otherwise for any other combination of neighbor states the cell will remain a zero.

The rules are simple enough for a child to understand, yet the game of life leads to an endless number of different patterns, and to significant complexity. We see gliders, puffers, guns, 'oscillating' particles with different rates of translation and spontaneous particle emission from some oscillating patterns. We have even seen a pattern that resembles a particle exchange process.

How many different 2D geometric cellular automata's can be constructed from all possible rules? This number is unimaginably large. For simple binary cells, with 8 neighboring cells there are 8+1 cells that influence a given cell (previous state of a cell can influence it's next state), which leads to $2^{512}$ possible binary combinations or approximately $10^{154}$ different CA's, of which the game of life is but one. In general, for an Nth Dimensional Geometric CA with (m) neighbors, there are $2^k$ possible rules available for the Cellular Automata, where $k = 2^{(m+1)}$. Assuming our universe is a simple 3D geometric CA, then there are $2^{134,217,728}$ possible rules to choose from! You can give up trying to find the rules that govern our universal CA by simple trial and error.

In the early 1900's Max Plank (ref. 11) defined a set of fundamental scales based on his then newly discovered quantum of energy h (E = h v). Three fundamental constants G (Gravitational Constant), h (Plank's Quantum of Action), and c (the speed of light) were assembled in a set of equations that define natural physical units independent of any man made units. The Plank length is $1.6 \times 10^{-35}$ meters and the Plank time is $5.4 \times 10^{-44}$ seconds. The Plank length has often been suggested as the fundamental quantization scale of our universe. This suggests that the 'size' of a cell in the CA computer is the Plank length, and that the Plank time is the period of the cellular automata 'clock' (the smallest possible time period for a change in the state of the CA). Note: The cells actually have no real size, since they represent storage locations for numbers. Instead, this represents the smallest distance that you can increment as you move along a ruler. We will see that the actual quantization scale of the CA is much finer than the Plank Scale.

Even in conventional modern physics, there is growing evidence accumulating that suggests that the plank units of distance and time somehow represent the quantization scale of space-time itself. In CA theory, quantization is automatic! We will see (EMQG section) that the quantization scale of the universal CA is much finer than the Plank Scale. There are also some other physical units that can be derived from the three fundamental constants listed above which includes: Plank Energy, Plank Temperature, Plank Mass, Plank Speed, and Plank Wavelength, which also represent fundamental universal limits to these parameters.

All physical things are the result of CA processes; including space, time, forces and matter (particles). Although, the exact rules of the CA that is our universe is unknown at this time, some very general physical conclusions can be drawn from the CA model. For example, matter is constructed from elementary particles, which move in space during some period of time. Particles interact via forces (exchange particles) which bind particles together to form atoms and molecules. Quantum field theory tells us that forces result from particles called vector bosons, which are readily exchanged between charged particles, and that these exchanges cause momentum changes and accelerations that we interpret as forces. Elementary particles and forces on the CA consist of oscillating information patterns, which are numbers changing state dynamically in the cells of the CA. These numerical information patterns roam around from cell to cell in given directions.

The shifting rates of a particle or information pattern, relative to some other particle, is interpreted by us as the state of relative motion of the particle. As we have seen, particles can interact by exchanging particles, which are also information patterns. Exchange particles are readily emitted at a given fixed rate by the source particle, and absorbed by target particle. When these force particles are absorbed, the internal state changes, which we interpret as a change in the particle momentum. The result of this process is that the shifting rates or motion of the matter particle changes by undergoing a positive or negative acceleration with respect to the source. This is what we observe as a fundamental unit of force. Of course, when we observe forces on the classical scale, the astronomical number of particle exchanges occurring per second blurs the 'digital' impact nature of the force exchange, and we perceive a smooth force reaction. All 'motion' is relative in CA theory, since all cells are identical and indistinguishable. In other words, we cannot know the specific cell locations that a particle occupies.

The closest CA elements that correspond to our space and time are the empty cells and the clock cycles that elapse. But this correspondence is not exact, as we shall find. The cells, which are storage locations for numbers, really form a low-level basis of the physical concept of space. Because the information patterns can roam freely in various directions that are determined by the dimensionality of the CA, we interpret this freedom of motion as space. Similarly, while matter patterns are in motion, a definite time period elapses. We can only sense the elapse of time when matter is in motion, by the changing state of the CA. The ultimate cause of change is the CA clock, and the common rules that govern the cells. But, it is important to realize that the internal clock required for the CA to function is not the same as our measure of time in our universe. Our time is based on physical phenomena only. This fact is the origin of much confusion on the nature of time in physics. We will investigate this important concept later.

From these and other considerations, CA theory restores a great unity to all of physics. Where there used to be different phenomena described by different physical theories, now there is only one theory. Furthermore, CA theory is not only able to describe the way the universe works, but it also allows us to understand *how* it works in detail. Is there any evidence in the current laws of physics to support the idea that our universe really is a Cellular Automata computer simulation? The following sections will provide some rather speculative and sometimes circumstantial evidence to support this position. First we introduce another important player in CA theory and EMQG, the virtual particles of the quantum vacuum.

## 3. THE QUANTUM VACUUM AND IT'S RELATIONSHIP TO CA THEORY

*Philosophers:     "Nature abhors a vacuum."*

By definition of the word 'vacuum', the vacuum is supposed to be totally empty. One makes a vacuum by removing all the gases and matter from a container. After this is done, you might conclude the vacuum is empty. In fact, the vacuum would be far from empty! As we said, in order to make a complete vacuum one must remove all matter from an enclosure. However, this is still not good enough. One must not forget to lower the temperature down to absolute zero. This is required in order to remove all thermal electromagnetic radiation, which come in the form of photon particles that would spoil the perfect vacuum.

However, Nernst correctly deduced in 1916 (ref. 23) that empty space is still not completely devoid of all radiation after this is done. He predicted that the vacuum is still permanently filled with an electromagnetic field propagating at the speed of light, called the zero-point fluctuations (or sometimes called vacuum fluctuations). This was later confirmed by the full quantum field theory developed in the 1920's and 30's. Later, with the development of QED (ref. 40), it was realized that all quantum fields should contribute to the vacuum state, like virtual electrons and positron particles, for example. Of course Newton was totally unaware of the nature of the quantum vacuum in his day.

Modern quantum field theory actually requires that the perfect vacuum to be teeming with activity, as all types of quantum virtual fermion (matter) particles and virtual bosons (force particles) from the various quantum fields appear and disappear spontaneously. These particles are called 'virtual' particles because they result from quantum processes that have short lifetimes, and are generally undetectable. One way to look at the existence of the quantum vacuum is to consider that quantum theory forbids the absence of motion, as well as the absence of propagating fields (exchange particles).

According to QED, the quantum vacuum consists of the virtual particle pair creation/annihilation processes (for example, electron-positron pairs), and the zero-point-fluctuation (ZPF) of the electromagnetic field (virtual photons). The existence of virtual particles of the quantum vacuum is essential to understanding the famous Casimir effect (ref. 12), an effect predicted theoretically by the Dutch scientist Hendrik Casimir in 1948. The Casimir effect refers to the tiny attractive force that occurs between two neutral metal plates suspended in a vacuum. He predicted theoretically that the force 'F' per unit area 'A' for plate separation D is given by:

$$F/A = -\pi^2 h c /(240 D^4) \quad \text{Newton's per square meter} \quad \text{(Casimir Force 'F')} \quad (3.1)$$

This minute force can be traced to the disruption of the normal quantum vacuum virtual photon distribution between two nearby metallic plates. Certain photon wavelengths (and therefore energies) in the low wavelength range are not allowed between the plates, because these waves do not 'fit'. This creates a negative pressure due to the unequal energy distribution of virtual photons inside the plates as compared to outside the plate region. The pressure imbalance can be visualized as causing the two plates to be drawn together by radiation pressure. Note that even in the vacuum state, virtual photons carry energy and momentum.

Recently, Lamoreaux made (ref. 13) accurate measurements for the first time on the theoretical Casimir force existing between two gold-coated quartz surfaces that were spaced 0.75 micrometers apart. Lamoreaux found a force value of about 1 billionth of a Newton, agreeing with the Casimir theory to within an accuracy of about 5%.

The virtual particles of the quantum vacuum is central to our understanding of the quantum processes behind Newton's laws. We therefore present a brief review of some of the theoretical and experimental evidence for the existence of the virtual particles of the quantum vacuum:

(1) The extreme precision in the theoretical calculations of the hyper-fine structure of the energy levels of the hydrogen atom, and the anomalous magnetic moment of the electron and muon; these calculations are based on the existence of virtual particles within the framework of QED theory (ref. 40). These effects have been calculated to a very high precision (approx. 10 decimal places), and these values have also been verified experimentally to an unprecedented accuracy. This is a great achievement for QED, which is essentially a perturbation theory of the electromagnetic quantum vacuum. Indeed, this is one of greatest achievements of all theoretical physics.

(2) Recently, vacuum polarization (the polarization or rearrangement of electron-positron pairs near a real electron particle) has been observed experimentally by a team of physicists led by David Koltick (ref. 24). Vacuum polarization causes a cloud of virtual particles to form around the electron in such a way as to produce an electron charge screening effect. This is because virtual positrons tend to migrate towards the real electron, and the virtual electrons tend to migrate away. A team of physicists fired high-energy particles at electrons, and found that the charge screening effect of this cloud of virtual particles was reduced the closer a particle penetrated towards the electron. They reported that the effect of the higher charge for the penetration of the electron cloud with energetic 58 giga-electron volt particles was equivalent to a fine structure constant of 1/129.6. This agreed well with their theoretical prediction of 1/128.5 of QED. This can be taken as verification of the vacuum polarization effect predicted by QED, and further evidence for the existence of the quantum vacuum.

(3) The quantum vacuum explains why cooling alone will never freeze liquid helium, no matter how low the temperature. Unless pressure is applied, the quantum vacuum energy fluctuations prevents it's atoms from getting close enough to trigger solidification.

(4) For fluorescent strip lamps, the random energy fluctuations of the virtual particles of the quantum vacuum cause the atoms of mercury (which are in their exited state) to spontaneously emit photons by eventually knocking them out of their unstable energy orbital. In this way, spontaneous emission in an atom can be viewed as being directly caused by the random state of the surrounding quantum vacuum.

(5) In electronics, there is a limit as to how much a radio signal can be amplified. Random noise signals are always added to the original signal. This is due to the presence of the virtual particles of the quantum vacuum as the real radio photons from the transmitter propagate in space. The vacuum fluctuations add a random noise pattern to the signal by slightly modifying the energy content of the propagating radio photons.

(6) Recent theoretical and experimental work done in the field of Cavity Quantum Electrodynamics (the observation of exited atoms surrounded by a conducting cavity) suggests that the orbital electron transition time for excited atoms can be affected by the state of the virtual particles of the quantum vacuum immediately surrounding the excited atom in a cavity, where the size of the cavity modifies the spectrum of the virtual particles.

In the weight of all this evidence, we believe that today very few physicists doubt the existence of the virtual particles of the quantum vacuum. Yet, to us, it seems strange that the quantum vacuum should barely reveal it's presence. We feel that is strange that we only know about the vacuum's existence through some rather obscure and hard to measure physical effects. After all, the observable particles of ordinary real matter constitutes a minute fraction of the total population of virtual particles of the quantum vacuum at any given instant of time. In other words if we could observe the detailed numeric cellular automaton operation of our universe, the interactions of the virtual particles of the quantum vacuum are, by far, the most common process we would find. A real, ordinary matter particle would be extremely hard to find in this mess. In fact, it would be extremely difficult to follow a real matter process such as a hydrogen atom, since the electron following it's orbital around the nucleus would be constantly destroyed and recreated by countless interactions with the virtual positrons of the quantum vacuum.

Instead, we believe that the quantum vacuum plays a ***much more prominent role*** in all of physics then is currently believed. We maintain that the effects of the quantum vacuum are felt in *virtually all* physical activities. We believe that inertia is completely due to the electrical force effects between real matter and the electrically charged virtual particles of the quantum vacuum. In fact, Newton's three laws of motion can be understood to originate directly from the force effects due to the presence of electrically charged virtual particles of the quantum vacuum (ref. 35).

Is there any relationships that exist between Cellular Automata model and the existence of the quantum vacuum? Recall that the quantum vacuum implies that almost *all* of space is filled with virtual particle processes. Through studies of simple 2D geometric CA's (such as the Conway's Game of Life), we found that most (random) initial numeric states or 'seed' patterns in the cells (often from small localized initial patterns, with all the other remaining cells being in the zero state) are observed to evolve into a complex soup of numeric activity, everywhere. This activity in the 2D CA is very reminiscent of the virtual activities believed to be happening in the quantum vacuum. In the game of life you can even see events that even look suspiciously like random 'particle' collisions, particle annihilation, and particle creation after a sufficiently long simulations. Of course this is not hard evidence that our universe is a vast CA, but it is highly suggestive. We now present some other (somewhat) circumstantial evidence to support the view that our universe is a vast CA simulation.

## 4. EVIDENCE FOR BELIEVING THAT WE DO LIVE INSIDE A VAST CA

Is there any evidence, direct or indirect, that we are indeed living in a universe that is a vast mathematical Cellular Automaton simulation? Here we list some of the important circumstantial evidence for believing that this is true. Note that much of this material is speculative in nature.

## 4(A) THE BIG BANG (START OF SIMULATION AT T=0) AND CA THEORY

*"I want to know how God created this world (Universe)"*   *- A. Einstein*

If the universe is a vast CA computer simulation, then it stands to reason there must have been a point where the simulation was first started. We believe that this occurred ≈15 billion years ago (our time), in accordance to the standard, hot big bang theory. We like to refer to this creation event as the 'GREAT INFORMATION EXPLOSION'. The event is shown in a highly schematic form in figure 12. It is important to realize that the creation of the numeric state of our universe, if it were to be done *now* in a single step or act of creation with out evolution, would be *very* difficult to accomplish. All the galaxies, stars, and planets, and the wide variety of life forms must be specified for all the cell states in the cellular automaton, which for our universe is something on the order of $10^{100}$ cells per cubic meter of space!

We currently believe that our universe contains on the order of a few billion galaxies, and many of these galaxies have something on the order of 100 billion stars in it. Currently, there is also evidence for the possibility that a certain percentage of these stars have one or more planets circling around them. Each star and planet has it's own unique orbit, chemical composition, temperature, rotation rate, size, atmosphere, landscape and possibly even life forms. In the process of creating our universe, it is far more economical to start with just the "right" rules of the cellular automaton so that complexity can naturally arise. In this way, stars and planets (possibly the evolution of life itself) are the natural byproducts of the evolution of the CA over a vast number of clock cycles.

Basically you must start with the right CA model, the right mathematical rules, and the right initial cell patterns; and then let the natural evolution of the CA run its due course. It is also more "interesting" to start this process, and than "see" what comes out of it after a lot of computation. In fact, that is what the purpose of the computer is anyway. The purpose of our CA computer universe is to compute our particular universe! To summarize, CA theory *absolutely requires* that our universe be an evolutionary process, to be fully deterministic, and to start with a fairly simple beginning state. Of course many cells and clock cycles would be required to produce any significant complexity, but that is just what has happened in our universe. With $10^{100}$ cells per cubic meter of space, and $10^{43}$ 'clock cycles' per second of our measure of time and therefore 10 billion years of operation, we believe that almost anything is possible from a simulation this large.

Computation is obviously a mathematical operation based on logical operations. Is there evidence that our universe is mathematical in nature? There is plenty of evidence for this, which we explore next.

## 4(B)    WHY OUR UNIVERSE IS MATHEMATICAL IN NATURE

*"Why is it possible that mathematics, a product of human thought that is independent of experience, fits so excellently the objects of physical reality."*
                                                                                                                                  *- Albert Einstein*

It is clear to most physicists that all the known laws of physics are mathematical in nature. Many physicists like Einstein for example, have commented on this mysterious fact as the quote above signifies. Yet no good explanation has been given as to why this should be so for our universe. This fact is made even more mysterious when one considers that mathematics is strictly an invention or byproduct of intellectual activity.

In some sense mathematics is like art and music. For example, the mathematical concepts of infinity, the imaginary numbers, and the Mandelbrot set in the complex plane are all mathematical objects that are invented by mathematicians. In mathematics, you start with virtually any set of self-consistent axioms, and formulate new mathematics based on these postulates. Mathematics is strictly a *creative* process. Yet our universe definitely operates in a mathematical way. Every successful physical theory has been formulated in the language of pure mathematics, and a good theory can even predict new phenomena that was not expected from the original premises of the theory.

If the universe is a cellular automaton, then there is a clear explanation as to why the universe is mathematical in nature. Quite simply put, everything in our universe is numerical information, which is governed by mathematical rules that specify how the numbers change as the computation progresses. In short, "***the universe is numbers***"**,** as was once proclaimed by the great Greek philosopher and mathematician Pythagoras. The design of the cellular automaton must have required intelligence, which was applied to the cellular automaton in the form of the mathematical rules for the cells. CA theory claims that all the laws of physics that we know today are mathematical descriptions of the underling, discrete mathematical nature of the numeric patterns that are present in our universal cellular automaton. Fredkin (ref. 2) once proposed that the universe should be modeled with a single set of cellular automaton rules, which will model all of microscopic physics exactly. He called this CA 'Digital Mechanics'. The laws of physics in this form are discrete and deterministic, and would replace the existing differential equations (based on the space-time continuum) for modeling all phenomena in physics.

In the rush to discover the theory of everything (or simply *The Theory* as it is now known) we should not be looking for a final set of partial differential equations which describes everything. Instead, we should be looking for the correct structure of the universal CA, and the corresponding set of logical rules that govern it's operation, and the initial state.

## 4(C) WHY IS THE LIGHT SPEED THE MAXIMUM SPEED YOU CAN GO?

*"Experiment has provided numerous facts justifying the following generalization: absolute motion of matter, or, to be more precise, the relative motion of weighable matter and ether, cannot be disclosed. All that can be done is to reveal the motion of weighable matter with respect to weighable matter…"*
*H. Poincare (1895)*

Special Relativity theory is founded on Newton's laws and these two basic postulates:

***(1) The velocity of light in a vacuum is constant and is equal for all observers in inertial frames (An inertial frame is one in which Newton's law of inertia is obeyed).***

***(2) The laws of physics are equally valid in all inertial reference frames.***

The special theory of relativity implies that the speed of light is the limiting speed for any from of motion in our universe. Furthermore, light speed appears constant no matter what inertial frame an observer chooses. However, nowhere in special relativity theory or any other theory that we are aware of, is there an explanation as to why this might be so. It is simply a postulate, based on physical observations such as the null results from the Michelson-Morley experiment. The second postulate also implies that there are no experiments that can be performed that will reveal which observer is in a state of 'absolute rest'. The second law alone is an alternative way of stating the Galilean transformations, which are an extension of Newton's laws of motion. With the addition of postulate 1, we are led to the Lorentz Transformations, which replace the Galilean transformations.

The second postulate of special relativity states that the laws of physics are equally valid in all inertial reference frames. Stated in a weaker form, there are no preferred reference frames to judge absolute constant velocity motion (called inertial frames). This latter form is easily explained in CA theory, by remembering that all cells and their corresponding rules in the cellular automata are absolutely identical everywhere. Motion itself is an illusion, and really represents numeric information transfers from cell to cell. To assign meaning to motion in a CA, one must relate information pattern flows from one numeric pattern group with respect to another group, since the actual cell locations are totally inaccessible by experiment. Therefore motion requires reference frames. Unless you have access to the absolute location of the cells, all motion must remain relative in CA theory. In other words, there is *no* reference frame that is accessible by *experiment* that can be considered as the absolute reference frame for constant velocity motion.

In a cellular automaton, the clock rate specifies the time interval in which all the cells are updated, and acts as the synchronizing agent for the cells. Matter is known to consist of atoms and molecules, which themselves consist of elementary particles bound together by forces. An elementary particle in motion is represented in CA theory by a shifting numeric information pattern, that is free to 'roam' from cell to cell. Recall that space consists of cells or storage locations for numbers in the cellular automata, and particles (number patterns) freely 'move' in this cell space.

From these simple ideas, we will show that there must be a maximum rate that numeric information patterns are able to achieve on a CA. This is due to the following two reasons:

- First, there is fixed, constant rate in which cells can change state.
- Secondly, information can only be transferred sequentially, from one cell to adjacent cell, and the fastest rate is only one cell at every 'clock cycle' (figure 9).

These are simply due to the limitations of the structure of cellular automata computer model. Recall that the CA provides the most massively parallel computer model known. It is the CA's high degree of parallelism that is responsible for these limitation, because a particular cell state can only be affected by its immediate neighbors. Information can only evolve after each 'clock' period, and numeric information patterns can only arrive at a distant location by shifting from cell to adjacent cell. This result in a *definite **maximum speed limit*** for transfer of information patterns (particles) on the CA This maximum speed limit might represents light velocity (figure 9), which is the fastest speed any particle can go. In this way, we can view light motion as being a messenger from the absolute and regular world of the cells and 'clock cycles' existing on the CA. In fact, light velocity can be expressed in two different sets of measurement units:

(1) Light has a velocity measured in ordinary physical units of 299 792 458 meters per second. The meter and the second can be measured with ordinary clocks and rulers. The meter and second do not correspond to fundamental features of our universe.
(2) Light has a velocity specified in absolute CA units of 1 cell per 'clock cycle' or 1 PVU where PVU is a Plank Velocity Unit. The number of cells separating two cell locations on the CA is an abstract measure of distance. Similarly, the number of elapsed 'clock cycles' between two CA events is an abstract measure of the elapsed time. Cells and 'clock cycles' are fundamental aspects of our universe.

***NOTE: According to EMQG this maximum speed limit is actually the 'low-level' light velocity, defined as the velocity of photons in between scattering encounters with the***

*virtual particles, measured in the absolute CA units of measure (cells and 'clock cycles'). The electrical scattering process with the charged virtual particles of the quantum vacuum reduces the 'low-level' light velocity to the value we observe for light in the vacuum. This process is somewhat like how transparent water reduces the velocity of light in the vacuum, which is responsible for the index of refraction of water. Section 18 describes the details of this process.*

This maximum speed limit can be calculated if the precise quantization scale of space and time on the cellular automata level is known. Let us assume for now that the quantization of space and time corresponds *exactly* to the plank distance and the time scales (not true in EMQG theory). This means that the shifting of one cell represents a change of one fundamental plank distance $L_P$: $1.6 \times 10^{-35}$ meters, and that the time required for the shift of one cell is one fundamental plank time $T_P$: of $5.4 \times 10^{-44}$ seconds. Let us further assume that a photon represents the fastest of all the information patterns that shifts around in the CA. In fact, we propose that the photon information pattern is *only* capable of shifting one cell per clock period, and not at any other rate, and therefore exits at one speed with respect to the cells. The value for the speed of light can then be derived simply as the ratio of (our) distance over (our) time for the information pattern transfer rate. The maximum information transfer velocity is thus:

$$V_P = L_P / T_P = 3 \times 10^8 \text{ meter/sec} = c \qquad (4.1)$$

Therefore $V_P$ is equal to 'c', the speed of light. The velocity of light can also be expressed as one plank velocity, which is defined in units of plank length divided by plank time. (There are plank units for mass, temperature, energy, etc. as detailed in ref. 11). Thus the fastest rate that the photon can move (shift) is an increment of one cell's distance, for every 'clock cycle'.

## 5. QUANTUM MECHANICS, PARTICLE PHYSICS AND CA THEORY

*"By getting to smaller and smaller units, we do not come to fundamental units, or indivisible units, but we do come to a point where division has no meaning."*

- *Werner Heisenberg*

In the ultimate quest for the fundamental structure of matter, we have discovered that matter can be broken down to smaller and smaller components. This quest started with the ancient Greeks, who postulated that matter was ultimately composed of indivisible objects called atoms. Later it was discovered that atoms have a structure, in the form of a nucleus surrounded by a cloud of electrons (leptons). The nucleus was found to consist of neutrons and protons in the early 20th century. More recently in the 1960's it was discovered that neutrons and protons are composed of a specific combinations of quarks.

All this has now been well established theoretically and experimentally in the form of the highly successful 'Standard Model' of particle physics. The Standard model describes these levels of structure of matter in detail along, with the corresponding force exchange particles or 'bosons'. Currently we have penetrated four levels of structure in the composition of matter. Are there any new layers to be peeled off? The standard model has many limitations, which are well known (ref. 37). For example, the standard model proposes 18 varieties of quarks and six lepton varieties, and a dozen force carrying particles (if you include there anti-particles). It is hard to believe that so many 'elementary' particles are required to build the universe. Furthermore, of the three families of elementary particle specified by the standard model, only the first family is actually required to build ordinary matter. The other two families only show themselves in sophisticated particle collision experiments, and seem to contribute nothing to the structure of the universe.

Furthermore there are a large number of 'adjustable' fundamental constants that are determined experimentally in the standard model, which cannot be traced to deeper theoretical principles. One of the biggest limitations of the standard model is it's inability to include gravity in a common framework, based on the graviton particle exchange process. For an excellent review of the problems with the Standard Model refer to ref. 44. This has led many physicist to propose alternative unification schemes that include gravity, and that involve a lot fewer particles than those of the standard model. Yet in spite of this, it should be noted that there (currently) exists no firm evidence for the existence of any internal structure for leptons (electrons, etc.) and for the quarks. The lack of experimental evidence is no doubt due to our current inability to probe very small distance scales with our current generation of particle accelerators.

The central problem that is addressed by the standard model is; what is an elementary particle and what are the forces that the particle feels? A particle physicist once remarked that elementary particles behave more like mathematical objects than like familiar point-like objects such as grains of sand, for example. It is well known result in particle physics that elementary particles are able to transform from one species of particle to another. For example, an unbound neutron can transform (or decay) into a proton and a neutrino. An electron and positron particle (considered elementary in the standard model) can combine, annihilate, and produce an energetic photon. At times elementary particles seem to be spread-out in some sort of oscillatory wave function, and at other times they behave like point-like objects. Virtual particles can be spontaneously created and readily annihilated in the vacuum. No material grain of sand (or any other 'mechanical' model) have properties like these.

None of these processes seem familiar, when based on our everyday experience about the way that ordinary objects behave. Related to the question of particles and forces is the question; what is motion? Motion must be defined in terms of space and time, i.e. space divided by time. What is space and time? Therefore, to understand the fundamental nature of a particle definitely requires an understanding of space and time. Einstein was once asked what question would he pose to the 'Creator', if could only ask him a single question. Einstein replied that it would sufficient to ask the Creator *'what is an electron'*.

According to cellular automata theory, elementary particles must be numeric information patterns that 'shift' (shifting replacing the concept of motion) around from cell to cell in a kind of a quantized CA 'cell space'. CA theory requires that everything results from the dynamic information on the CA, which changes state according to some specific set of mathematical rules. Recall that the familiar notions of space, time, and matter do *not* actually represent basic elements of reality, and that all physical phenomena turn out to be the end result of a vast amount of numerical information being processed. All elements of reality are due to numerical information being processed on the 'cells' at incredibly 'high speeds' (with respect to our measure of time). How does an elementary particle emerge from this sort of activity?

Incorporated in the standard model is (relativistic) quantum theory, which states that all elementary particles have wave functions. When a particle is detected it appears as a point-like interaction, usually a change in the energy level of an electron in some atom, which gives off a photon (or possibly an outer electron being totally ejected from the atom, thus ionizing the atom). When the particle is not disturbed there is much evidence to support the idea that it exists as some sort of oscillating wave, with a specific wavelength called the 'De Broglie' wavelength: $\gamma = h/(mv)$, where h is plank's constant, m is the particle mass, and v is ***the relative velocity*** of the particle. To our knowledge, nobody has actually 'observed' the wave function directly, although there is good indirect evidence available for it's existence (Note: Some physicist look upon the wave function as a statistical prescription for determining where a particle is, and they believe that the wave function says nothing about the actual characteristics of a single particle. We will proceed with the view that the wave function somehow represents a real characteristic of a single particle.)

Part of the reason for our lack of ability to directly observe the wave function is that the amplitude of the wave function physically represents only the probability of finding a point-like interaction at a certain place and time. This probabilistic nature of matter has led some physicists (like Einstein) to question the completeness of quantum theory. From the perspective of CA theory, it is very obvious that the probabilistic nature of matter particles represents our ignorance of the exact numeric state of the matter particle and it's immediate surroundings.

It can be shown that if the universe were truly random at the numeric cellular level, then a computing device would take an infinite number of operations to reproduce perfect random behavior. From CA theory, it is readily apparent that quantum wave function is *not complete* as a total description of the physical reality of the particle. Instead, the probabilities we measure must represent our ignorance of the exact numerical state of the CA. Furthermore, we only have information patterns available to us to probe other numeric information patterns. What we learn from this process is only how the particles interact, and from this sort of behavior we have to infer the particle's fundamental nature.

The difficulty in determining the exact nature of a particle arises because we are incapable of 'measuring' the exact (numeric) state of a particle. To do this requires the ability to read out the numeric contents of the cells of the CA that contain the particle pattern. Therefore in CA theory, the wave function must be incomplete as the full description of the exact state of an elementary particle. When we bear in mind that particles exist on the Cellular Automata, we can understand why Heisenberg was really correct to say (in the quote at the beginning of this section) that *'By getting to smaller and smaller units, we do not come to fundamental units, or indivisible units, but we **do** come to a point where division has no meaning'*.

5.1 TWO TENTATIVE CA COMPATIBLE MODELS FOR THE WAVEFUNCTION

Currently, we have done some preliminary work on two possible models of the quantum wave function of a non-relativistic quantum particle with a DeBroglie wavelength given by $\gamma = h/(mv)$. These two tentative models for the CA-based quantum wave function are presented below. Both models are rather speculative, and both have problems. We present these models here in order to stimulate interest in discovering the nature of the quantum wave function on the cellular automaton.

**(1)     THE DIRECT OSCILLATING NUMERIC WAVE FUNCTION MODEL**

In our first approach, the wave function somehow represents the *'oscillation'* of the CA numeric information pattern directly on the CA (numeric patterns that vary periodically, and return to the original pattern after a fixed number of clock cycles). In this model the wavelength of the wave function corresponds directly to the wavelength of the periodic, fluctuating numeric information wave pattern existing on the CA cells. In other words, the wave function is a coherent, and a purely numeric pattern that represents the particle in absolute CA space and time.

Oscillating numeric patterns are very common in the game of life (an interesting, simple geometric 2D CA), and can also be found in many other CA models. In fact according to the Wolfram classification scheme for cellular automata, all CA's can be divided roughly into 4 classes (ref. 6), with Class II CA being cellular automata that settle down into isolated periodic structures. A Class I CA settles down to a constant field, Class III to a uniformly chaotic field, and Class IV to isolated structures showing complicated internal behavior.

This model readily explains why the wave function has ***not*** been detected directly by experiment (diffraction and interference experiments aside, since they are indirect observations over many experimental runs). It is obvious that no one can detect the presence of the numeric oscillating cells directly through experiment, since the numbers cannot be 'read' out from the cells. Only the effects of one particle against another can be perceived. This model also explains the probabilistic nature of the wave function. Since we cannot know the numeric state of an information pattern at a given time, we cannot predict the exact future evolution of a particle pattern and it's interactions mathematically, even though we believe that we know all there is to know about the particle from the use of our experimental apparatus.

The wavelength of wave function is given by the DeBroglie wavelength 'h/(mv)', where h is a constant, 'm' represents the mass of the particles, and 'v' the velocity. However, *what reference frame* is used to gauge the velocity 'v'? According to special relativity, the velocity must be the *relative* velocity between the observer and the quantum particle, since absolute velocity does not exist. Therefore the state of relative motion of the observer is extremely important to the value of the De Broglie wave length. Whether the particle moves, or the observer moves makes no difference in relativity, it is only the relative velocity that matters.

How can the velocity of the observer affect the wavelength of the oscillation of a numeric information pattern on the CA? The answer is surprisingly simple. The observer ***cannot*** *affect the wavelength*! The wavelength of the wave function that a real observer 'sees' is a ***Doppler shifted wavelength of an absolute wavelength already existing on the CA,*** that depends on the relative velocity between the observer and the particle. Recall that there are two different sets of space and time measurement units in the CA, the observable relativistic space-time and the absolute cells and 'clock cycles' of the CA. We take the position that a particle has an absolute wavelength, meaning that the wave contains a specific number of cells which separate the numeric oscillation cycles. Therefore if you move towards a particle at a high velocity, the particles wavelength does *not* actually decrease as you approach it. It's wavelength remains constant on the CA, which is determined by the internal (energy) characteristic of the particle. The real observer sees this wavelength as a blue Doppler shifted wave (compared to when the relative velocity is zero).

There are some difficulties with this model. It is not clear how the mass of the particle enters into this picture. Increasing mass causes decreasing wavelength. Also in simple 2D CA models like the game of life, it is very rare to find large coherent patterns on the CA. However a very low energy photon, like a radio photon (somewhere in the AM radio band for example), has a wavelength the size of a house! How can this huge number of cells oscillate coherently on the CA, especially recalling that there are at least $10^{100}$ cells in a cubic meter of space, and that a cell can only communicate with it's neighbor cells? Furthermore, how does a wave function of this size collapse (apparently instantaneously) when an observation of the radio photon is made?

Related to this question is the infamous and well known non-locality problem of quantum mechanics. Apparently two quantum entangled pairs of photons are correlated over great distances, where an observation of one photon causes the other to take on the same state (instantaneously) in spite of the vast distances that might separate them. In fact, the choice of what state to measure for one photon can even be delayed while the other has already been in flight half way across the universe!

Einstein posed this question in his famous EPR thought experiment, stating that quantum mechanics was incompatible with the basic premises of special relativity because the particles would have to have the capability of faster than light communication. Later this clash between classical theories and quantum theory was put in mathematical form by Bell in the 60's through his famous EPR inequality. Later still, experiments by Aspect and others in the 80's seem to confirm that it is the quantum mechanical picture that is correct. The question still remains (if you assume that the wave functions are entangled), how can these two entangled particles communicate information about their mutual state faster than light?

*Note: We will see that in EMQG theory the <u>measured light speed</u> does <u>not</u> correspond to the low-level speed of a photon in between the charged virtual particle scattering processes. Therefore, the speed that numerical information patterns can evolve on the CA is <u>much faster</u> than the speed of light (by some large but unknown amount). Recall that the low-level light velocity represents the simple shifting of information from cell to adjacent cell in every clock cycle, which is the maximum speed possible for information changes (section 7). Therefore, if there is an actual wave function collapse (this is still disputed), then the collapse can happen at speeds much greater than light velocity since the collapse could presumably work on the numeric level. This is because numeric information can be processed much faster than the measured light velocity (the measured light speed depends on the exact nature of photon scattering, and therefore depends on the index of refraction of the quantum vacuum). Therefore the wave function collapse (if is truly real) can progress at speeds much faster than light speed, but definitely not infinitely fast.*

### (2) THE QUANTUM VACUUM WAVE-LIKE DISTURBANCE MODEL

Our second approach utilizes the idea that since **inertial mass 'm'** is involved in the DeBroglie wavelength 'h/(mv)', than there exists the possibility that the DeBroglie wavelength may result from interactions of the quantum mass particle with the immediate surrounding virtual particles of the quantum vacuum. The virtual particles are an important player in quantum inertia theory, and in the concept of inertia mass (section 8). In this model the wave function results from interactions of a real electrically charged particle with the electrically charged virtual particles of the quantum vacuum, causing some wave-like disturbance of the immediate surrounding virtual particles of the quantum vacuum. According to Quantum Inertia (section 8), inertial mass results from the local electrical interactions of the charged quantum particles that make up a mass with the electrically charged virtual fermions of the quantum vacuum. Therefore, the wave function might result from a kind of 'bow wave' caused by a small 'point-like' (small information pattern) in motion through the surrounding 'sea' of virtual particles in the quantum vacuum. An analogy of this process is motor boat causing a bow wave when it moves along the surface of the water, where the vacuum acts like a fluid that is disturbed by the motion of the charged particle.

The motion of a 'point-like' particle might somehow induce a periodic oscillation of the immediate virtual particles of the quantum vacuum in some unknown fashion (For inertial mass, coherent forces occur only when acceleration is involved, and <u>not</u> for constant velocity motion). In this way, the quantum wave function represents the periodic fluctuation of the virtual particles of the quantum vacuum near a 'point-like' particle. This model also readily explains the probabilistic nature of the wave function. This results from our inability to measure the exact state and interactions of the vast numbers of charged virtual particles of the quantum vacuum that would be involved in such a process.

This model is somewhat reminiscent of the DeBroglie / Bohm pilot wave theory, where a point-particle is guided by an unobservable quantum wave. Here we are substituting the unobservable quantum wave in this model with the 'wave-like' disturbance of the virtual particles. Like in the Bohm model, the disrupted virtual particles (being electrically charged) can act back on the electrical charged point-like particle that originated the wave disturbance in the first place, and the vacuum helps *guide* it's motion. In this way, a double slit will cause interference of the virtual particle wave structure, which in turn influence the location of where the 'point-like' information pattern can be found.

This approach to quantum theory also has it's difficulties. For example, it can be shown the Bohm quantum pilot wave that guides the particle has non-local aspects to it, which are hard to resolve in a wave-like disturbance model proposed here.

5.2   INDISTINGUISHABILITY OF PARTICLES IN THE SAME QUANTUM STATE

One of the most unfamiliar particle attributes of elementary particles is quantum indistinguishability. Quantum mechanics teaches us that electrons in the same quantum state (or in other words having the same quantum numbers) are absolutely identical, and indistinguishable from each other. You cannot mark one electron so that it is different than another. An electron is currently described by quantum mechanics as a particle with quantum numbers like: mass, charge, spin, position, and momentum, which are represented as numbers in the wave function description of the electron. It is these properties alone that specify all there is to know about the electron. The electron has no size or shape.

Equality is strictly a mathematical concept. In mathematics, the equality 1+1=2 is an exact relationship. In classical physics, no two marbles can be constructed to be exactly the same. When it comes to elementary particles, however, two quantum particles can be **exactly** the same. According to quantum mechanics, two electrons in the same state of motion (and spin) are absolutely identical and indistinguishable. The cellular automata model can explain this remarkable fact by stating that the two electrons in the same state have *exactly* the same numeric information pattern, and thus described by the same quantum wave function. Therefore, they are truly, mathematically identical. In constructing a universe, it is actually very desirable to have building blocks that are identical, and exactly repeatable so that large, complex structures such as ourselves can be easily formed from the known repeatable sub-units or particles.

We now shift our focus from the small picture, and take a long look at the big picture; namely the nature of space and time on the CA. We will start with a brief review of Einstein's Special and General theory of Relativity, which teaches us to look at space and time as being relative, united, and forming a relativistic 4D space-time that can be affected by both motion and by gravitational fields.

## 6. EINSTEIN'S SPECIAL AND GENERAL RELATIVITY THEORIES

*" Common sense is the layer of prejudice laid down in the mind prior to the age of 18"*

*- A. Einstein*

If our universe is a vast Cellular Automaton, then all physical phenomena must result from the strictly local interactions inside the CA. The very nature of the cellular automata model is totally forbids any instantaneous action at a distance, since information can only be sent sequentially from cell to adjacent cell in any given direction, only at each and every 'clock cycle'. This means that there can be no action at a distance in all the laws of physics. Einstein was the first to abolish action at a distance in special relativity with his famous velocity of light postulate. He also removed gravitational action at a distance in general relativity, by replacing Newton's instantaneous gravitational force law with space-time curvature.

We generally conclude that *special relativity is already manifestly compatible with the CA model.* We will show in the next section why this is so. Special relativity is one of the most successful theories of physics, and along with quantum theory forms one of the two great pillars of modern physics. However, it has failed to account for *why* the universe has a maximum speed, which has still remained as one of the two postulates of special relativity. CA theory provides a simple explanation for this. In fact, the CA model *demands* that the universe have a maximum speed limit! In addition to this, the second postulate regarding the relativity of inertial frames (constant velocity motion) can also be seen as a simple consequence of the basic structure of the CA.

However, general relativity as it is currently formulated, is *not* compatible with CA theory. First, general relativity is formulated with the classical continuum concept for matter-energy, and is also formulated with a 4D space-time continuum. Both of these fields are not generally compatible with the CA model, or with quantum theory. Secondly, there is no known local action that couples a large mass to the surrounding space-time curvature. What is it about a large mass that causes space-time curvature around it? In general relativity, there exists a global tensor field called the 4D space-time metric, which merely describes the amount of the curved 4D space-time. Because of the relative nature of space-time (observers in free fall near the earth live in flat space-time), it is very difficult to conceive how relativistic 4D space-time can work on a CA. How does the principle of equivalence work on a CA? Why does the inertial mass is equal to the gravitational mass, especially since they are defined differently?

General relativity has failed to make any progress towards the understanding of inertia. Inertia is introduced in general relativity exactly as was conceived by Newton in his famous inertia law: F=MA. Associated with Newton's formulation of inertia are the problems introduced by Mach's principle, which is a loose collections of ideas and paradoxes that have to do with accelerated or rotating motion. Mach argued that motion would appear to be devoid of any meaning in the absence of some surrounding matter, and that the local property of inertia must somehow be a function of the cosmic distribution of all the matter in the universe. Mach's principle has remained as an *untestable* philosophical argument, even within the scope of general relativity.

We have found that general relativity must be revised in order to be compatible with CA theory. These modifications of general relativity came about from a new understanding of inertia and the principle of equivalence. Inertia has been found to be a strictly local quantum process, involving an electromagnetic interaction between the charged particles that make a mass and the charged virtual particles of the vacuum. This model for inertia is manifestly compatible with CA theory, and also automatically resolves Mach's paradox. We call this new theory of inertial 'Quantum Inertia' or QI. This new theory of inertia also explains the origin of the Einstein principle of equivalence, which is not really a fundamental principle of nature, but due to similar quantum processes occurring in accelerated frames and gravitational fields. Therefore the principle of equivalence is also manifestly compatible with Cellular Automata.

To summarize, general relativity is reformulated with a new approach to inertia called Quantum Inertia, which explains the origin of the principle of equivalence, and in a form that is now manifestly compatible with CA theory. This new theory is called 'ElectroMagnetic Quantum Gravity' or EMQG (ref. 1). It is a quantum theory of gravity, because matter is treated as quantum particles, and 4D space-time is quantized and results from pure quantum particle processes. Furthermore, the action between a large mass and the surrounding 4D space-time is clearly understood. Therefore, gravity also turns out to be manifestly compatible with our Cellular Automata principle. We now briefly review relativity theory, totally from the perspective of CA theory. We will keep the mathematical details to a minimum. For a more complete treatment of this subject refer to our paper on Special Relativity and CA (ref. 3).

## 7.    SPECIAL RELATIVITY AND CELLULAR AUTOMATA

**"... space by itself and time by itself, are doomed to fade away into mere shadows ...."**

                                                                                          - H. Minkowski

Special Relativity theory is founded on two important postulates. Let us review these basic postulates and their intimate relationship to CA theory.

**POSTULATES OF SPECIAL RELATIVITY**

**(1) The velocity of light in a vacuum is constant and is equal for all observers in inertial frames (inertial frame is one in which Newton's law of inertia is obeyed).**

**(2) The laws of physics are equally valid in all inertial reference frames.**

These postulates are used by Einstein to derive the famous Lorentz transformations, a set of equations that relate space and time measurements between different observers in different inertial frames. The first postulate is the famous statement of the constancy of light for all observers moving at different constant velocities. Therefore, the motion of a light source does not affect the light velocity. The second postulate implies that there are no absolute reference frames in the universe that can be used to gauge constant velocity motion. All inertial frames are equally valid in describing velocities. In a general sense, all the laws of physics are also equally valid in all inertial frames. Some of the important inescapable consequences of special relativity are:

(1) Comparisons of space and time measurements between observers in different inertial frames are governed by the Lorentz transformations.
(2) The universe is four dimensional, where 3D space and time now have to be united to a common 4D coordinate system.
(3) There is a maximum speed to which matter can obtain.
(4) Mass and energy are interchangeable.
(5) Momentum (and mass) is relative. Mass varies with the relative velocity between two inertial frames.
(6) Spatially separated events that are simultaneous in one inertial frame are not generally simultaneous in another inertial frame.

The special theory of relativity implies that the speed of light is the limiting speed for any from of motion in the universe. Furthermore light speed appears constant no matter what inertial frame an observer chooses. The second postulate also implies that there are no experiments that can be performed that will reveal which observer is in a state of 'absolute rest'.

The second postulate of special relativity states that the laws of physics are equally valid in all inertial reference frames. Stated in a weaker form, there are no preferred reference frames to judge absolute constant velocity motion (or inertial frames). This latter form is easily explained in CA theory, by remembering that all cells and their corresponding rules in the cellular automata are absolutely identical everywhere. Motion itself is an illusion, and really represents information transfers from cell to cell. To assign meaning to motion in a CA, one must relate information pattern flows from one numeric pattern group with respect to another group (the actual cell locations are inaccessible to experiment). Therefore, motion requires reference frames. Unless you are able access the absolute location of the cells, all motion remains relative in CA theory. In other words, there is *no* reference frame accessible by *experiment* that can be considered as the absolute reference frame for constant velocity motion. (The virtual particles of the quantum vacuum still do not allow us to reveal our (constant velocity) motion between two inertial frames. However this is not the case for observers in a state of acceleration, or for observers subjected to gravitational fields).

Since the contents of the cells or their locations are not physically observable to us, they cannot be used to help us setup a universal absolute reference frame for motion. However, there does exist a universal reference frame in the CA, and this frame is completely hidden from experimentation, which we call the 'CA absolute reference frame'. We associate with this frame an **absolute space** and **absolute time**. Everyday objects like this desk, which is a very large collection of elementary particles, occupies a specific volume of cells in CA space. These cell patterns are (most likely) shifting through our cell space at some specific rate. Therefore, there does exist a kind of Newtonian absolute space and absolute time scale, but these are hidden from the viewpoint of an observer living in the CA. We will find that in EQMG theory, the idea of absolute CA space and CA time becomes very important in considerations of inertial and gravitation frames. Even more important to observers, is the state of the virtual particles of the quantum vacuum. These virtual particles can act to produce forces for observers in a state of acceleration. Understanding and resolving the paradoxes introduced in Mach's principle, and understanding Einstein's weak principle of equivalence depends on the existence of virtual particles. We will return to this subject later, when we fully develop EQMG theory.

In order to provide a full explanation for the postulates of special relativity, a detailed model for matter and space is required for the cellular automata theory of our universe. Since this model has not been found yet, we can use some simple assumptions about the nature of matter in a CA. In cellular automata, the clock rate specifies the time interval in which all the cells are updated, and acts as the synchronizing agent for the cells. Matter is known to consist of atoms and molecules, which themselves consist of elementary particles bound together by forces. An elementary particle in motion is represented in CA theory by a shifting numeric information pattern, that is free to 'roam' from cell to cell. Recall that space consists of cells or storage locations for numbers in the cellular automata, and particles (number patterns) freely 'move' in this cell space.

From these simple ideas, it can be seen that there must be a maximum rate that number patterns are able to achieve. This is due to the following two reasons. First, there is fixed, constant rate in which cells can change state. Secondly in CA theory, information can only be transferred sequentially, from one cell to adjacent cell, and only one cell at a time per clock cycle. This is simply a limitation of the structure of the cellular automata computer model. The CA structure provides the most massively parallel computer model known. It is the CA's high degree of parallelism that is responsible for this limitation, because a particular cell state can only be affected by its immediate neighbors. Information can only evolve after each 'clock' period, and information can only shift from cell to adjacent cell. These facts result in a *definite* **maximum speed limit** for information transfers on the CA.

This maximum speed limit might represent light velocity, which is the fastest speed any particle can go. *(NOTE: Later, we shall see that this maximum speed actually represents the raw or 'low-level' light velocity, defined as the velocity of light in between encounters with virtual particles. We will see that the scattering of photons with the virtual particles of the quantum vacuum reduces the speed of the photons to the familiar observable light velocity (figure 6)).*

This maximum speed limit can be calculated if the precise quantization scale of space and time on the cellular automata level is known. Let us assume for now that the quantization of space and time corresponds *exactly* to the plank distance and the time scales. This means that the shifting of one cell represents a change of one fundamental plank distance $L_P$: $1.6 \times 10^{-35}$ meters, and that the time required for the shift of one cell is one fundamental plank time $T_P$: of $5.4 \times 10^{-44}$ seconds (figure 9). Let us further assume that a photon represents the fastest of all the information patterns that shifts around in the CA. In fact, we propose that the photon information pattern is *only* capable of shifting one cell per clock period, and not at any other rate, and therefore exits at one speed with respect to the cells. The value for the speed of light can then be derived simply as the ratio of (our) distance over (our) time for the information pattern transfer rate. The maximum information transfer velocity is thus:

$$V_P = \frac{L_P}{T_P} = 3 \times 10^8 \text{ meter/sec} = c$$

Therefore, $V_P = c$, the speed of light. The velocity of light can also be expressed as one plank velocity, which is defined in units of plank length divided by plank time. (There are plank units for mass, temperature, energy, etc. as detailed in reference 11).

Thus, the fastest rate that the photon can move (shift) is an increment of one cell's distance, for every clock cycle. If two or more clock cycles are required to shift information over one cell, then the velocity of the particle is lower than the speed of light. We now see that light is a messenger from the shadowy CA world, which is isolated from us by tremendously small distance scales of the CA . Photons 'move' in the *simplest possible way* on the CA, and it is this simple motion that is responsible for the rather odd behavior of light from our perspective. The photon patterns always shift over one cell at every 'clock cycle', regardless of the motion of the source. As Einstein pointed out, light is fundamental to our perception of space and time.

To summarize, in cellular automata theory the maximum speed simply represents the *fastest* speed in which the cellular automata can transfer information from place to place. Matter is information in the cellular automata, which occupies the cells. The cells themselves provide a means where information can be stored or transferred, and this concept corresponds to what we call the 'low level' discrete space. 'Low level' time corresponds to the time evolution of the state of the cellular automata, which is governed by the 'clock period'. To put it another way, the rate of transfer of information in any cellular automata is limited, and infinite speeds are simply not possible. Of course, this rules out action at a distance, which is why CA theory is manifestly compatible with special relativity.

In passing, it is interesting to note that in the famous 2D Geometric CA, called Conway's game of life, there exists a stable, coherent 'L' shaped pattern commonly known as a 'glider' pattern. This pattern is always contained in a 3 x 3 cell array, and the glider completes a kind of an internal 'oscillation' in four clock cycles. Thus, in four clock cycles it returns to it's initial 'L' shaped starting pattern. This glider travels in 2D cell space, at *one fixed speed*! It is also the fastest moving pattern known in Conway's game of life. The glider particle in some sense resembles the photon particle in our universe! It has an internal oscillation, and it only moves at one fixed velocity. However, the similarity ends here, because in the game of life, the glider only moves in four fixed directions.

It can be shown that the constancy of light velocity, and the principle of relativity (Einstein's first and second postulate) leads directly to the famous Lorentz Transformations, a set of equations that allows us to relate space and time measurements between two different inertial reference frames.

How can we translate the behavior of photons existing in CA absolute space and separate absolute time, into a statement concerning the *measurable* light velocity? In other words, into a statement based on an inertial observer's actual measurement data for his light velocity measurement. Furthermore, how do we compare these readings with respect to other inertial observers, who also use *actual measuring instruments*? A definition of a space and time measurement must be defined, along with a method of comparing these measurements among different observers in different inertial reference frames. This definition is required, because light velocity is defined as the measured distance that light moves, divided by the measured time that is required to cover this distance.

First we must define an inertial reference frame in far space, away from gravitational fields. Imagine a three dimensional grid of identically constructed clocks, placed at regular intervals measured with a ruler, in the three dimensional space (figure 11). Local observers are stationed at each of the clocks. Thus, the definition of an inertial frame is a whole set of observers uniformly distributed in space as we have described. All observers in a given reference frame agree on the position and the time of some event. Only one observer would actually be close enough to record the event (an event is defined as something that occurs at a single point in space, at a single instant in time). The data collected by all the observers are communicated to the others at a later time (by any means). Notice how light naturally enters in the definition of an inertial reference frame. Light is required by observers to literally 'see' the clock readings (figure 11).

Now we are in a position to evaluate the Lorentz transformations from our low-level CA definition of space, time, and constancy of light velocity. Light is an absolute constant in absolute CA space and time units. No matter what the state of motion of the source, whether it is an inertial source of even if it is accelerated, the light moves as an absolute constant that is unaffected by the source. We must now translate this statement about *measured* light velocity, into the actual reality of the CA with imaginary observers with highly specialized measuring instruments capable of measuring plank distance and time units (which is not possible in our reality). Let us introduce an absolute, discrete (3D space) integer array: $[x(k),y(k),z(k)]$, where information changes state at every $t(k)$. These units represent our absolute space and time measurements (but in practice, we cannot actually make these measurements). The origin is an arbitrarily chosen cell (which can be looked at as being at absolute rest on the CA). A shift of data from one cell in any space direction to next, for example from x(5) to x(6), represents one plank distance unit (pdu), and if this take one clock unit, it happens in one plank time unit (ptu). The velocity of light represents one plank velocity unit (pvu) in our absolute units. We intend to show that when two different inertial observers measure light velocity using *absolute* space and time units, both observers *measure* light velocity as being one plank velocity unit. However, space and time *measurements* between our two inertial observers, do *not* compare to our absolute units. We will show that this is the same situation we find in special relativity, for two observers with *real* measuring instruments in space-time.

Imagine two inertial observers with a relative velocity '$v_r$' in the CA absolute units. Both observers are in a state of constant velocity motion with respect to our absolute cell coordinate system. Observer 'A' contains a green light source and moves with absolute velocity $v_a$ with respect to the cell rest frame. Observer 'B' moves with absolute velocity $v_b$, and is moving away from our observer 'A' (so that $v_b > v_a$), and $v_r = v_b - v_a$. Both observers carry measuring instruments capable of measuring space and time in absolute units. Of course, this is not actually possible with real observers.

Observer 'A' measures the velocity of light of his green light source, with his measuring instruments (figure 10). He uses a ruler of length 'd' in absolute units, and measures the number of CA clock cycles it takes for the wave front of the green light to move the length of the ruler. Because observer 'A' is moving with velocity $v_a$, with respect to the absolute frame, his measurement of length and time are distorted. Recall that light simply shifts from cell to cell, in every clock unit immediately after leaving the source. His measurement of the length that the wave front moving across his ruler appears longer, because of his motion $v_a$. Thus, observer 'A' distance measurement appears longer by: '$d + v_a d$' pdu, where $v_a$ is less than one. (for example, if $v_a$ = ½ pvu, and d=1,000,000 pdu, then the distance measured is 1,000,000 + ½ 1,000,000 pdu). In comparison, an observer at absolute rest would measure a distance of 'd' pdu (figure 10). Similarly, the clock measures a longer time, because it takes longer for the wave front to reach the end of the receding ruler. Therefore, the time required to transverse the ruler is: '$d + v_a d$' ptu (in our example, the time taken for light to traverse the ruler is 1,000,000 + ½ 1,000,000 ptu). Thus, the *measured* light velocity in absolute CA units is: $(d + v_a d) / (d + v_a d) = 1$ pvu, the velocity of light. Similarly, for observer 'B' moving at velocity $v_b$, the measured velocity of the green light he receives in his reference frame is: $(d + v_b d) / (d + v_b d) = 1$ pvu, again equal to light velocity in absolute units. Thus, both observers conclude that light is a universal constant, equal to one pvu, no matter what the state of motion of the light source in an inertial frame! This is similar to the same situation in ordinary space-time.

What happens if observer 'A' sends his measurements to observer 'B' (by any means, carrier pigeon for example)? First, will observer 'B' conclude that the color of light received from 'A' is green? Secondly, will the measured distances and times be equal? It is obvious from the above analysis, that the measurements are not equal, unless $v_a = v_b$! Furthermore, observer 'B' concludes that the received light is shifted towards the red.

Why? Observer 'B' examines the light received from 'A'. A 'wave marker' passes by him, and he then finds that the next 'wave marker' appears to take a longer to arrive, compared to when both observers are both at absolute rest. Thus, the light appears to have a longer wavelength that is shifted towards the red, when compared to observer 'A' (figure 10). The actual spacing between 'wave markers' is constant, and was determined by the energy of observer 'A's light emitting equipment. Note that observer 'A's measurement of his light wavelength at velocity $v_a$ is actually different from the wavelength measurement when he is at absolute rest, when measured in absolute units!

Let us now examine the results of the same experiment with measurements made in *ordinary* space-time, with ordinary measuring equipment like clocks and rulers. A common reference is required to make comparison measurements of length and time, since the absolute coordinate system is *not* available. Based on our definition of reference frames (as a grid of observers), light becomes the natural choice for comparative space-time measurements. Observer 'A' decides to define length in terms of the green light from his light source, where one basic length unit (blu) ≡ 1000 wavelengths of green light, from which he has constructed a standard ruler of this length. Similarly, observer 'A' chooses to define the time of one basic time unit (btu) as the elapsed time required to receive 1000 cycles (or 1000 audible clicks from each wave crest, for example) of the green reference light, from which he constructs a calibrated standard clock. Observer 'B' has the identically constructed ruler and clock. Now, as before, observer 'B' has a relative velocity of $v_r$, with respect to 'A'. What happens when observer 'B' makes measurements on the incoming green light, sent by observer 'A'?

Now we do *not* have the luxury of absolute units to arbitrate between the two observers. Furthermore, no observer can now be regarded as being at absolute rest! Both observers have an equal right to formulate the laws of physics of motion in his own frame. Observer 'A' measures the light velocity as follows: The green light travels distance '$D_a$' in time $T_a$, and therefore the measured light velocity is: $c = D_a/T_a$. Observer 'B' uses his identically constructed standard clock and standard ruler to measure the incoming green light. Does his measuring instruments measure the velocity of light the same as 'A'. The answer is yes. Recall that observer 'A''s velocity does not affect the light velocity at all. It is an absolute constant, and cannot be affected by the source motion. Recall that observer 'A' specifies the wavelength of light, through his source apparatus. Once set, the wavelength of light propagates as a constant, not affected by the source (as described in our CA model of light above). Therefore, observer 'B' measures the light velocity as follows:

$$(D_a + kD_a) / (T_a + kT_a) = [(1 + k)/(1+k)] (D_a/T_a) = D_a/T_a = c \text{ as for observer 'A'}.$$

The motion of observer 'B' ruler adds a length of $kD_a$ cycles of light to his measurement distance, and adds the same $kT_a$ time delay, leaving the measured light velocity the same as 'A'. Observers 'A' and 'B' decide to compare their space and time measurements, with their identically constructed ruler and clock. Do these measurements agree? It is very clear that they do not!

Observer 'B' performs similar measurements, with identically constructed equipment on the incoming green light. Observer 'B' notices that the wavelength of the green light is shifted to the red, as we just discussed. Thus, his standard ruler of a length of one blu contains *less* than 1000 wavelengths of the incoming light, because each wavelength is longer than 'A's. Similarly, he notices that when he listens for 1000 audible clicks (which should correspond to one btu of time), more than one btu of time elapses on his identically constructed clock (because each click takes a longer time to arrive). When the results of these measurements are compared by any means (by carrier pigeon, for example) observer 'B' concludes that his time has been dilated, and his distances have contracted compared to observer 'A's measurement. Incidentally, if observer 'B' has the green light source and shines it towards 'A', observer 'A' would conclude the same thing.

How do we mathematically compute the values of these space-time comparisons? One may be tempted to apply a one dimensional Doppler-type analysis to deduce the quantity of space-time distortion. This, however, would *not* yield the correct answer. The above analysis is applicable for all the 3 dimensions that light can travel in space. Therefore, one must correct for light moving in all directions. This is precisely how Einstein derived the Lorentz transformations! In other words, the velocity of light measured in all directions of an expanding spherical wave front is what we take to be a constant. Thus, by showing that the velocity of light propagates as a constant in all directions in CA absolute space and time, we find that all inertial observers measure light velocity as a constant. However, they do not agree on the actual values of the space and time measurements. In this way, the principle of relativity leads us directly to the Lorentz transformation.

In summary, by postulating that on the lowest level of the CA, photons are information patterns 'moving' by a simple shifting from cell to adjacent cell at every clock 'cycle' in any given direction, we found that:

(1) Light propagates in an absolute, quantized 3D space, and separate 1D time (plank units) of the cellular automata, whose velocity is totally *unaffected by the source motion*. The light source determines the energy, and therefore the wavelength of the light. Once the light leaves the source, the wavelength and velocity is an absolute constant (totally decoupled from the source), specified in absolute CA units.
(2) In absolute CA space and CA time units, observers have an absolute velocity. The actual cell addresses of the information on the CA become the absolute 'rest' frame (which is not directly accessible by experiment). Hypothetical measurements in these absolute units yield light velocity and wavelength to be a constant, no matter what the state of motion of the source, where light has a unit velocity in plank velocity units (pvu's)..
(3) When two (or more) inertial observers with real measuring instruments are employed, and the measurements are made in the familiar 4D space-time defined by relativity theory, we have shown that all observers *measure* the velocity of light as a constant. However, when two (or more) inertial observers compare their space and time measurement (which is required to measure velocity, the measurements can be communicated by any means), they find that the measurements do not agree.

(4) We showed that the *measured* light velocity is constant in *all* space directions, which still remains only a postulate of special relativity. The Lorentz transformation directly follows from this through simple algebra. The Lorentz transformations form the central core of special relativity, and yields the familiar results of relativity such as: time dilation, Lorentz contraction, velocity addition, and so on.

In regards to inertial frames, one might be tempted to consider that the virtual particles of the quantum vacuum might act as some sort of an abstract universal reference frame. One might think that the virtual particles in the neighborhood of a point might be used to gauge your constant velocity motion, a frame that would have been unknown to Einstein when he formulated special relativity. However, the virtual particles have completely random velocities, move in completely random directions, and most importantly are short lived and unobservable. Furthermore, one cannot 'tag' the virtual particles with labels, and follow the progress of all the virtual particles in order to judge your own motion with respect to the average motion of the virtual particles! Therefore, it is impossible to tell your state of constant velocity motion with respect to the vacuum, unless a force or some other vacuum phenomena makes it's presence felt.

It is well known experimental result that the virtual particles introduce no new forces for inertial observers. However, this is definitely not the case for an accelerated frame, where we are now concerned with the state of the acceleration vectors of the virtual particles with respect to a Newtonian accelerated mass (F=MA). Now forces are present, which originate from the electromagnetic interaction of the quantum vacuum with the matter. This becomes the basis for the formulation of EMQG for accelerated reference frames, and also for gravitational reference frames!

Acceleration is a special motion, because an accelerated observer can detect his state of acceleration (inside a closed box, for example) by simply measuring the force exerted on him with an accelerometer. He does not need to compare his motion against some other reference frame to find out if he is accelerating. Newton was well aware of this fact, which led him to postulate the existence of 'absolute' space. Therefore, it appears that an accelerated test mass does *not* require another reference frame to gauge motion, and therefore acceleration has a special status in physics.

It will be shown through the new quantum principle of inertia (section 8), acceleration also has a special hidden reference frame that was unknown to both Einstein and Newton when they formulated their famous theories of motion. The reference frame in question here is the state of accelerated motion of the test mass with respect to the virtual particles of the quantum vacuum. However, it is not the velocity of the particles that sets up this abstract reference frame, it is the net statistical average **acceleration** of the virtual particles of the quantum vacuum near the test mass that forms the absolute reference frame. These concepts affect the physical meaning of inertial mass. Therefore we formulate a new framework to understand the meaning of inertial mass and Newtonian inertia in the next section.

## 8.     THE QUANTUM THEORY OF INERTIA

**"Under the hypothesis that ordinary matter is ultimately made of subelementary constitutive primary charged entities or 'partons' bound in the manner of traditional elementary Plank oscillators, it is shown that a heretofore uninvestigated Lorentz force (specifically, the magnetic component of the Lorentz force) arises in any accelerated reference frame from the interaction of the partons with the vacuum electromagnetic zero-point-field (ZPF). ... The Lorentz force, though originating at the subelementary parton level, appears to produce an opposition to the acceleration of material objects at a macroscopic level having the correct characteristics to account for the property of inertia."**

**- B. Haisch, A. Rueda, H. E. Puthoff**

According to CA theory, there must be a localized cellular explanation for all global phenomena such as acceleration and gravity. The Newtonian law of inertia is no exception. Inertia (and gravity) should originate from the small-scale particle interactions such that a global law emerges from the activity. Recall that CA theory is based on the local rules for the local cellular neighborhood, and these rules are repeated on a vast scale for all the cells in the universe. Many of our existing physical theories are general, global principles or general observations of nature. Both gravity and inertia have only been described successfully by "classical theories", applicable on global scales. In EQMG, both inertia and gravity have a detailed, particle level explanation based on the local "conditions" at the neighborhood of a given matter particle, and is thus manifestly compatible with the philosophy of a cellular automata theory and the principle of locality in special relativity.

In a recent theory (ref. 5) proposed by Haisch, Rueda, and Puthoff (known here as the HRP Theory of Inertia), it was shown that inertia comes from the buzz of activity of the virtual particles that fills even a perfect vacuum. It is this ever-present sea of energy that resists the acceleration of mass, and so creates inertia. Thus, they have found the low-level quantum description of inertia that is manifestly compatible with CA theory. Inertia is now described as being purely the result of quantum particle interactions. Haisch, Rueda, and Puthoff have come up with a new version of Newton's second law: F=MA. As in Newton's theory, their expression has 'F' for force on the left-hand side and 'A' for acceleration on the right. But in the place of 'M', there is a complex mathematical expression tying inertia to the properties of the vacuum. They found that the fluctuations in the vacuum interacting with the charge particles of matter in an accelerating mass give rise to a magnetic field, and this in turn, creates an opposing force to the motion. Thus, electromagnetic forces (or photon exchanges) is ultimately responsible for the force of inertia! The more massive an object, the more 'partons' it contains; and the more partons a mass contains means more individual (small) electromagnetic forces from the vacuum present and the stronger the reluctance to undergo acceleration. But, when a mass is moving at a **constant** velocity, inertia disappears, and there is no resistance to motion in any direction as required by special relativity.

In their theory, inertia is caused by the magnetic component of the Lorentz force which arises between what the author's call 'parton' particles in an accelerated reference frame interacting with the background vacuum electromagnetic zero-point-field (ZPF). The author's use the old fashion term originated by Feynman called the 'parton', which referred to the elementary constituents of the nuclear particles such as protons and neutrons. It is now known that Feynman's partons are quarks (ref. 37), and that the proton and neutron each contain three quarks of two types: called the 'up' and 'down' quarks.

We have found it necessary to make a small modification of HRP Inertia theory in our investigation of the principle of equivalence. In EMQG, the modified version of inertia is known here as the "Quantum Inertia", or QI. In EMQG, a new elementary particle is required to fully understand inertia, gravitation, and the principle of equivalence. **All** matter, including electrons and quarks, must be made of nature's most fundamental mass unit or particle which we call the 'masseon' particle. These particles contain one fixed, fundamental 'quanta' of both inertial and gravitational mass. The masseons also carry one basic, smallest unit or quanta of electrical charge as well, of which they can be either positive or negative. Masseons exist in particle or anti-particle form (called anti-masseon), that can appear at random in the vacuum as masseon/anti-masseon particle pairs of opposite electric charge. The earth consists of ordinary masseons (no anti-masseons), of which there are equal numbers of positive and negative electric charge varieties. The masseon particle model will be elaborated later. Instead of the 'parton' particles (that make up an inertial mass in an accelerated reference frame) interacting with the background vacuum electromagnetic zero-point-field (ZPF), we postulate that the real masseons (that make up an accelerated mass) interacts with the surrounding, virtual masseons of the quantum vacuum, electromagnetically. However, the detailed nature of this interaction is not known at this time. For example, why is it that for constant velocity motion the forces add to zero, but when acceleration is introduced the forces add up to Newton's inertial force? Since the answers to these questions are not known, we treat the Quantum theory of Inertia as a postulate of EMQG.

We will see that quantum inertia is deeply connected with the subject of quantum gravity. EMQG explains why the inertial mass and gravitational mass are identical in accordance with the weak equivalence principle. The weak equivalence principle translates to the simple fact that the mass (m) that measures the ability of an object to produce (or react to) a gravitational field ($F=GMm/r^2$) is the same as the inertial mass value that appears in Newton's F=ma. In EMQG, this is not a chance coincidence, or a given fact of nature, which is assumed to exist, *a prior*. Instead, equivalence follows from a deeper process occurring inside a gravitational mass due to interactions with the quantum vacuum, which are *very similar* in nature to the interactions involved in inertial mass undergoing acceleration.

Since this new quantum theory of the inertia has still not been fully developed or confirmed yet, we raise QI to the level of a postulate. This is assigned as the third postulate of EMQG theory. The virtual particles of the quantum vacuum can be considered to be a kind of absolute reference frame for accelerated motion only. This frame is simply represented as the resultant acceleration vector given by the sum of all the acceleration vectors of the virtual particles of the quantum vacuum in the immediate neighborhood of a given charged particle in the accelerated mass. This quantum vacuum reference frame gauges absolute acceleration. We do not need to measure our motion with respect to this frame in order to confirm that a mass is accelerated, we simply need to measure if an inertial force is present. We will see that this new, local quantum vacuum reference frame is the key to resolving the paradoxes in Mach's principle in regards to what reference frame nature uses to gauge masses that are accelerated.

### 9. BASIC MASS DEFINITIONS ACCORDING TO QUANTUM INERTIA

Based on quantum inertia and the quantum principle of equivalence there exists three *different* mass definitions for an object, listed below:

(1) **INERTIAL MASS** is the measurable mass defined in Newton's force law F=MA. This is considered as the absolute mass in EMQG, because it results from force produced by the relative (statistical average) acceleration of the charged virtual particles of the quantum vacuum with respect to the charged particles that make up the inertial mass. To some extent, the virtual particles of the quantum vacuum forms Newton's absolute reference frame. In special relativity this mass is equivalent to the rest mass.

(2) **GRAVITATIONAL MASS** is the measurable mass involved in the gravitational force as defined in Newton's law $F=GM_1M_2/R^2$. This is what is measured on a weighing scale. This is also considered as absolute mass, and is (almost) exactly the same as inertial mass. The same quantum process in inertia is also occurring in gravitational mass.

(3) **LOW LEVEL GRAVITATIONAL 'MASS CHARGE'** which is the origin of the pure gravitational force, is defined as the force that results through the exchange of graviton particles between two (or more) quantum particles. This type of mass analogous to 'electrical charge', where photon particles are exchanged between electrically charged particles. Note that this force is very hard to measure because it is masked by the background quantum vacuum electromagnetic force interactions, which dominates over the graviton force processes.

These three forms of mass are not necessarily equal! It turns out (section 17.3) that the inertial mass is almost exactly the same as gravitational mass, but not perfectly equal. All quantum mass particles have all three mass types defined above. But bosons (only photons and gravitons are considered here) have only the first two mass types. This means that photons and gravitons transfer momentum, and do react to the presence of inertial frames and to gravitational fields, but they do not emit or absorb gravitons. Gravitational fields affect photons, and this is linked to the concept of space-time curvature, described in detail later (section 18). It is important to realize that gravitational fields deflect photons (and gravitons), but not by force particle exchanges directly. Instead, it is due to a scattering process (section 18).

You might think that if a particle has energy, it automatically has mass; and if a particle has mass, then it must emit or absorb gravitons. This reasoning is based on Einstein's famous equation E=mc$^2$, which is derived purely from considerations of inertial mass (and Einstein's principle of equivalence extended to gravitational fields). In his famous thought experiment, a photon is emitted from a box, causing a recoil to the box in the form of a momentum change, and from this he derives his famous E=mc$^2$. In quantum field theory this momentum change is traceable to a fundamental QED vertex (ref. 40), where a electron (in an atom in the box) emits a photon, and recoils with a momentum equivalent to the photon's momentum '$m_p c$". We have analyzed Einstein's thought experiment from the perspective of EMQG and concluded that the photon behaves as if it has an effective inertial mass '$m_p$' given by: $m_p = E/c^2$ in Einstein's light box. For simplicity, lets consider a photon that is absorbed by a charged particle like an electron at rest. The photon carries energy and is thus able to do work. When the photon is absorbed by the electron with mass '$m_e$', the electron recoils, because there is a definite momentum transfer to the electron given by $m_e v$, where v is the recoil velocity. The electron momentum gained is equivalent to the effective photon momentum lost by the photon $m_p c$. In other words, the electron momentum '$m_e v_e$' received from the photon when the photon is absorbed is equivalent to the momentum of the photon '$m_p c$', where $m_p$ is the effective photon mass. If this electron later collides with another particle, the same momentum is transferred. The rest mass of the photon is defined as zero. Thus, the effective photon mass is a measurable inertial mass. Note that the recoil of the light box is a backward acceleration of the box, which works against the virtual particles of the quantum vacuum. Thus, when one claims that a photon has a real mass, we are really referring to the photon's ability to impart momentum. This momentum can later do work in a quantum vacuum inertial process.

Does the photon have an effective gravitational mass? By this we mean; does it behave as if it carries a measurable gravitational mass in a gravitational field like the earth (as given by $E/c^2$)? The answer is yes! For example, when the photon moves parallel to the earth's surface, it follows a parabolic curve and deflects downwards. You might guess that this deflection is caused by the graviton exchanges originating from the earth acting on this photon, and that this deflection is the same as that inside an equivalent rocket accelerating at 1g. The amount of deflection is equivalent, but according to Einstein this is a direct result of the space-time curvature near the earth and in the rocket. Our work on the equivalence principle has shown however, that this is not true. The photon deflection is caused by a different reason, but ends up giving the <u>same</u> result. In the rocket, the deflection is simply caused by the accelerated motion of the rocket floor, which carries the observer with it. This causes the observer to perceive a curved path (described as curved space-time). In a gravitational field, however, the deflection of light is real, and caused by the <u>scattering</u> of photons with the downward accelerating virtual particles. The photon scatters with the <u>charged</u> virtual particles of the quantum vacuum, which are accelerating downwards (statistically). The photon moving parallel to the surface of the earth undergoes continual absorption and re-emission by the falling virtual (electrically charged) particles of the quantum vacuum. The vacuum particles induce a kind of '<u>Fizeau-like</u>' scattering of the photons (Note: this scattering is present in the rocket, but does not lead to photon deflection because only the rocket and observer are accelerated). The photons are scattered because of the electromagnetic interaction of the photons with the falling charged virtual particles of the vacuum. Since the downward acceleration of the quantum vacuum particles is the same as the up-wards acceleration of the floor of the rocket, the amount of photon deflection is equivalent. Under the influence of a gravitational field, photons take on the same downward component as the accelerating (charged) virtual particles of the vacuum. This, of course, violates the constancy of the speed of light; which we will explore further in section 18. We will see that light velocity can be specified in two different sets of units, ordinary, measurable space-time and absolute CA space and time units. For now, one should note that downward acceleration component contributed from the downward accelerating vacuum that is picked up by the photons is only during the time the photons are absorbed by the quantum vacuum particles (and thus exist temporarily as electrically charged virtual particles). In between virtual particle scattering, the photon motion is still strictly constant in the CA absolute units of measure.

A similar line of reasoning as above applies to the motion of the graviton particle. The graviton has inertial mass because like the photon, it can transmit a momentum to another particle when absorbed in the graviton exchange process during a gravitational force interaction (although considerably weaker then photon exchanges). Like the photon, the graviton deflects when moving parallel to the floor of the rocket (from the perspective of an observer on the floor) and therefore has inertial mass. The graviton also has a gravitational mass (like the photon) when it moves parallel to the earth's surface, where it deflects under the influence of a gravitational field. Again, the graviton deflection is caused by the scattering of the graviton particle with the downward acceleration of the virtual 'mass-charged' particles of the quantum vacuum through an identical 'Fizeau-like' scattering process described above. Unlike the photon however, the scattering is caused by the 'mass charge' interaction (or pure graviton exchanges) of the quantum vacuum virtual particles, and not the electric charge as before. The end result is that the graviton has an effective gravitational mass like the photon. Again a graviton does not exchange gravitons with another nearby graviton, just as a photon does not exchange photons with other photons.

To summarize, both the photon and the graviton do not carry low level 'mass charge', even though they both carry inertial and gravitational mass. The graviton exchange particle, although responsible for a major part of the gravitational mass process, does not itself carry the property of 'mass charge'. Contrast this to conventional physics, where the photon and the graviton both carry a non-zero mass given by $M=E/C^2$. According to this reasoning, the photon and the graviton both carry mass (since they carry energy), and therefore both must have 'mass charge' and exchange gravitons. In other words, the graviton particle not only participates in the exchange process, it also undergoes further exchanges while it is being exchanged! This is the source of great difficulty for canonical quantum gravity theories, and causes all sorts of mathematical renormalization problems in the corresponding quantum field theory (ref. 42). Furthermore, in gravitational force interactions with photons, the strength of the force (which depends on the number of gravitons exchanged with photon) varies with the energy that the photon carries! In modern physics, we do not distinguish between inertial, gravitational, or low level 'mass charge'. They are assumed to be equivalent, and given a generic name 'mass'. In EMQG, the photon and graviton carry measurable inertial and gravitational mass, but neither particle carries the 'low level mass charge', and therefore do not participate in graviton exchanges.

We must emphasize that gravitons do not interact with each other through force exchanges in EMQG, just as photons do not interact with each other with force exchanges in QED. Imagine if gravitons did interact with other gravitons. One might ask how it is possible for a graviton particle (that always moves at the speed of light) to emit graviton particles that are also moving at the speed of light. For one thing, this violates the principles of special relativity theory. Imagine two gravitons moving in the same direction at the speed of light that are separated by a distance d, with the leading graviton called 'A' and the lagging graviton called 'B'. How can graviton 'B' emit another graviton (also moving at the speed of light) that gets absorbed by graviton 'A' moving at the speed of light? As we have seen, these difficulties are resolved by realizing that there are actually three different types of mass. There is measurable inertial mass and measurable gravitational mass, and low level 'mass charge' that cannot be directly measured. Inertial and gravitational mass have already been discussed and arise from different physical circumstances, but in most cases give identical results. However, the 'low level mass charge' of a particle is defined simply as the force existing between two identical particles due to the exchange of graviton particles only, which are the vector bosons of the gravitational force. Low level mass charge is not directly measurable, because of the complications due to the electromagnetic forces simultaneously present from the quantum vacuum virtual particles.

It would be interesting to speculate what the universe might be like if there were no quantum vacuum virtual particles present. Bearing in mind that the graviton exchange process is almost identical to the photon exchange process, and bearing in mind the complete absence of the electromagnetic component in gravitational interactions, the universe would be a very strange place. We would find that large masses would fall faster than smaller masses, just as a large positive electric charge would 'fall' faster then a small positive charge towards a very large negative charge. There would be no inertia as we know it, and basically no force would be required to accelerate or stop a large mass.

## 10. APPLICATIONS OF QUANTUM INERTIA

Quantum Inertia is now applied to the understanding of Mach's principle, and can also account for Newton's three Laws of Motion (section 10.2). Central to this new understanding is that the state of ***relative acceleration*** (only) of the virtual particles of the quantum vacuum with respect to a mass, acts like Newton's absolute space to gauge particle acceleration. In fact, inertial forces are actually *caused* by the interaction of matter particles with the surrounding vacuum.

## 10.1 MACH'S PRINCIPLE

*"... it does not matter if we think of the earth as turning round on its axis, or at rest while the fixed stars revolve around it ... The law of inertia must be so conceived that exactly the same thing results from the second supposition as from the first."*

*E. Mach*

Ernst Mach (ca 1883) proposed that the inertial mass of a body does not have any meaning in the absence of the rest of the matter in the universe. In other words, acceleration requires some other reference frame in order to determine accelerated motion. Thus, it seemed to Mach that the only reference frame possible was that of the average motion of all the other masses in the universe. This implied to Mach that the acceleration of an object must somehow be dependent on the sum total of all the matter in the universe. To Mach, if all the matter in the universe were removed, the acceleration, and thus the force of inertia would completely disappear since no reference frame is available to determine the actual acceleration.

A spinning elastic sphere bulges at the equator due to the centrifugal force. The question that Mach asked was how does the sphere 'know' that it is spinning, and therefore must bulge. If all the matter in the universe was removed, how can we be sure that it really rotates? Therefore, how would the sphere 'know' that it must bulge or not? Newton's answer would have been that the sphere somehow felt the action of Newtonian absolute space. Mach believed that the sphere somehow 'fells' the action of all the cosmic masses rotating around it. To Mach, centrifugal forces are somehow gravitational in the sense that it is the action of mass on mass. To Newton, the centrifugal force is due to the rotation of the sphere with respect to absolute space. To what extent that Einstein's general theory of relativity incorporates Mach's ideas is still a matter of debate (ref. 29). EMQG (through the quantum inertia principle) takes a similar view as Newton, where Newton's absolute space is replaced by the virtual particles of the vacuum. Mach was never unable to develop a full theory of inertia based on his idea of mass affecting mass.

Mach's ideas on inertia are summarized as follows:

(1) A particle's inertia is due to some (unknown) interaction of that particle with all the other masses in the universe.
(2) The local standards of non-acceleration are determined by some average value of the motion of all the masses in the universe.
(3) All that matters in mechanics is the relative motion of all the masses.

Quantum inertia theory fully resolves Mach's paradox by introducing a new universal reference frame for gauging acceleration: the net statistical average acceleration vector of the virtual particles of the quantum vacuum with respect to the accelerated mass. In other words, the cause of inertia is the interaction of each and every particle with the quantum vacuum. Inertial force actually *originates* this way. It turns out that the distant stars do affect the local state of acceleration of our vacuum here through the long-range gravitation force. Thus, our local inertial frame is slightly affected by all the masses in the distant universe. However, in our solar system the local gravitational bodies swamp out this effect. (This long-range gravitational force is transmitted to us by the graviton particles that originate in all the matter in the universe, which will distort our local net statistical average acceleration vector of the quantum virtual particles in our vacuum with respect to the average mass distribution). Thus, it seems that Mach was correct in saying that acceleration here depends somehow on the distribution of the distant stars (masses) in the universe, but the effect he predicted is minute.

## 10.2 THE QUANTUM NATURE OF NEWTON'S THREE LAWS OF MOTION

*'... and the Newtonian scheme was based on a set of assumptions, so few and so simple, developed through so clear and so enticing a line of mathematics that conservatives could scarcely find the heart and courage to fight it.'*

*- Isaac Asimov*

Here we briefly outline the connection between quantum inertia and Newton's laws of motion. Reference 35 gives a much more detailed account on the relationship between Newtonian mechanics and Cellular Automata theory. We are now in a position to understand the quantum nature of Newton's classical laws of motion. According to the standard textbooks of physics (ref. 16) Newton's three laws of laws of motion are:

**1. An object at rest will remain at rest and an object in motion will continue in motion with a constant velocity unless it experiences a net external force.**

**2. The acceleration of an object is directly proportional to the resultant force acting on it and inversely proportional to its mass. Mathematically: $\Sigma F = ma$, where 'F' and 'a' are the vectors of each of the forces and accelerations.**

**3. If two bodies interact, the force exerted on body 1 by body 2 is equal to and opposite the force exerted on body 2 by body 1. Mathematically: $F_{12} = -F_{21}$.**

Newton's first law explains what happens to a mass when the resultant of all external forces on it is zero. Newton's second law explains what happens to a mass when there is a nonzero resultant force acting on it. Newton's third law tells us that forces always come in pairs. In other words, a single isolated force cannot exist. The force that body 1 exerts on body 2 is called the action force, and the force of body 2 on body 1 is called the reaction force.

In the framework of EMQG theory, Newton's first two laws are the direct consequence of the (electromagnetic) force interaction of the (charged) elementary particles of the mass interacting with the (charged) virtual particles of the quantum vacuum. Newton's third law of motion is the direct consequence of the fact that all forces are the end result of a boson particle exchange process.

Newton's First Law of Motion:

The first law is a trivial result, which follows directly from the quantum principle of inertia (postulate #3). First a mass is at relative rest with respect to an observer in deep space. If no external forces act on the mass, the (charged) elementary particles that make up the mass maintain a *net acceleration* of zero with respect to the (charged) virtual particles of the quantum vacuum through the electromagnetic force exchange process. This means that no change in velocity is possible (zero acceleration) and the mass remains at rest. Secondly, a mass has some given constant velocity with respect to an observer in deep space. If no external forces act on the mass, the (charged) elementary particles that make up the mass also maintain a *net acceleration* of zero with respect to the (charged) virtual particles of the quantum vacuum through the electromagnetic force exchange process. Again, no change in velocity is possible (zero acceleration) and the mass remains at the same constant velocity.

Newton's Second Law of Motion:

The second law *is* the quantum theory of inertia discussed above. Basically the state of *relative* acceleration of the charged virtual particles of the quantum vacuum with respect to the charged particles of the mass is what is responsible for the inertial force. By this we mean that it is the tiny (electromagnetic) force contributed by each mass particle undergoing an acceleration 'A', with respect to the net statistical average of the virtual particles of the quantum vacuum, that results in the property of inertia possessed by all masses. The sum of all these tiny (electromagnetic) forces contributed from each charged particle of the mass (from the vacuum) is the source of the total inertial resistance force opposing accelerated motion in Newton's F=MA. Therefore, inertial mass 'M' of a mass simply represents the total resistance to acceleration of all the mass particles.

Newton's Third Law of Motion:

According to the boson force particle exchange paradigm (originated from QED) all forces (including gravity, as we shall see) result from particle exchanges. Therefore, the force that body 1 exerts on body 2 (called the action force), is the result of the emission of force exchange particles from (the charged particles that make up) body 1, which are readily absorbed by (the charged particles that make up) body 2, resulting in a force acting on body 2. Similarly, the force of body 2 on body 1 (called the reaction force), is the result of the absorption of force exchange particles that are originating from (the charged particles that make up) body 2, and received by (the charged particles that make up) body 1, resulting in a force acting on body 1. An important property of charge is the ability to readily emit and absorb boson force exchange particles. Therefore, body 1 is both an emitter and also an absorber of the force exchange particles. Similarly, body 2 is also both an emitter and an absorber of the force exchange particles. This is the reason that there is both an action and reaction force. For example, the contact forces (the mechanical forces that Newton was thinking of when he formulated this law) that results from a person pushing on a mass (and the reaction force from the mass pushing on the person) is really the exchange of photon particles from the charged electrons bound to the atoms of the person's hand and the charged electrons bound to the atoms of the mass on the quantum level. Therefore, on the quantum level there is really is no contact here. The hand gets very close to the mass, but does not actually touch. The electrons exchange photons among each other. The force exchange process works both directions in equal numbers, because all the electrons in the hand and in the mass are electrically charged and therefore the exchange process gives forces that are equal and opposite in both directions.

## 11. SPECIAL RELATIVITY - INERTIAL MASS AND INERTIAL FORCE

**"In contrast to the Newtonian conception, it is easy to show that in relativity the quantity force, is not codirectional with the acceleration it produces ... It is also easy to show that these force components have no simple transformation properties ...."**

- M. Hammer

Quantum Inertia (QI) provides us with a new understanding of Newtonian momentum. We will show that it is only *inertial force* (and all forces in general) that is truly a ***fundamental*** concept of nature, not momentum nor the conservation of momentum law. The Newtonian momentum, which is defined by 'mv', is simply a bookkeeping value used to keep track of the inertial mass 'm' (defined as F/A) in the state of constant velocity motion 'v' ***with respect to another mass*** that it might collide with at some future time. In this way, momentum is a relative quantity. Momentum simply represents information (with respect to some other mass) about what will happen in later (possible) force reactions. This fits in with the fact that *inertial* mass cannot be measured for constant velocity mass in motion (in outer space for example, away from all other masses) without introducing some *test* acceleration. If a mass is moving at a constant velocity, there are *no* forces present from the vacuum. Furthermore, since momentum involves velocity, it requires some other inertial reference frame in order to gauge the velocity 'v'. The higher the velocity that a mass 'm' achieves, the greater will be the subsequent deceleration (and therefore the greater the subsequent inertial force present) during a later collision (when it meets with some another object). If the velocity doubles with respect to a wall ahead, for example, then the deceleration doubles in a later impact. Before doubling the velocity, the acceleration $a_0 = (v_0 - 0)/t$; and after doubling, $a = (2v_0 - 0)/t = 2a_0$. Therefore we find that $f = 2f_0$, the force required from the wall (assuming the time of collision is the same). Similarly, if the mass is doubled, the force required from the wall doubles, or $f=2f_0$. Recall that inertial force comes from the ***opposition of the quantum vacuum to the acceleration of mass*** (or deceleration as in this case). Similarly, the kinetic energy '$1/2mv^2$' of a mass moving at a constant relative velocity 'v', it is also a bookkeeping parameter (defined as the product of underline{force} and the underline{time} that a force is applied). This quantity keeps track of the subsequent energy reactions that a mass will have when later accelerations (or decelerations) occur with respect to some other mass. It is important to remember that it is the ***quantum vacuum force*** that acts against an inertial mass to oppose any change in its velocity that is truly fundamental.

We therefore conclude that according to principles of QI theory, the inertial force is absolute. We have also seen that acceleration ***can*** be considered absolute. By this we mean that it is only the acceleration 'a' of a mass 'm' with respect to the net statistical average acceleration of the virtual particles of the quantum vacuum that accounts for inertial force. Therefore, we conclude that inertial mass can also be considered to be absolute, and follows the simple Newtonian relationship 'M=F/A'. Since inertial force, acceleration, and mass can all be considered to be absolute in this framework, we must closely reexamine the principles of special relativity in regards to the variation of inertial mass with the relative velocity of another inertial frame. Relativity is based on the premise that all constant velocity motion is relative, and also on the postulate of the constancy of light velocity. According to special relativity (which restricts itself to frames of constant velocity, called inertial frames), the inertial mass 'm' is relative, and varies with the relative velocity 'v' with respect to a constant velocity observer, in accordance with the following formula: $m = m_0 / (1-v^2/c^2)^{1/2}$. Here m is defined as the inertial mass measured in the other frame with velocity v, and $m_0$ is defined as the rest mass (inertial mass measured in the same frame as the mass) and 'c' is the velocity of light. It appears on the surface that QI and special relativity are not compatible in regards to the meaning of inertial mass. From the point of view of quantum inertia, Einstein's definition of inertial mass cannot be ***fundamentally*** correct, because it is not related to the quantum vacuum process described above for inertia. This is because we cannot associate the relative velocity 'v' directly to any quantum vacuum process. Recall that it is only the acceleration 'a' of a mass 'm', with respect to the net statistical average acceleration vectors of the virtual particles of the quantum vacuum that is the source of inertial mass.

Most special relativistic textbook accounts of inertial force and mass are based on the so-called 'conservation of momentum approach' (ref. 18). The conservation of momentum is assumed to be a fundamental aspect of nature. In order for momentum to be conserved with respect to all constant velocity reference frames, the mass must vary. To see this, recall that momentum is defined as mass times velocity, or 'mv', and that the momentum is important in a collision only because it provides bookkeeping of the mass and relative velocity. The ***relative*** velocity between the two colliding masses will determine the amount of deceleration in the impact as follows: $a=(v_f - v_i)/\Delta t$, were $v_f$ is the final velocity, and $v_i$ the initial velocity. Also, the mass is important because the subsequent force (and therefore energy $E = F \Delta t$) is determined by F= m a through the quantum vacuum process described above. The more mass particles contained in a mass, the greater the resistance to the acceleration of the mass. Therefore, the product of mass and velocity is an indicator of the amount of ***future*** energy to be expected in a collision (or interaction) of the two masses. The total incoming momentum is defined as the momentum of the in-going masses ($m_1v_1 + m_2v_2$), the total out-going momentum is ($m_1v_1' + m_2v_2'$). Here the two masses $m_1$ and $m_2$ are moving at velocities $v_1$ and $v_2$ before the collision, with respect to an observer, and velocity $v_1$' and $v_2$' after the collision. In Newtonian mechanics, the total momentum is conserved for any observer in a constant velocity reference frame. Therefore, ($m_1v_1$ + $m_2v_2$) = ($m_1v_1$' + $m_2v_2$'), even though different observers in general will disagree with each of the relative velocities of a pair of masses that are colliding. This is what we mean by conservation of momentum. In special relativity, if we do not modify the definition of inertial mass, we would find that different observers in different constant velocity frames *disagree* on the conservation of momentum for colliding masses. However, it can be shown (ref. 18) that if the mass of an object 'm' (from the point of view of an observer in constant velocity motion 'v' with respect to a mass $m_0$, measured by an observer at relative rest) is redefined as follows:

$m = m_0 / (1-v^2/c^2)^{1/2}$ , then the *total* momentum of the collision remains conserved as in Newtonian mechanics.

How does special relativity treat the definition of inertial mass and inertial force? Since Einstein was aware that acceleration is not invariant in different inertial frames, he knew that Newton's law had to be modified.

Einstein had to modify Newton's inertial law during his program to revise all physics in order to be relativistic, and was not aware of the existence of the quantum vacuum at that time. When Einstein considered this law, he found that in addition to incorporating his new relative mass definition formula above, he had to contend with relative accelerated motion. Contrary to popular belief, special relativity *does* address the problem of accelerated motion, which can be measured by any observer in an inertial reference frame. Therefore, in order to allow different observers in different states of constant velocity motion to measure inertial forces, Newton's law of motion must be changed. Since space and time are involved in measuring acceleration relative to an observer, and therefore acceleration must also be relative.

As we have seen in our analysis, *inertial mass as absolute*. Furthermore, there exists absolute acceleration of a mass, which is defined as the state of acceleration of the matter particles making up that mass, with respect to the (net statistical) average acceleration of the virtual particles of the quantum vacuum. (Note: Since virtual particles interact with each other, not all the individual accelerations of the virtual particles with respect to the mass will be the same, hence the statistical nature of this statement). What about the applied force? As the force is applied, an acceleration results which causes the velocity of the mass to increase. What if the velocity of the mass approaches light speed with respect to the applied force? Is the force still as effective in further increasing the velocity of the mass?

## 12. RELATIVISTIC MASS VARIATION FROM PARTICLE EXCHANGES

Classical physics is based on the assumption that forces between two bodies can act upon each other instantaneously through direct contact. Furthermore, the resulting action is independent of the relative velocity of the two bodies from which the force acts. When Einstein proposed special relativity, he abolished all action-at-a-distance, where forces can act instantaneously. However, forces were still treated by Einstein within the framework of classical physics. In his program to make classical physics relativistic, he accepted Newton's law of inertial force without modification (for example, Newton's law F=MA was still taken as being fundamentally correct).

Modern physics now treats *all* forces as a quantum particle exchange process. (Note: The gravitational force is a special case where two exchange particles are involved, section 17. There we will find that two different force exchange particles are involved simultaneously: the photon and the graviton particle). As an example, consider the electromagnetic force exchange process, which involves the photon particle as described by Quantum Electrodynamics (ref. 1). Here the charged particles (electrons, positrons) act upon each other through the exchange of force particles, which are photons. In the language of computer science, electromagnetic forces can be viewed as being 'digital'. What appears as a smooth force variation is really the end result of countless numbers of photon exchanges, each one contributing a 'quanta' of electromagnetic force (or 'kick', which is an impulse of force acting for a nearly instantaneously small period of time).

To see how the exchange process works for electromagnetic forces, we will examine the classical Coulomb force law in the rest frame of two stationary charges. The electric force from the two charged particles decrease with the inverse square of their separation distance (the inverse square law: $F = kq_1q_2/r^2$, where k is a constant, $q_1$ and $q_2$ are the charges, and r is the distance of separation). QED accounts for the inverse square law by the existence of an exchange of photons between the two electrically charged particles. The number of photons emitted by a given charge (per unit of CA time) is fixed and is called the charge of the particle. Thus, if the charge doubles, the force doubles because twice as many photons are exchanged during the force interaction. This force interaction process causes the affected particles to accelerate either towards or away from each other depending on whether the charge is positive or negative (different charges transmit photons with a slightly different wave functions).

It is interesting to note that certain cellular automata patterns exhibit behaviors like charge. For example, in the famous 2D geometric CA called Conway's game of life there exists a class of CA patterns called 'guns', which constantly emit a steady stream of 'glider' patterns indefinitely. This CA emission process is constant without any degradation of the original gun pattern. This resembles the charge property possessed by electrons, where photons are constantly emitted without any change of state of the electron.

The strength of the electromagnetic force depends on the quantity of the electric charge, and also depends on the distance of separation between the charges in the following way: each charge sends and receives photons from every direction. But, the number of photons per unit area, emitted or received, decreases by the factor $1/4\pi r^2$ (the surface area of a sphere) at a distance 'r', because the photon emission process take place in all directions. Thus, if the distance doubles, the number of photons exchanged decreases by a factor of four. This process can be easily visualized on a 3D geometric CA. Imagine that an electron is at the center of a sphere and sends out virtual photons in all directions. Imagine that a second electron somewhere on the surface of a sphere at a distance 'r' from the emitter, absorbing some of the exchange photons. The absorption of the exchange photons causes an outward acceleration, and thus a repulsive force. If the charge is doubled on the electron, there is twice as many photons appearing at the surface of the sphere, and twice the force acting on the electron. Thus, this accounts for the linear product of charge terms in the inverse square law. In QED, photons do not interact with photons (by a force exchange interaction). As a result, in-going and out-going photons do not affect each other during the exchange process.

We will go into more detail on the consequences of the particle exchange process for gravity, and the connection with cellular automata later (section 14). For now we are interested in the consequences of the force exchange process for special relativity, where the exchange particle has a <u>finite, and fixed velocity of propagation</u> (the speed of light, with the exception of the weak nuclear force where some bosons carry mass). To our knowledge, no one has examined the consequences of particle exchanges from the point of view of forces acting on each other in different inertial frames, where exchange particle propagates at the speed of light. At the time that Einstein developed special relativity, the force exchange process was unknown. The basic idea we want to develop here is that the quantity of force transmitted between two objects very much depends on the received <u>flux rate</u> of the exchange particles. In other words, the number of particles exchanged per unit of <u>time</u> represents the magnitude of the force transmitted between the particles. For example, imagine that there are two charged particles at relative rest in an inertial reference frame. There are a fixed number of particles exchanged per second at a separation distance 'd'. Now imagine that particle B is moving away at a slow constant relative velocity 'v' with respect to particle A. If the relative velocity v<<c the exchange process appears almost the same as when the two particles are at rest. This is because the velocity of light is very high when compared to 'v', and the flux rate is unaffected. Now imagine that the relative recession velocity v -> c, which is comparable to the velocity of the exchange particle. Does the received flux rate of particle B get altered from the perspective of particle B's frame? The answer is yes, and this follows from another result of special relativity: the Lorentz Time Dilation!

It is clear from Lorentz time dilation that the timing of the exchange particle will be altered when there is a very high relative velocity away from the source. Recall the Lorentz time dilation formula of special relativity: $t = t_0 / (1 - v^2/c^2)^{1/2}$, which states that the timing of events varies with relative velocity 'v'. If the timing of the exchange particles is altered, then the flux rate is altered as well, since flux has units of numbers of particles per unit time.

Now assume that particle A emits a flux of $\Phi_a$ particles per second, as seen by an observer in particle A's rest frame. When the force exchange particles are transmitted to particle B, particle B sees the flux rate decrease because of time dilation. Therefore, we find that particle B receives a smaller quantity of exchange particles per second $\Phi_b$ then when the particles are at relative rest. Thus, particle A acts like it transmits a smaller flux rate $\Phi_b$, such that $\Phi_b = \Phi_a \ (1 - v^2/c^2)^{1/2}$. Since the force due to the particle exchange is directly proportional to the flux of particles exchanged, we can therefore write:

$$F = F_0 \ (1 - v^2/c^2)^{1/2}$$

where is $F_0$ is the magnitude of the force when particle A and B are at relative rest, and F is the resulting smaller force acting between particle A and B when the receding relative velocity is 'v'. Thus, we can conclude that when a force acts to cause an object to recede away from the source of the force, the force <u>reduces</u> in strength. With a similar line of reasoning, we find that the force increases in strength when a force acts to cause an object to move towards the source of the force.

We are now in a position to see the apparent relationship between the inertial mass and velocity. Since all forces are due to particle exchanges, we can use the method developed above to study the forces acting between to inertial frames. First, at relative rest where v=0, we have F= $F_0$. The rest mass '$m_0$' is defined by Newton's law: $F_0 = m_0 \ a$, where 'a' is a test acceleration that is introduced to measure the inertial rest mass. Now, assume that there is a relative velocity 'v' between the applied force and the mass 'm', which causes the mass to recede. Therefore, we can write:

$$F = F_0 \ (1 - v^2/c^2)^{1/2} = m \ a$$

where the force is reduced in magnitude for the reasons discussed above, and the mass 'm' is considered absolute (or m= $m_0$, as in Newtonian Mechanics). In EMQG, we believe this equation represents the actual physics of the force interaction. However, if one takes the position that the force does not vary with velocity, but that the mass is what actually varies, then the above equation can be interpreted as:

$$F = F0 = m \ a = m_0 \ (1 - v^2/c^2)^{-1/2} \ a \ , \ and \ \ m = m_0 \ (1 - v^2/c^2)^{-1/2}$$ as given by Einstein.

So we see that we are in a situation where it is experimentally impossible to distinguish between the following two approaches: inertial mass variation with high velocity (Einstein) versus the force variation with high velocity (EMQG). What velocities can a mass achieve through the application of an accelerating force? According to our analysis above, the answer is that the limiting speed is the speed of the exchange particles, or light velocity. At this limit, the accelerating force effectively becomes zero!

It is, however, convenient to associate the variation of force with an increase in relativistic mass as Einstein proposed, for two important reasons. First, this restores the conservation of total momentum in collisions for all inertial observers (in fact, this is how Einstein derived his famous mass-velocity relationship). Secondly, if a mass is accelerated to the relativistic velocity 'v' with respect to observer 'A' by some given force, and this force is removed, there will be no way to determine the subsequent energy release when a collision occurs later. In other words, when this mass collides with another object, a rapid deceleration occurs with a large release of energy (which is force multiplied by time). This energy release is greater then what can be expected from Newton's laws. In fact, the large energy release is due to the effective increase in the force during the collision due to increased numbers of force exchange particles acting to reduce the speed of the colliding mass.

The force $F = F_0 (1 - v^2/c^2)^{1/2}$ tends to zero as the velocity v -> c. This means that any force becomes totally ineffective as the mass is accelerated to light velocity with respect to the source. As we have seen, this is attributed to the force resulting from exchange particles, which become totally ineffective in propagating from the source to the receiver, as the velocity of the receiver with respect to the source approach the velocity of the exchange particle. In order to clarify these ideas, we will analyze an actual experiment that was performed to confirm relativistic mass increase effect.

## 13. EQUIVALENCE OF MASS AND ENERGY: $E = M c^2$

One of the most important results of special relativity is the equivalence of mass and energy. This is represented in perhaps the most famous formula in all of physics: $E=Mc^2$. This formula implies that photons carry mass, since they carry energy. Photons are capable of transferring energy from one location to another, as by solar photons for example. Do photons really have mass?

You might think that if a particle has energy, it automatically has mass; and if a particle has mass, then it must emit or absorb gravitons. This reasoning is based of course, on $E=mc^2$. Einstein derived this formula from his famous light-box thought experiment (ref. 18). In his thought experiment, a photon is emitted from a box, causing the box to recoil and thus to change momentum. In quantum field theory this momentum change is traceable to a fundamental QED vertex, where a electron (in an atom in the box) emits a photon, and recoils with a momentum equivalent to the photon's momentum '$m_p c$". Therefore, we can conclude that the photon behaves as if it has an effective inertial mass '$m_p$' given by: $m_p = E/c^2$ in Einstein's light box. For simplicity, lets consider a photon that is absorbed by a charged particle like an electron at rest. The photon carries energy and is thus able to do work. When the photon is absorbed by the electron with mass '$m_e$', the electron recoils, because there is a definite momentum transfer to the electron given by $m_e v$, where v is the recoil velocity. The electron momentum gained is equivalent to the effective photon momentum lost by the photon $m_p c$. In other words, the electron momentum '$m_e v_e$' received from the photon when the photon is absorbed is equivalent to the momentum of the photon '$m_p c$', where $m_p$ is the effective photon mass. If this electron later collides with another particle, the same momentum is transferred. The rest mass of the photon is defined as zero. Thus, the effective photon mass is a measurable inertial mass.

**Note**: the recoil of the light box is a backward acceleration of the box, which works against the virtual particles of the quantum vacuum. Thus, when one claims that a photon has a real mass, we are really referring to the photon's ability to impart momentum. This momentum can later do work in a quantum vacuum inertial process. We will see in section 9 that although the photon carries inertial mass (and gravitational mass), it does not posses any low-level mass charge.

Einstein's derivation of $E=mc^2$ was unnecessarily complex (ref. 18) because of his reluctance to utilize results from quantum theory. Although he was one of the founders of the (old) quantum theory, he remained skeptical about the validity of the theory throughout his whole career. In EMQG, we treat the energy-mass equivalence as a ***purely*** quantum process, and not as a result of special relativity. Although Einstein derived this law when he developed special relativity, it can be derived purely from quantum theory. As we hinted, the ability of a photon to transfer momentum (and thus carry energy) can be traced to a QED vertex (ref. 40), where a packet of momentum is transferred from the photon to an electron. Let us assume that the effective mass of the photon is $m_c$. Furthermore, the photon has a velocity c, momentum p, energy E, a wavelength $\lambda$, and a frequency $\nu$. Therefore, by using the properties of the photon below (where h is plank's constant):

P=mc   (from classical physics)                             (CLASSICAL)    (A)
$c=\nu\lambda$   (definition of frequency & wavelength)     (CLASSICAL)    (B)
$E=h\nu$   (from Plank's energy-frequency law)              (QUANTUM)      (C)
$\lambda=h/p$ (from DeBroglie wavelength law)               (QUANTUM)      (D)

Therefore, $c/\nu = h/p = h/(mc)$ (using equations B, D, and A).

Therefore, $c/(E/h) = h/(mc)$ (using equation C), or $E=mc^2$. Thus, a very simple derivation of the energy-mass relationship is possible with the help of quantum mechanics. Therefore we conclude that $E=mc^2$ results from ***both*** the basic postulates of relativity theory and quantum theory, and should ***not*** be treated as being only a result of special relativity theory.

## 14. ALL FORCES MUST BE CAUSED BY FORCE PARTICLE EXCHANGES

The theory that best describes the quantization of the electromagnetic force field is called Quantum Electrodynamics (QED). Here the charged particles (electrons, positrons) act upon each other through the exchange of force particles, which are called photons. The photons represent the quantization of the classical electromagnetic field. In classical electromagnetic theory, the force due to two charged particles decreases with the inverse square of their separation distance (Coulomb's inverse square law: $F = kq_1q_2/r^2$, where k is a constant, $q_1$ and $q_2$ are the charges, and r is the distance of separation). QED accounts for this inverse square law by postulating the exchange of photons between the charged particles. The number of photons emitted and absorbed by a given charge (per unit of CA time) is fixed and is called the charge of the particle. Thus, if the charge doubles, the force doubles because twice as many photons are exchanged during the force interaction. This force interaction process causes the affected particles to accelerate either towards or away from each other depending on if the charge is positive or negative (because different charges transmit photons with slightly different wave functions). Certain known cellular automata roughly exhibit behaviors like this. For example, in the famous 2D geometric CA called Conway's game of life there exist a large variety of CA patterns types generally called 'guns'. They are constantly emitting a steady stream of 'gliders' as they travel through CA space. This emission process is constant without any degradation to the original gun pattern. This resembles the property possessed by electrons called charge, where photons are constantly emitted without any degradation change to the emitting electron.

The strength of the electromagnetic force varies as the inverse square of the distance of separation between the charges in the following way: each charge sends and receives photons from every direction. But, the number of photons per unit area, emitted or received, decreases by the factor $1/4\pi r^2$ (the surface area of a sphere for a 3D geometric CA) at a distance 'r' due to the photon emission pattern spreading in all directions. Thus, if the distance doubles, the number of photons exchanged decreases by a factor of four. This process can easily be visualized on a 3D geometric CA. Imagine that an electron particle is at the center of a sphere sending out virtual photons in all directions. Imagine that another electron is on the surface of a sphere at a distance 'r' from the emitter, which absorbs some of these photons. The absorption of these photons causes an outward acceleration, and thus a repulsive force. If the charge is doubled on the central electron, there is twice as many photons appearing at the surface of the sphere, and twice the force acting on the other electron. This accounts for the linear product of charge terms in the numerator of the inverse square law. In QED, photons do not interact with each other (through force exchanges). As a result, in-going and out-going photons do *not* affect each other during the exchange process, by the exchange of force particles.

Someone that is not fully versed in modern quantum field theory may question why two oppositely charged particles can be attracted to each other, while each is absorbing an exchange particle. On face value, classical thinking would imply that the momentum transfer would cause the particles to always move apart! The typical QED textbooks 'explain' this fact by the mathematics of momentum transfer at the vertices of the associated Feynman fundamental process (ref. 40). Certainly, classical models cannot explain this process, nor can classical models explain why photons are constantly emitted without degradation to the original electron, simply because all that is involved is a purely numerical CA process. We speculate that in the context of CA theory, the constant emission of photons (which maintains the charge of a particle) happens without degradation of the original electron pattern. This is possible because the original electron is a 'numeric' pattern which can remain stable indefinitely during this emission process (we have seen CA counterparts to this process in the familiar 'Game of Life' CA). Similarly, we speculate that the absorption of vector boson information patterns alters the internal state of the numeric pattern in such a way as to change the state of motion of that pattern (or causes acceleration). We believe that there must be a lot of *hidden activity* in the Feynman fundamental vertices of QED (ref. 40), and that the details are hidden from the physicist because of the purely numeric aspect of this process. In fact, in 'Conway's game of life', we discovered a CA pattern that is called a 'loop' which evolves into something resembling a two particle exchange process. Two larger internal oscillating CA patterns are seen to move apart while 'glider' particles (which are small, high-speed oscillating patterns reminiscent of photons) are exchanged. This pattern is something like a CA prototype pattern of a particle exchange process leading to a force. However, we found that this is not a perfect model, because the gliders are not emitted in every direction. Also, the particle exchange gives a constant velocity outward motion (and not accelerated motion as required). To date, no one has found a perfect particle exchange process that looks identical to real physical particles in any Cellular Automata simulations. However, we believe that something like this is happening on the plank scale in real particle exchanges in our universe.

## 15. GENERAL RELATIVITY, ACCELERATION, GRAVITY, AND CA

**"The general laws of physics (and gravitation) are to be expressed by equations which hold good for all systems of coordinates."**

**- Albert Einstein**

From the perspective of EMQG, Einstein's gravitational field equations are a set of observer dependent equations for observers that are subjected to gravity and/or to acceleration. These equations are based on *measurable* 4D space-time. The core of Einstein's theory is the principle of equivalence and the principle of general covariance. General Covariance allows any observer in any state of motion (and coordinate system) to describe gravity and acceleration with tensor equations that take on the same form.

CA theory places little significance to an observer unless the observer interferes with the interaction being measured. In a CA, physical processes continue without regards to the presence of an observer, where events unfold in absolute space and time. We will reconcile these two, very different views of gravity later in section 17. First, we review the central concepts of general theory of relativity.

**POSTULATES OF GENERAL RELATIVITY**

General relativity is a classical field theory founded on all the postulates and results of special relativity, as well as on the following new postulates:

**(1) PRINCIPLE OF EQUIVALENCE (STRONG) - The results of any given physical experiment will be precisely identical for an accelerated observer in free**

**space as it is for a non-accelerated observer in a perfectly uniform gravitational field. A weaker form of this postulate states that: objects of the different mass fall at the same rate of acceleration in a uniform gravity field.**

**(2) PRINCIPLE OF COVARIANCE - The general laws of physics can be expressed in a form that is independent of the choice of space-time coordinates and the state of motion of an observer.**

As a consequence of postulate 1, the inertial mass of an object is equivalent to it's gravitational mass. Einstein uses this principle to encompass gravity and inertia into his single framework of general relativity in the form of a metric theory of acceleration and gravity, based on quasi-Riemann geometry.

These postulates, and the additional assumption that when gravitational fields are present nearby, space-time takes the form of a quasi-Riemannian manifold endowed with a metric curvature (of the form $ds^2 = g_{ik} dx^i dx^j$) led Einstein to discover his famous gravitational field equations given below:

$$R_{ik} - \frac{1}{2} g_{ik} R = \frac{8\pi G}{c^2} T_{ik} \quad \ldots \text{Einstein's Gravitational Field Equation} \quad (15.1)$$

where, $g_{ik}$ is the metric tensor, $R_{ik}$ is the covariant Riemann curvature tensor. The left-hand side of the above equation is called 4D space-time curvature (the Einstein tensor or $G_{ik}$), which is the mathematical statement of space-time curvature that is reference frame independent and generally covariant. The right hand side ($T_{\alpha\beta}$) is the stress-energy tensor which is the mathematical statement of the special relativistic treatment of mass-energy density, G is Newton's gravitational constant, and c the velocity of light.

For comparison purposes, we will now present the EMQG equations (which are derived in ref. 1) for the classical gravitational field where the gravitational field is not *too strong*, or *too weak*:

$$\nabla^2 \phi - \frac{1}{c^2} \frac{\partial^2 \phi}{\partial t^2} = 4\pi G \rho(x,y,z,t) \quad \ldots \text{EMQG Gravitational Potential} \quad (15.2)$$

where $\phi$ represents the classical Newtonian potential and $\rho(x,y,z,t)$ represents the **absolute** mass density distribution (that can be time varying) as measured from an observer at relative rest from the center of mass. (This is a modified Poisson's equation, where the first term corresponds to the Poisson term, and the second term corresponds to the delay in the propagation of the graviton particles originating from the mass distribution). In EMQG, all distance units are in expressed in absolute cellular automata space units in a 3D rectangular cell grid, and time as a count of the elapsed clock cycles. In other words, space is measured by counting the number of cells between two points (cells). Time is measured by counting the number of clock cycles that has elapsed between two events. The acceleration vector **a** for an average virtual particle at point (x,y,z) in CA space from the center of mass can be obtained from the gravitational potential φ at this point by the following:

$$\mathbf{a} = \nabla \phi \quad \ldots \text{EMQG Virtual Particle Acceleration Field} \quad (15.3)$$

Einstein's law of gravitation cannot be arrived at by any 'rigorous' proof. The famous physicist S. Chandrasekhar writes (ref. 28):

*"... It seems to this writer that in fact no such derivation exists and that, at the present time, no such can be given. ... It is the object of this paper to show how a mixture of physical reasonableness, mathematical simplicity, and aesthetic sensibility lead one, more or less uniquely, to Einstein's field equations."*

The principle of equivalence (in its strong form) is incorporated in the above framework by the assertion that all accelerations that are caused by either gravitational or inertial forces are **metrical** in nature. More precisely, the presence of acceleration caused by either an inertial force or a gravitational field modifies the geometry of space-time such that it is a quasi-Riemannian manifold endowed with a metric.

Furthermore, point particles move in gravitational fields along geodesic paths governed by the equation:

$$\frac{d^2 x^i}{ds^2} + \Gamma_{jk}^{i} \frac{dx^j}{ds} \frac{dx^k}{ds} = 0 \quad \ldots \text{Equation for the geodesic path} \quad (15.4)$$

The most striking consequence of general relativity is the existence of curved 4D space-time (specified by the metric tensor $g_{ik}$). In EMQG theory, the meaning of the geodesic is very simple; it is the path taken by light or matter through the falling virtual particles undergoing acceleration, in the absence of any other external forces. We will see that curvature can be completely understood at the particle level as purely as the interaction of virtual particles and real particles, which move in absolute CA space and time. Furthermore, we will see that the principle of equivalence is a pure particle interaction process, and not a fundamental rule of nature. Before we can show this, we must carefully review the principle of equivalence from the context of general relativity theory.

## 16.   THE PRINCIPLE OF EQUIVALENCE AND GENERAL RELATIVITY

*"I have never been able to understand this principle (principle of equivalence) ... I suggest that it be now buried with appropriate honors."*
                                             **- Synge:  Relativity- The General Theory**

It should be noted that Einstein did not explain the origin of inertia in general relativity, nor the reason why the exists equivalence between accelerated and gravitational frames. Instead he relied on the existing Newtonian theory of inertia. Inertia was described by Newton in his famous law: F=MA; which states that an object resists being accelerated. A force (F) is required to accelerate an object of mass (M) to an acceleration (A). Since acceleration is a form of motion, it would seem that a reference frame is required in order to gauge this motion. But this is not the case in Newtonian physics. All observers agree as to which frame is actually accelerating by finding out which frames has a force associated with it. Only non-accelerated frames are relative. Einstein did not elaborate on this anomaly, or provide a reason why the inertial and gravitation masses are equal. This still remains as a postulate in his theory. The principle of equivalence has been tested to great accuracy. The equivalence of inertia and gravitational mass has been verified to an accuracy of one part in about $10^{-15}$ (ref 24).

Einstein's general theory of relativity is considered a "classical" theory, because matter, space, and time are treated as continuous classical variables. It is known however, that matter is made of discrete particles, and that forces are caused by particle exchanges as described by quantum field theory. A more complete theory of gravity should encompass a detailed quantum process for gravity involving particle interactions only. We will return to both special and general relativity in later sections, where a completely new interpretation and formulation of general relativity is given in the context of EMQG theory.

Inertia ought to be explained at the particle level as well, and should somehow be tied in to gravity on the particle level in a deep way according to the principle of equivalence. Yet the gravitational force must result from particle exchanges (gravitons) between particles that posses 'mass charge'. Somehow from all this, a curved 4D space-time must result from these particle activities, Until recently there has been no adequate explanation on the particle level for inertia and gravity. The next section summarizes original work on resolving these problems, which we call EMQG.

## 17. ELECTRO-MAGNETIC QUANTUM GRAVITY

*"Subtle is the lord…"*
                                                                                    *- Einstein*

In this section, we provide a brief review of EMQG. Reference 1 gives a much more detailed account. We will focus on the hidden quantum physics behind the equivalence of 4D space-time in accelerated and gravitational fields, and provide a new view of quantum gravity.

17.1   INTRODUCTION

Quantum Gravity is generally defined as the unification of quantum field theory with Einstein's general relativity theory. It should describe the behavior of gravitational forces at quantum distance scales, in enormous gravitational fields, and for cosmological distance scales. Various attempts at this unification have not been completely successful in the past, because these theories do not grasp the true nature of inertia, or the hidden physical processes behind Einstein's principle of equivalence. In developing a theory of quantum gravity, one might ask which of the existing approaches to quantum gravity is more relevant or fundamental, quantum field theory or Einstein's theory of general relativity? Currently, it seems that both of these theories are not compatible with each other (ref. 42).

Based on a postulate that the universe operates like a Cellular Automata (CA) computer, we assume that quantum field theory is in closer touch to the actual workings of the universe. General relativity is taken as a global, classical description of space-time, gravity and matter. General relativity reveals the large-scale patterns and organizing principles that arise from the hidden quantum processes existing on the tiniest distance scales. Quantum field theory reveals to us that forces originate from a quantum particle exchange process, which transfers momentum from one quantum particle to another. The exchange process is universal, and applies to electromagnetic, weak and strong nuclear forces, and also for gravitational force, as we shall see. The generic name given to the force exchange particles is the 'vector boson'.

We have developed a quantum theory of gravity called ElectroMagnetic Quantum Gravity (or EMQG) based on the Cellular Automata computer model, which describes gravity as a pure particle exchange process. We summarize the important results of this theory here, the details being available in reference 1. We will be dealing with the photon (the exchange particle for electromagnetic force) and the graviton (the exchange particle for the *pure* gravitational force). What is unique about EMQG is that gravitation involves ***both*** the photon and graviton exchange particles, where the photon particle plays a very important, and dominating role in gravitational interactions on the earth!

In order to formulate a theory of quantum gravity, a mechanism must be found that produces the gravitational force, while somehow linking to the principle of equivalence of inertial and gravitational mass. In addition, this mechanism should naturally lead to 4D space-time curvature and should be compatible with the principles of general relativity theory (ref. 42). Nature has another long-range force called electromagnetism, which has been described successfully by the principles of quantum field theory. This well-known theory is called Quantum ElectroDynamics (QED), and this theory has been tested for electromagnetic phenomena to an unprecedented accuracy. It is therefore reasonable to assume that gravitational force should be a similar process, since gravitation is also a long-range force like electromagnetism. However, a few obstacles lie in the way, which complicate this line of reasoning.

First, gravitational force is observed to be always attractive! In QED, electrical forces are attractive and repulsive. As a result of this, there are an equal number of positive and negative charged virtual particles in the quantum vacuum (section 17) at any given time because virtual particles are always created in equal and opposite charged particle pairs. Thus, there is a balance of attractive and repulsive forces in the quantum vacuum, and the quantum vacuum is electrically neutral, overall. If this were not the case, the virtual charged particles of one charge type in the vacuum would dominate over all other physical interactions.

Secondly, QED is formulated in a relativistic, flat 4D space-time with no curvature. In QED, electrical charge is defined as the fixed rate of emission of photons (strictly speaking, the fixed probability of emitting a photon) from a given charged particle. Electromagnetic forces are caused by the exchange of photons, which propagate between the charged particles. The photons transfer momentum from the source charge to the destination charge, and travel in flat 4D space-time (assuming no gravity). From these basic considerations, a successful theory of quantum gravity should have an exchange particle (graviton), which is emitted from a mass particle at a fixed rate as in QED (ref. 41). The 'mass charge' or just plain mass replaces the idea of electrical charge in QED, and the graviton momentum exchanges are now the root cause of gravitational force. Yet, the graviton exchanges must somehow produce disturbances of the normally flat space and time, when originating from a very large mass.

Since mass is known to vary with velocity (special relativity), one might expect that 'mass charge' must also vary with velocity. However in QED (ref. 40) the electromagnetic force exchange process is governed by a fixed, universal constant ($\alpha$) which is not affected by anything like motion (more will be said about this point later). Should this not be true for graviton exchange in quantum gravity as well? It also strange that gravity, which is also a long-range force, is governed by same form of mathematical law as found in Coulomb's Electrical Force law. Coulomb's Electric Force law states: $F = KQ_1Q_2/R^2$, and Newton's Gravitational Force law: $F=GM_1M_2/R^2$. This suggests that there is a deep connection between gravity and electromagnetism. Yet, gravity supposedly has no counterpart to negative electrical charge. Thus, there seems to be no such thing as negative 'mass charge' for gravity, as we find for electrical charge. Furthermore, QED also has no analogous phenomena as the principle of equivalence. Why should gravity be connected with the principle of equivalence, and thus inertia, and yet no analogy of this exists for electromagnetic phenomena?

To answer the question of negative 'mass charge', EMQG postulates the existence of negative 'mass charge' for gravity, in close analogy to electromagnetism. Furthermore, we claim that this property of matter is possessed by all anti-particles that carry mass. Therefore anti-particles, which are opposite in electrical charge to ordinary particles, are also opposite in 'mass charge'. In fact, negative 'mass charge' is not only abundant in nature, it comprises nearly half of all the mass particles (in the form of 'virtual' particles) in the universe! The other half exists as positive 'mass charge', also in the form of virtual particles. Furthermore, all familiar ordinary (real) matter comprises only a vanishing small fraction of the total 'mass charge' in the universe! Real anti-matter seems to be very scarce in nature, and no search for it in the cosmos has revealed anything to date.

Both positive and negative 'mass charge' appear in huge numbers in the form of virtual particles, which quickly pop in and out of existence in the quantum vacuum (section 3), everywhere in empty space. The existence of negative 'mass charge' is the key to the solution to the famous problem of the cosmological constant (ref. 36), which is one of the great unresolved mysteries of modern physics. Finally, we propose that the negative energy, or the antimatter solution of the famous Dirac equation of quantum field theory is also the *negative 'mass charge' solution*.

The question raised above regarding the principle of equivalence is much more difficult to answer. The principle of equivalence is one of the founding postulates of general relativity theory. Stated in a weaker form; objects of different mass fall at the same rate of acceleration in a uniform gravity field. Alternatively, it means that the inertial mass (mass defined by Newton's law of motion: $m_i = F_i/g$) is exactly equal to the gravitational mass (mass defined by Newton's universal gravitational law $m_g = (F_g r^2) / (GM)$ ). The equivalence principle requires that $m_i = m_g$. Why two such different physical definitions for these two mass types should give the same numerical result has remained a deep and unsolved mystery in modern physics, and deserves an explanation. This paper provides a solution to this difficult question, and is also an invitation to explore a new approach towards a full theory of quantum gravity. EQMG is based on a new understanding of both inertia and the principle of equivalence, which exists on the quantum particle level and is hidden from view.

The principle of equivalence has become a major pillar of modern physics, and has been tested under a wide variety of gravitational field strengths and distance scales. It has been tested with different material types (ref. 21). It has been tested to high degrees of precision (up to 3 parts in $10^{-12}$ for laboratory bodies, and 1 part in $10^{-12}$ for the acceleration of the moon and earth towards the sun). Yet, the principle of equivalence has remained only as a postulate of general relativity. It cannot be proven from fundamental principles. Some of the better literature on general relativity have drawn attention to this fact, and admit that no explanation can be found as to; "why our universe has a deep and mysterious connection between acceleration and gravity" (ref. 22). After all, while standing on the surface of the earth, gravity appears like a static force holding your mass to the surface. Yet, when your standing in an accelerated rocket moving with an acceleration of 1 g, the principle of equivalence states that there is an equivalent force exerted against the rocket floor by your inertial mass which is caused by the dynamic accelerated motion of the rocket. Why should there be such a deep connection between what appear to be two completely different physical phenomena?

The equality of inertial and gravitational mass has been deduced strictly by observation, and by actual experience. But is the principle of equivalence exact? Since the principle of equivalence cannot be currently traced to deeper physics, we can never say that these two mass types are *exactly* equal. EMQG reveals the hidden phenomena that account for the principle of equivalence.

Although we are not aware of any experiment that contradicts the equivalence principle, we have theoretical reasons to believe that it is not perfect. We find that the equivalence principle follows from lower level physical processes that gives near perfect equivalence. *Mass equivalence* arises from the equivalence of the electromagnetic forces generated between the net statistical average acceleration vectors of the (charged) matter particles inside a mass interacting with the surrounding virtual particles of the quantum vacuum inside an accelerating rocket. This is the *same* force occurring between the (charged) matter particles of a mass near the earth and the surrounding virtual particles of the quantum vacuum. Equivalence is *not* perfect, however, and breaks down when the accuracy of the measurement is extremely high!.

Thus we maintain that gravity also involves accelerated motion of a sort, like in a rocket undergoing uniform acceleration. However, under a gravitational field like the earth, this accelerated motion is hidden from our direct view. The motion that is hidden from us turns out to be the relative *accelerated motion of the particles of the quantum vacuum, which are falling*. The quantum vacuum consists of short-lived particles called virtual particles, which fills the surrounding space, everywhere (section 3). It is the hidden motion of virtual particles that turns out to be responsible for the equivalence between inertial and gravitational mass. We offer, for the first time, a detailed derivation of the principle of equivalence on the quantum scale, based on a few simple postulates. Since the most important physical process involved in gravity (and the principle of equivalence) is the ordinary electromagnetic force, we therefore call this new quantum gravity theory **'Electro-Magnetic Quantum Gravity'** or EMQG. Another reason for choosing this name is that the graviton particle turns out to have *almost the identical characteristics* as the photon particle.

Existing quantum gravity theories that include gravitons as the exchange particle do not properly address the problem of variation of mass with velocity predicted by special relativity, which is very obvious when masses achieve velocities that are approaching the speed of light. How can mass vary with velocity, when according to the principles of quantum field theory, mass (or 'mass charge') is the property of a particle to emit and absorb gravitons. Does this mean that the graviton emission rate varies with speed? However, if you are an observer in the same reference frame as a high velocity mass, then the mass appears to you to be the same as the rest mass, which would leave the graviton emission rate unchanged. What is going on here? The graviton emission cannot be relative. We cannot have different observers disagree on the graviton emission process! In fact, we believe that the graviton emission rate *must* be constant, and independent of velocity as in the precedent set by QED for the photon emission process of electrical charge. EMQG provides a new understanding of gravitation, and is also testable, because it predicts new experimental results that cannot be explained by conventional theory.

Here we briefly summarize the results of our new CA compatible theory of gravity called Electromagnetic Quantum Gravity or EMQG theory (ref. 1). We have found that both Special and General Relativity must be modified to be compatible with EMQG theory. Gravity is one of the four basic forces of nature. The highly successful standard model of particle physics does not account for gravity. The standard model addresses the electromagnetic, weak and strong nuclear forces within the framework of quantum field theory. In quantum field theory, forces are thought to originate as the exchange of force particles (vector bosons) which are represented by the quanta of the associated classical field. In EMQG, it is found that two fundamental particle exchange processes are responsible for gravity; one particle exchange being very familiar, while the second particle exchange type has been postulated but not yet been successfully detected. The particles involved are the photon and graviton exchange particles. We will see that the photon and the graviton are almost identical in their physical properties, except for their relative strengths. The particle exchange process (in general) fits very well into the general framework of CA theory without much modification. The boson acts like the go between particle, shifting from cell to cell until it is absorbed by a destination particle. This transfers an acceleration (or force) without action at a distance. We will now examine the nature of forces as particle exchanges on the CA in more detail.

17.2    GRAVITATION ORIGINATES FROM GRAVITON EXCHANGES

For gravitational forces, it is experimentally observed that the force originating from two particles possessing mass decreases with the inverse square of their separation distance, and is given by Newton's inverse square law: $F = Gm_1m_2 / r^2$, where G is the gravitational constant, $m_1$ and $m_2$ are the masses, and r is the distance of separation. For electromagnetic forces, it is also experimentally observed that the force originating from two particles possessing charge also decreases with the inverse square of their separation distance, and is given by Coulomb's inverse square law: $F = kq_1q_2 / r^2$, where k is Coulomb's constant, $q_1$ and $q_2$ are the masses, and r is the distance of separation. It can be seen that the two force laws are very similar in form. QED theory accounts for Coulomb's law by the photon exchange process. Following the lead from the highly successful QED, EMQG postulates that we replace the concept of electrical 'charges' exchanging 'photons' with the idea that 'mass charges' exchange gravitons. Hence, gravitational mass at a fundamental level is simply the ability to emit or absorb gravitons, and pure low-level gravitational mass is interpreted as 'mass charge'.

For gravity, instead of photons, there are gravitons, which are the force exchange particles of gravity. Like charge, it is the property called mass-charge that determines the number of exchange gravitons. The larger the mass, the greater the number of gravitons exchanged. Like electromagnetism, the strength of the gravity force decreases with the inverse square of the distance. This conceptual framework for quantum gravity has been around for some time now (ref. 41), but how are we to merge these simple ideas to be compatible with the framework of general relativity? We must be able to explain the Einstein's Principle of Equivalence and the physical connection between inertia, gravity, and curved space-time all within the general framework of graviton particle exchange. General Relativity is based on the idea that the forces experienced in a gravitational field and the forces due to acceleration are equivalent, and both are due to the space-time curvature.

In classical electromagnetism, if a charged particle is accelerated towards an opposite charged particle, the rate of acceleration depends on the electrical charge value. If the charge is doubled, the force doubles, and the rate of acceleration is doubled. If quantum gravity were to work in the exact same way, we would expect that the rate of acceleration of a mass near the earth would double if the mass doubles. The reason for this expectation is that the exchange process for gravitons should be very similar to electromagnetism. In other words, if the 'mass-charge' is doubled, the gravitational force is doubled. The only difference between the two forces is that gravity is a lot weaker by a factor of about $10^{-40}$. The weakness of the gravitational forces might be attributed either to the very small interaction cross-section of the graviton particle as compared to the photon particle, or to a very weak coupling constant (the absorption of a single graviton causing a minute amount of acceleration), or both.

Unfortunately, if the graviton exchange process worked exactly like QED, it would not reproduce the known nature of gravity. First, there is the problem of variation of mass with velocity as described by special relativity as $m = m_0 (1-v^2/c^2)^{-1/2}$. At face value, this would mean that the number of gravitons exchanged depends on **velocity** of the gravitational mass, which does not easily fit into the framework of a QED approach to quantum gravity. Secondly, if two masses are sitting on a table with mass 'M' and mass '2M', the forces against the tabletop varies with the mass, just as you would expect in a QED-like exchange of graviton particles. If the mass doubles, the force on the table doubles. Yet, the rate of acceleration is the same for these two masses in free fall. Why? Since twice the number of gravitons is exchanged under mass '2M', you would expect twice the force, and therefore mass '2M' would arrive early. Matter has inertia, and this complicates everything. In almost all quantum gravity theories inertia appears as a separate process that is 'tacked' on in an ad hoc manner. The principle of equivalence merely raises this relationship between inertia and gravitation to the status of a postulate without providing any deeper insights.

All test masses accelerate at the same rate ($g=9.8$ m/sec$^2$ on the earth) no matter what the value of the test mass is. This is a direct consequence of the principle of equivalence. Mathematically, this follows from Newton's two **different** force laws:

$F_i = ma$     ..... (Inertial force)

$F_g = \dfrac{GmM}{r^2}$ ..... (Gravitational force)

In free fall, an object (mass m) in the presence of the earth's pull (mass M) is force free, i.e. $F_i = F_g$. Note that the same mass value 'm' appears in these two mass definition formulas (for some *mysterious* reason).

Therefore, $ma = \dfrac{GmM}{r^2}$   or,  $a = \dfrac{GM}{r^2}$ ..... Equivalence Principle

Thus, the rate of acceleration does not depend on the test mass m. All test masses accelerate at the same rate. Thus, inertia and gravity are intimately connected in deep way because the measure of mass m is the same for acceleration as for gravity. What is mass? In EMQG, gravitational mass originates from a low-level graviton exchange process originating from 'mass charge', where the emission rate is constant. In fact, mass is quantized in exactly the same way as electric charge in QED. (There exists a fundamental unit of mass charge that is carried by the masseon particle, the lowest quanta of mass).

Recall the new quantum theory for inertia given in the previous section. We have found an explanation for inertia based on low-level quantum processes. The quantum source of the force of inertia is the resistance to acceleration offered by the virtual electrically charged particles of the quantum vacuum. What is unique about EMQG theory, is that the same virtual electrically charged particle processes are also present near a large gravitational mass. It is the interactions of these electrically charged virtual particles with the real electrically charged particles in the mass that accounts for the bulk of the gravitational force, and for the principle of equivalence. Yet we have retained the same simple QED type model for the fundamental low-level gravitational interactions through the graviton exchange process.

This new quantum theory of gravity is called Electro-Magnetic Quantum Gravity, primarily because gravity involves a strong electromagnetic component (secondly because of the similarities between the graviton and photon). It is based on low-level quantum descriptions of inertia and gravity involving electromagnetic photon exchanges and graviton exchanges, and thus totally compatible with CA theory.

## 17.3    EMQG AND THE PRINCIPLE OF EQUIVALENCE

*"The principle of equivalence performed the essential office of midwife at the birth of general relativity, but, as Einstein remarked, the infant would never have got beyond its long clothes had it not been for Minkowski's concept [of space-time geometry]. I suggest that the midwife be now buried with appropriate honors and the facts of absolute space-time faced."*                                                      *- Synge*

The principle of equivalence means different things to different people, and to some it means nothing at all as can be seen in the quotation above. The equality of inertial and gravitational mass is only known to be true strictly through observation and experience. Is this equivalence exact, though? Since the principle of equivalence cannot be currently traced to deeper physics, we can never say that these two mass types are *exactly* equal. Currently, we can only specify the accuracy to which the two mass types have been shown *experimentally* to be equal.

How is the principle of equivalence defined? Well, there are two main formulations of the principle of equivalence. The strong equivalence principle states that the results of any given physical experiment will be precisely identical for an accelerated observer in free space as it is for a non-accelerated observer in a perfectly uniform gravitational field. A weaker form of this postulate restricts itself to the laws of motion of masses only. In other words, the laws of motion of identical masses on the earth are identical to the same situation inside an accelerated rocket (at 1g). Technically, this holds only at a point near the earth. It can be stated that objects of the different mass fall at the same rate of acceleration in a uniform gravity field.

According to EMQG, if a large mass is present, the mass emits huge numbers of graviton particles, and distorts the surrounding virtual particles of the quantum vacuum. In an accelerated frame, there are very few gravitons, and the quantum vacuum is not affected. However, an observer in the accelerated frame 'sees' the quantum vacuum accelerating with respect to his frame, and hence the space-time distortion. However, the quantum vacuum *still remains undisturbed*. Thus in EMQG, the equivalence principle is regarded as being a coincidence due to quantum vacuum appearing the same for accelerated observers and for observers in gravitational fields.

Recently, some theoretical evidence has appeared as to why the strong equivalence principle does not hold in general. First, if gravitons can be detected experimentally with a new and sensitive graviton detector (which is not likely to be possible in the near future), we would be able to distinguish between an inertial frame and a gravitational frame with this detector. This is possible because inertial frames would have virtually no graviton particles present, whereas the gravitational fields like the earth have enormous numbers of graviton particles. Thus, we have performed a physics experiment that can detect whether you are in a gravitational field or an accelerated frame. Secondly, recent theoretical considerations of the emission of electromagnetic waves from a uniformly accelerated charge, and the lack of radiation from the same charge subjected to a static gravitational field leads us to the conclusion that the strong equivalence principle does not hold for radiating charged particles (ref. 20).

As for the weak equivalence principle, we can now only specify the accuracy as to which the two different mass types have been shown *experimentally* to be equal in an inertial and gravitational field. In EMQG, we will show that the equivalence principle follow from lower level physical processes and the basic postulates of EMQG. We will see that mass equivalence arises from the equivalence of the force generated between the net statistical average acceleration vectors of the matter particles inside a mass interacting with the surrounding quantum vacuum virtual particles inside an accelerating rocket. The *same* force occurs between the matter particles and virtual particles for a mass near the earth. We will find that equivalence is *not* perfect, and breaks down when the accuracy of the measurement approaches $10^{-40}$!

Basically, the equivalence principle arises from the *reversal* of the net statistical average acceleration vectors between the charged matter particles and virtual charged particles in the famous Einstein rocket, with the same matter particles and virtual particles near the earth. To fully understand the hidden quantum processes in the principle of equivalence on the earth, we examine the behavior of test masses and the propagation of light near the earth. Equivalence is shown to hold for both stationary test masses and for free-falling test masses.

MASSES INSIDE AN ACCELERATED ROCKET AT 1g

In figure 1 there are two different masses at rest on the floor of an rocket which is accelerated upwards at 1 g far from any gravitational sources. The floor of the rocket experiences a force under the mass '2M' that is twice as great as for the mass 'M'. In Newtonian physics, the inertial mass is defined in precisely this way, the force 'F' that occurs when a mass 'M' is accelerated at rate 'g' as given by F=Mg. The quantum inertia explanation for this is that the two masses are accelerated with respect to the net average statistical motion of the virtual particles of the vacuum by the rocket. Since mass '2M' has twice the particle count as mass 'M', the sum of all the little electromagnetic forces between the virtual vacuum and the particles of mass '2M' is twice as great as compared to mass 'M', i.e. for mass 'M', $F_1$=Mg and for mass '2M', $F_2$=2Mg=2$F_1$. Because the particles that make up the masses do not maintain a net zero acceleration with respect to the virtual particles, a force is always present from the rocket floor (figure 1).

In figure 2, the two different masses (M and 2M) have just been released and are in free fall inside the rocket. According to Newtonian physics, no forces are present on the two masses since the acceleration of both masses is zero (the masses are no longer attached to the rocket frame). The two masses hit the rocket floor at the same time. The quantum inertia explanation for this is trivial. The net acceleration between all the real particles that make up both masses and virtual particles of the vacuum is a net (statistical average) value of zero. The rocket floor reaches the two masses at the same time, and thus unequal masses fall at the same rate inside an accelerated rocket.

MASSES INSIDE A GRAVITATIONAL FIELD (THE EARTH)

In figure 3, there are the same two masses (2M and M) which are at rest on the surface of the earth. The surface of the earth experiences a force under mass '2M' that is twice as great as for that under mass 'M'. The reason for this is that the two stationary masses do not maintain a net acceleration of zero with respect to the net statistical average acceleration of the virtual particles in the neighborhood. This is because the virtual particles are all accelerating towards the center of the earth ($\mathbf{a}$=GM/$\mathbf{r}^2$) due to the graviton exchanges between the real particles consisting of the earth and the virtual particles of the vacuum. Since mass '2M' has twice the particles as mass 'M', the sum of all the tiny electromagnetic forces between the virtual particles of the vacuum and the real particles of mass '2M' is twice as great as that for mass 'M'. Thus, a force is required from the surface of the earth to maintain these masses at rest, mass '2M' having twice the force of mass 'M'. The physics of this force is the same as for figure 1 in the rocket, but with the acceleration frames of the virtual charged particles and the real charged particles of the mass being reversed (with the exception of the direct graviton induced forces on the masses, which is negligible). Equivalence between the inertial mass 'M' on a rocket moving with acceleration 'A', and gravitational mass 'M' under the influence of a gravitational field with acceleration 'A' can be seen to follow from Newton's laws as follows:

$F_i$ = M(A)          ...inertial force opposes the acceleration A of the mass 'M' in rocket.
$F_g$ = M(GM$_e$/$r^2$) ...gravitational force, where $\mathbf{GM_e/r^2}$ is now virtual particle acceleration.

Under gravity, the magnitude of the gravitational field acceleration $A=GM_e/r^2$, which is the same as the magnitude of the acceleration of the rocket. From the reference frame of an average accelerated virtual particle on earth, a virtual particle 'sees' the real particles of the stationary mass M accelerating in exactly the same way as an average stationary virtual particle in the rocket 'sees' the accelerated mass particles in the rocket. In other words, the vacuum state appears the same from both of these reference frames. We have illustrated equivalence in a special case; between an accelerated mass M and a stationary gravitational mass 'M'. Equivalence holds because $GM_e/r^2$ represents the net statistical average downward acceleration vector of the virtual particles with respect to the earth's center, and is **equal** to the acceleration of the rocket. Newton's law of gravity was rearranged here to emphasis the form F=MA for *gravitational mass* so that we can see that the **same** electromagnetic force summation process for real particles of the mass occurs under gravity as it does for accelerated mass. Thus the same processes at work in inertia are also present in gravitation.

This example shows why both the masses of figure 1 are equivalent to the masses in figure 3. The force magnitude is the same because the calculation of the force involves the same sum of all the little electromagnetic forces between the virtual charged particles and the real particles of the mass. The only difference in the physics of the masses in figure 1 is that the relative motions of all the tiny electromagnetic force vectors are reversed. The other difference is that large numbers of graviton particles (that originate from the earth's mass) slightly unbalances perfect equivalence between the masses falling on the earth. The larger mass has the largest graviton flux.

**Note:** There is a *very small* discrepancy in the equivalence principle for unequal masses in free fall near the earth which is caused by the excess graviton exchange force for the heavier mass. This discrepancy in the free fall rate of test masses near the earth is extremely minute in magnitude because there is a ratio of about $10^{40}$ in the field strength existing between the electromagnetic and gravitational forces. In principle it could be measured by extremely sensitive experiments, if two test masses are chosen with a very large mass difference.

In figure 4 two different masses are in free fall near the surface of the earth, and no external forces are present on the two masses. The two masses hit the earth at the same time. The net statistical average acceleration of the real particles that make up the masses and virtual charged particles of the vacuum is still zero, because this process is dominated by the electromagnetic force (the direct graviton exchanges are negligible). The electromagnetic forces between the virtual particles and the matter particles of the test mass dominates the interactions, because the electromagnetic force is $10^{40}$ times stronger than the graviton component. Although mass '2M' has twice the gravitational force due to twice the number of graviton exchanges, this is totally swamped out by the electromagnetic interaction, and the accelerated virtual particles and the test masses are in a state of electromagnetic equilibrium as far as acceleration vectors are concerned. Both masses fall at the same rate (neglecting the slight imbalance of the note above).

## 18. EMQG AND THE PROBLEM OF SPACE-TIME CURVATURE

*"The relativistic treatment of gravitation creates serious difficulties. I consider it probable that the principle of the constancy of the velocity of light in its customary version holds only for spaces with constant gravitational potential."*

- Albert Einstein (in a letter to his friend Laub, August 10, 1911)

4D space-time curvature is relative in general relativity, and depends on the state of motion of the observer. It is very difficult to see how this can arise from the cellular automaton. In the development of quantum inertia, EMQG has yielded a new and completely different approach to 4D space-time curvature.

In this section, we contrast the two different approaches to the problem of space-time curvature and the propagation of light in a gravitational field: Einstein's General Relativity theory and EMQG theory. First we will look at Einstein's gravitational field equations for a spherical mass called the Schwarzschild metric, which describes the amount of 4D space-time curvature near the earth. Next, we summarize the EQMG theory of space-time curvature, based on the Cellular Automata theory and Quantum Inertia.

### 18.1 GENERAL RELATIVISTIC 4D SPACE-TIME CURVATURE

General relativity accounts for the motion of light under all scenarios (section 15) for a large spherical mass. General Relativity postulates space-time curvature in order to preserve the constancy of the light velocity in an accelerated frame or in a gravitational field. The solution of Einstein's gravitational field equation for the case of spherical mass distribution is called the Schwarzchild metric (ref. 19). This is a complete mathematical description of the space-time curvature near the large spherical mass in spherical coordinates in differential form called the 4D space-time metric. From this, it can be show (ref. 19) that the comparison of time measurements between a clock outside a gravitational field (called proper time) to a clock at distance r from the center of a spherical mass distribution (called the coordinate time) do not agree. Time dilation occurs between a clock on the earth compared to a clock positioned at infinity, where clocks on the earth run slower then those at infinity. Similarly, distance measurements on the earth are distorted when compared to infinity. This is 4D space-time curvature. This curvature can literally be seen, by examining the path that light takes moving parallel to the earth's surface. According to the equivalence principle light deflects in an accelerated rocket, and therefore should deflect on the earth.

## 18.2  EMQG AND 4D SPACE-TIME CURVATURE

In order to understand space-time curvature and the principle of equivalence in regards to the equivalence of all light motion in an accelerated rocket compared with that on the surface of the earth, we must examine the effects of the background virtual particles on the propagation of light. The big question to consider here is this:

***Does the general downward acceleration of the virtual particles of the quantum vacuum near a large mass affect the motion of nearby photons? Or is the deflection of photons truly the result of an actual space-time geometric curvature (which holds down to the tiniest of distance scales), as required by the constancy of the light velocity in Einstein's special relativistic postulate?***

What hinges on this important question is whether our universe is truly geometrically curved and therefore a curved, Minkowski 4D space-time continuum on the lowest possible distance scales; or whether curved 4D space-time geometry merely results from the activities of quantum mass particles interacting with quantum vacuum virtual particles. EMQG takes the second view. According to postulate 4 of EMQG theory, light takes on the same general acceleration as the net statistical average value of quantum vacuum virtual particles, through a 'Fizeau-like' scattering process. By this we mean that the photons are frequently absorbed and re-emitted by the electrically charged virtual particles, which are (on the average) accelerating towards the center of the large mass. When a virtual particle absorbs the real photon, a new photon is re-emitted after a small time delay in the same general direction as the original photon. This process is called photon scattering (figure 5). We will see that photon scattering in the presence of downward accelerating, and electrically charged virtual particles of the quantum vacuum is of central importance to the understanding of the origin of 4D space-time curvature.

The velocity of light in an ordinary moving medium is already known to differ from its value in an ordinary stationary medium. Fizeau (1851) demonstrated this experimentally with light propagating through a current of water flowing with a constant velocity. Later (1915), Lorentz identified the physics of this phenomena as being be due to his microscopic electromagnetic theory of photon propagation. Einstein attributed this to the special relativistic velocity addition rule. In EMQG, we propose that in gravitational fields (and in accelerated motion) the moving water of Fizeau's experiment is now replaced by the accelerated virtual particles of the quantum vacuum. Like in the Fizeau experiment, photons scatter by the accelerated motion of the virtual particles of the quantum vacuum.

Imagine what would happen if Fizeau placed a clock inside his stream of moving water. Would the clock keep time properly, when compared to an observer with an identically constructed clock placed outside the moving water? Of course not! The very idea of this seems almost ridiculous. Yet we are expected to believe that the flow of virtual particles does not affect clocks and rulers under the influence of a gravitational field, as compared to the identical circumstance in far space. If Einstein knew the nature of the quantum vacuum at the time he proposed general relativity theory, he might have been aware of this connection between gravity, space-time curvature, and accelerated virtual particles.

In our review of special relativity, we have seen (section 7) the importance of the propagation of light in understanding the nature of space and time measurements. Recall that the definition of an inertial frame in space is a vast 3D grid of identically constructed clocks placed at regular intervals with a ruler. Therefore, we will closely examine the behavior of light near the earth.

We will now take a bold step and assume that propagation of photons moving upwards decelerates (in absolute CA units, the mechanism will be revealed later) on the surface of the earth according to: $c - gt = c(1 - gh/c^2)$. This holds only for *very short distances and times*. Technically this is true only at a point, which means that this equation must be written in differential form. We ignore the special relativistic postulate of the constancy of light velocity for now, and address this problem later. This means that photons continuously vary their velocity (the velocity of light is still an absolute constant between vacuum scattering events) by scattering with the falling virtual particles, as they propagate upwards. The scattering process will be described in detail later. If this picture is true, why is it that we do not observe this variation in light velocity in actual experiments on the earth?

First we must carefully understand what is meant by light velocity. Velocity is *defined* as distance divided by time, or $c=d/t$. Light has very few observable characteristics in this regard: we can measure velocity c (the ratio of d/t); frequency $\nu$; wavelength $\lambda$; and we can also measure velocity by the relationship $c=\nu\lambda$. It is important to note that all these observables are related. We know that $\nu = 1/t$ (t is the period of one light cycle) and $\lambda=d$ (the length of one light cycle). Thus, $c=d/t$ and $c=\nu\lambda$ are equivalent expressions. If we transmit green light to an observer on the ceiling of a room on the earth, and he claims that the light is red shifted, it is impossible for him to tell if the red shift was caused by the light velocity changing, or by space and time distortions which causes the timing and length of each of the light cycles to change. For example, if the frequency is halved, or $\nu_f = (1/2)\nu_i$ and the wavelength doubles $\lambda_f = 2\lambda_i$ (and you were not aware of both changes), then the velocity of light remains unchanged ($c=\nu\lambda$). However, if the velocity of light halved, and you were not aware of it, then you could conclude that the frequency is halved, $\nu_f = (1/2)\nu_i$ and the wavelength doubles $\lambda_f = 2\lambda_i$. To illustrate this point, we will now examine what happens if an observer on the floor feeds a ladder (which represents the wave character of light) with equally spaced rungs to an observer on the ceiling, where each observer cannot see what the other observer does with the ladder.

Imagine a perfect ladder with equally space rungs of known length being passed up to you at a known velocity, such that it is impossible to tell the motion of the ladder other than by observing the rungs moving past you. If the rung spacing are made larger, you would conclude that either the ladder is slowing down, or that the spacing of the ladder rungs was increased. But it would be impossible to tell which is which. Let us assume that you make a measurement on the moving rungs, and observe a spacing of 1 meter between any two rungs. Then you observe that two rungs move past you every second. You therefore conclude the velocity of the ladder is 2 m/sec. Now, suppose that the ladder is fed to you at half speed or at 1 m/sec, and that you are not aware of this change in velocity. You could conclude that the velocity halved from your measurements, because you now observe that one rung appears in view for every second that elapses instead of two rungs, and that the velocity was thus reduced to 1 m/sec. However, you could just as well conclude that your space and time was altered, and that the velocity of the ladder is constant or unaffected. Since you observe only one rung in view per second instead of the usual two rungs, you could claim that the rung spacing on the ladder is enlarged (red-shifted) or doubled by someone, and that the velocity still remains unaltered. From this, you conclude that the frequency is halved, and that time measurements that will be based on this ladder are now dilated by a factor of two.

Which of these two approaches is truly correct? It is impossible to say by measurement, unless you know before hand what trait of the ladder was truly altered. For photons, the same problem exists. No known measurement of photons in an accelerated rocket or on the surface of the earth can reveal whether space and time is affected, or whether the velocity of light has changed. In EMQG theory, the variable light velocity approach is chosen for several reasons. First, the *equivalence of light motion in accelerated and gravitational frames now becomes **fully understood*** as a dynamic process having to do with motion (for gravity, hidden virtual particle motion), just as we found for ordinary matter in motion. Secondly, the *physical basis of the curvature* of Minkowski 4D space-time near a large mass now becomes clear. It arises from the interaction of light and matter with the background accelerated virtual particle processes. This process can be visualized as a fluid flow (for acceleration only) affecting the motion of light and matter. Finally, *the physical **action** that occurs between the earth and the surrounding space-time curvature now becomes clearly understood*. The earth acts on the virtual particles of the quantum vacuum through graviton exchanges, causing them to accelerate towards the earth. The accelerated virtual particles act on light and matter to produce curved 4D space-time effects. The physical process involved is photon scattering. In order to clarify the photon scattering approach to space-time curvature, we will now review the conventional physics of light scattering in real moving and real non-moving transparent matter such as water or glass. After this brief review, we will examine photon scattering due to the virtual particles of the quantum vacuum.

18.3    FIZEAU EFFECT:  THE SCATTERING OF PHOTONS IN REAL MATTER

It is well known result of classical optics that light moves slower in glass then in air. Furthermore, the velocity of light in air is slower than that of its vacuum velocity. It also has been known for over a century that the velocity of light in a moving medium differs from its value in the same, stationary medium. Fizeau demonstrated this experimentally in 1851. For example, with a current of water (with refractive index of the medium of n=4/3) flowing with a velocity V of about 5 m/sec, the relative variation in the light velocity is $10^{-8}$ (which he measured by use of interferometry). Fresnel first derived the formula in 1810 with his ether dragging theory. The resulting formula relates the longitudinal light velocity '$v_c$' moving in the same direction as a transparent medium of an index of refraction 'n' defined such that 'c/n' is the light velocity in the stationary medium, which is moving with velocity 'V' (with respect to the laboratory frame), where c is the velocity of light in the vacuum:

Fresnel Formula: $v_c = c/n + (1 – 1/n^2) V$ (18.31)

Why does the velocity of light vary in a moving (and non-moving) transparent medium? According to the principles of special relativity, the velocity of light is a constant in the vacuum, with respect to all inertial observers. What Einstein proposed this postulate, he was not aware of the fact the vacuum is not empty. However, he was aware of Fresnel's formula and derived it by the special relativistic velocity addition formula for parallel velocities (to first order). According to special relativity, the velocity of light relative to the proper frame of the transparent medium depends only on the medium. The velocity of light in the stationary medium is defined as 'c/n'. Recall that velocities u and v add according to the formula:

$u+v = (u + v) / (1 + uv/c^2)$ (18.32)

Therefore, we can write:

$v_c = [ c/n + V ] / [ 1 + (c/n) (V)/c^2 ] = (c/n + V) / ( 1 + V/(nc) ) \sim c/n + (1 – 1/n^2) V$ (18.33)

This is the Fresnel formula.

The special relativistic approach to deriving the Fresnel formula does not say much about the actual quantum processes going on at the atomic level. At this scale, there are several explanations for the detailed scattering process in conventional physics. Because light scattering is central to EMQG theory, we will investigate some of these different approaches in more detail below.

### (1) CLASSICAL PHOTON SCATTERING THEORY IN MATTER

The Feynman Lectures on Physics gives one of the best accounts of the classical theory of the origin of the refractive index and the slowing of light through a transparent material like glass (ref. 31, chap. 31 contains the mathematical details). We will summarize the important points of the argument below:

(1) The incoming source electromagnetic wave (light) consists of an oscillating electric and magnetic field. The glass consists of electrons bound elastically to the atoms, such that if a force is applied to an electron the displacement from its normal position will be proportional to the force.

(2) The oscillating electric field of the light causes the electron to be driven in an oscillating motion, thus acting like a new radiator generating a new electromagnetic wave in the same direction as the source wave. This new wave is always delayed, or retarded in phase. This delay results from the time delay required for the bound electron to oscillate to full amplitude. Recall that the electron carries mass, and therefore inertia, and therefore time is required to move the electron.

(3) The total resulting electromagnetic wave is the sum of the source electromagnetic wave plus the new phase-delayed electromagnetic wave, where the total resulting wave is phase-shifted.

(4) The resulting phase delay of the electromagnetic wave is the cause of the reduced velocity of light in a medium such as glass.

### (2)   LORENTZ SEMI-CLASSICAL THEORY OF PHOTON SCATTERING

The microscopic theory of the light propagation in matter was developed as a consequence of Lorentz's non-relativistic, semi-classical electromagnetic theory. We will review and summarize this approach to photon scattering, which will not only prove useful for our analysis of the Fizeau effect, but will provide insight into the 'Fizeau-like' scattering of photons near large gravitational fields in EMQG theory.

To understand what happens in photon scattering inside a moving medium, imagine a simplified one-dimensional quantum model of the propagation of light in a refractive medium. The medium consisting of an idealized moving crystal of velocity 'V', composed of evenly spaced point-like atoms of spacing 'l'. When a photon traveling between atoms at a speed 'c' (vacuum light speed) encounters an atom, that atom absorbs it and another photon of the same wavelength is emitted after a time lag '$\tau$'. In the classical wave interpretation, the scattered photon is out of phase with the incident photon. We can thus consider the propagation of the photon through the crystal is a composite signal. As the photon propagates, part of the time it exists in the atom (technically, existing as an electron bound elastically to some atom), and part of the time as a photon propagating with the undisturbed light velocity 'c'. When it exists as a bound electron, the velocity is 'V'. From this, it can be shown (ref. 30) that

$$v_c = \frac{(c/n) + (1 - 1/n) V (1 - V/c)}{1 - (1 - 1/n)(V/c)} \sim c/n + (1 - 1/n^2) V \quad \text{(first order in V/c)} \quad (18.34)$$

Again, this is the Fresnel's formula. Thus this simplified non-relativistic atomic model of the propagation of light through matter explains the Fresnel formula to first order in V/c through the simple introduction of a scattering delay between photon absorption and re-emission.

### (3)   QUANTUM FIELD THEORY OF PHOTON SCATTERING IN MATTER

The propagation of light through a transparent medium is a very difficult subject in quantum field theory (or QED). It is impossible to compute the interaction of a collection of atoms with light exactly. In fact, it is impossible to treat even one atom's interaction with light exactly. However, the interaction of a real atom with photons can be approximated by a simpler quantum system. Since in many cases only two atomic energy levels play a significant role in the interaction of the electromagnetic field with atoms, the atom can be represented by a quantum system with only two energy eigenstates. In the book "Optical Coherence and Quantum Optics" a thorough treatment of the absorption and emission of photons in two-level atoms is given (ref. 32, Chap. 15, pg. 762). When a photon is absorbed, and later a new photon of the same frequency is re-emitted by an electron bound to an atom, there exists a time delay before the photon re-emission. According to QED, a *finite time* is required before re-emission of the photon.

18.4     SCATTERING OF PHOTONS IN THE QUANTUM VACUUM

The above analysis can now be used to help us understand how photons travel through the virtual particles of the quantum vacuum. First we investigate the propagation of photons in the vacuum in far space, away from all gravitational fields. The virtual particles all have random velocities and move in random directions, and have random energies $\Delta E$ and life times $\Delta t$, which satisfies the uncertainty principle: $\Delta E \, \Delta t > h/(2\pi)$. Imagine a real photon propagating in a straight path through the virtual particles in a given direction. The real photon will encounter an equal number of virtual particles moving in a certain direction, as it does from the exact opposite direction. The end result is that the quantum vacuum particles do not contribute anything different than if all the virtual particles were at relative rest. Thus, we can consider the vacuum as some sort of stationary matter medium, with a very high density.

Is the progress of the real photon delayed as it travels through the quantum vacuum, where it encounters many electrically charged virtual particles? The answer to this question depends on whether there is a time delay between the absorption, and subsequent re-emission of the photon by a given virtual particle. Based on our arguments above, we postulate that the photon is delayed as it travels through the quantum vacuum (EMQG Postulate #4). The uncertainty principle definitely places a lower limit on this time delay. In other words, according to the uncertainty principle the time delay cannot be exactly equal to zero! Our examination of the physics literature has not revealed any previous work on the time delay analysis of photons propagation through the quantum vacuum, or any evidence to contradict our hypothesis of photon vacuum delay (presumably because of the precedent set by Einstein's postulate of light speed constancy).

We will take the position that the delays due to photon scattering through the quantum vacuum reduces the 'raw light velocity $c_r$' (defined as the photon velocity between vacuum particle scattering) to the average light velocity 'c' in the vacuum of 300,000 km/sec that we observe in actual experiments. Furthermore, we propose that the quantum vacuum introduces a vacuum index of refraction 'n' such that $c = c_r / n$. What is the raw light velocity? It is unknown at this time, but must be significantly larger than 300,000 km/sec.

The vacuum index of refraction 'n' must be very large because of the high density of virtual particles in the vacuum. What happens if the entire quantum vacuum is accelerated? How does the motion of a photon get affected? These questions turn out to have a deep connection to space-time curvature.

18.5    PHOTON SCATTERING IN THE ACCELERATED QUANTUM VACUUM

Anyone who believes in the existence of the virtual particles of the quantum vacuum (which carry mass), will acknowledge the existence of an accelerated state of virtual particles of the quantum vacuum near any large gravitational field. Gravitons from the real particles on the earth exchange gravitons with the virtual particles, causing them to accelerate downward. The virtual particles of the quantum vacuum (now accelerated by a large mass) acts on light (and matter) in a similar manner as a stream of moving water acts on light in the Fizeau effect. How does this work mathematically? Again, it is impossible to compute the interaction of an accelerated collection of virtual particles of the quantum vacuum with light exactly.

Since the average distance between virtual charged particles is very small, the photons (which are always created at velocity $c_r$) spend most of the time existing as some virtual charged particle undergoing downward acceleration. Because the electrically charged virtual particles of the quantum vacuum are falling in their brief existence, the photon *effectively* takes on the same downward acceleration as the virtual vacuum particles (ref. 1). In other words, because the index of refraction of the quantum vacuum 'n' is so large, and $c = c_r/n$ we can write in equation 16.52:

$$v_c(t) = c_r/n + (1 - 1/n^2) gt = c + gt = c (1 + gt/c) \text{ if } n >> 1. \quad (18.51)$$

Similarly, for photons going against the flow (upwards): $v_c(t) = c (1 - gt/c)$     (18.52)

This formula for the variation of light velocity near a large gravitational field leads to the correct amount of general relativistic space-time curvature (ref. 1).

Einstein, himself briefly considered the hypothesis of variable light velocity near gravitational fields shortly after releasing his paper on the deflection of light in gravitational fields, as can be seen in the quotation at the beginning of this section. It would be interesting to contemplate what Einstein might have concluded if he new about the existence of virtual particles undergoing downward acceleration near a massive object (or in accelerated frames). Since Einstein was aware of the work by Fizeau on the effect of light velocity by a moving media, he might have been able to explain the origin of space-time curvature at the quantum level.

Now let us imagine that two clocks that are identically constructed, and each calibrated with a highly stable monochromatic light source in the same reference frame. These clocks keep time by using a high-speed electronic divider circuit that divides the light output frequency by "n" such that an output pulse is produced every second. For example, the light frequency used in the clock is precisely calibrated to $10^{15}$ Hz; this light frequency is converted in to an electronic pulse train of the same frequency, where it is divided by $10^{15}$ to give an electronic pulse every second. Another counter in this clock increments every time a pulse is sent, thus displaying the total time elapsed in seconds on the clock display. Now, let us place these two clocks in a gravitational field on earth with one of them on the surface, and the other at a height "h" above the surface. The clocks are compared every second to see if they are still running in unison by exchanging light signals. As time progresses, the clocks loose synchronism, and the lower clock appears to run slower. According to general relativity, light always maintains a constant speed, and space-time curvature is responsible for the difference in the timing of the two clocks. Recalling the accelerated Fizeau-like quantum vacuum fluid, we can derive the same time dilation effect by assuming that the light velocity has <u>exactly</u> the same downward acceleration component of the background falling quantum vacuum virtual particles.

18.6    SPACE-TIME CURVATURE FROM SCATTERING THEORY

We can now see that in order to formulate a theory of gravity involving observers with measuring instruments (such as clocks and rulers) we must take into account how these measurements are affected by the local conditions of the quantum vacuum. Our analysis above shows that quantum vacuum can be viewed as a Fizeau-like fluid undergoing downward acceleration near a massive object, which affects the velocity of light. Indeed, not only is the velocity of light affected, it is *all* the particle exchange processes including graviton exchanges. Therefore, we find that the accelerated Fizeau-like 'quantum vacuum fluid' effects all forces. This has consequences for the behavior clocks, which are constructed with matter and electromagnetic forces. After all, nobody questions the fact that a clock that is submerged in moving water cannot keep proper time with respect to an external clock. Similarly, a clock near a gravitational field also cannot be expected to keep proper time with respect to an observer outside the gravitational field, if all the particles that make up the clock are subjected to an 'accelerated vacuum flow'.

The accelerated Fizeau-like 'quantum vacuum fluid' moves along radius vectors directed towards the center of the earth, and thus has a specific direction of action. Therefore, the associated space-time effects should also work along the radius vectors (and not parallel to the earth). This is precisely the nature of curved 4D space-time near the earth, as we will see.

For the case of light moving parallel to the floor on the earth, the path that light takes is the end result of a tremendous number of photon-virtual particle scattering (figure 5). Again, in between virtual particle scattering, the light velocity is constant and 'straight'. However the total path is curved as shown in figure 5. The path the light takes is called a geodesic in general relativity. In EMQG, this path simply represents the natural path that light takes through the accelerated vacuum. For the case of light moving parallel to the floor of the accelerated rocket (figure 6), the path for light is also the result of virtual particle scattering, but now the quantum vacuum is not in a state of relative acceleration.

Therefore, the path is straight for the observer outside the rocket. The observer inside the rocket sees a curved path simply because he is accelerating upwards.

We are now in a position to show why Einstein's gravitational theory takes the form that it does. Because of the continuously varying frequency and wavelength of the light with height, Einstein interpreted this as a variation of space and time with height. We postulated that the scattering of light with the falling vacuum changes the light velocity in absolute CA units, which cause the *measurements of space and time* to be affected. As we have already seen, these two alternative explanations ***cannot*** be distinguished by direct experimentation. This is why the principle of the constancy of light velocity is still a postulate in general relativity (through the acceptance of special relativity).

We can now understand the concept of a geodesic proposed by Einstein. The downward acceleration of the virtual electrically charged particles of the quantum vacuum serves as an effective 'electromagnetic guide' for the motion of light (and for test masses) through space and time. This 'electromagnetic guide' concept replaces the 4D space-time geodesics that guide matter in motion in relativity. For light, the guiding action is through the electromagnetic scattering process. For matter, the electrically charged virtual particles guide the particles of a mass by the electromagnetic force interaction that results from the relative acceleration. Because the quantum vacuum virtual particle density is quite high, but not infinite (at least about $10^{90}$ particles/m$^3$), the quantum vacuum acts as a very effective reservoir of energy to guide the motion of light or matter.

The relative nature of 4D space-time can now be easily seen. Whenever the background virtual particles of the quantum vacuum are in a state of *relative acceleration* with respect to an observer, the observer *lives* in curved 4D space-time. Why should the reader accept this new approach, when both approaches give the same result? The reason for accepting EMQG is that the action between a large mass and 4D space-time curvature becomes clear. The reason that 4D space-time is curved in an accelerated reference is also clear. The relative nature of curved 4D space-time also becomes very obvious. An observer inside a gravitational field would normally live in a curved 4D space-time. If he decides to free-fall, he cancels his relative acceleration with respect to the quantum vacuum, and 4D space-time is restored to flat 4D space-time for the observer. The principle of general covariance no longer becomes a principle, but merely results for the deep connection between the quantum vacuum state for accelerated frames and gravitational frames. Last, but not least, the principle of equivalence is completely understood as a reversal of the (net statistical) relative acceleration vectors of the charged virtual particles of the quantum vacuum, and real particles that make up a test mass. We have seen EMQG at work for spherically symmetrical and non-rotating masses. What about the nature of the virtual particle acceleration field around an arbitrary mass distribution in any state of motion?

## 19. EMQG GRAVITATIONAL FIELD EQUATIONS

By treating the quantum vacuum as a continuous fluid surrounding a mass, we can account for the motion of the virtual particles of the quantum vacuum near an arbitrarily shaped large mass with absolute mass density ρ. We further assume that there is a large enough mass involved which will significantly disturb the nearby quantum vacuum particles, and that the graviton flux is not so high as to disrupt the principle of equivalence. If the graviton flux is extremely high, it can compete with the normal electromagnetic forces in the vacuum and disturb equivalence. Furthermore, since the density of the virtual particles in the quantum vacuum is so high (at least $10^{90}$ particles per cubic meter), the variation of the virtual particle acceleration from point to point in space can be considered as a classical continuous field. Therefore, the methods of vector calculus can be used with the assumption that CA space and separate CA time form a perfect continuum, and that the acceleration of the virtual particles from place to place is a mathematical vector field. The EMQG field equations are formulated in absolute CA space and time units, and thus not directly observable (space-time effects mask these results).

We will start by reviewing the classical equations of gravitation as given by Newton and Poisson and see how these relate to EMQG theory. Central to development of the EMQG equations of virtual particle motion is the concept of forces as particle exchanges.

## 19.1    THE CLASSICAL NEWTONIAN GRAVITATIONAL FIELD

Here is a brief review of the classical laws of gravitation based on the classical concept of a force field. The field concept can be traced to Newton's instantaneous law of gravity for two point masses repeated for the large collection of particles in the mass.

Newton's law of gravitation:    $F = \dfrac{G M_1 M}{r^2}$     (19.1)

which is the mathematical form of Newton's gravitational law for the force F between two point masses $M_1$ at (x,y,z) and M at (x', y',z') directed along the line between the two points, r is the distance between the two particles; $r = [(x-x')^2 + (y-y')^2 + (z-z')^2]^{1/2}$ , and G is Newton's gravitational constant   $G=6.673 \times 10^{-11}$ $m^3$ $kg^{-1}$ $s^{-2}$. This can be stated in a concise form as follows:

***"Every particle in the universe attracts every other particle with a force which is directly proportional to the product of the two masses and inversely proportional to the square of the distance between them; the direction of the force being in the line joining the two mass particles."*** **This force acts instantaneously.**

According to the particle exchange paradigm, this law works the same way as Feynman's photon exchange process in Coulomb's law of electrical attraction. Newton's inverse square law is a result of the geometry of the graviton exchange process. The exchange particle flux density spreads out on the surface of a sphere (area = $4\pi r^2$), and the flux is directly proportional to the product of the magnitudes of the masses. Stated in terms of particles, the product of the number of particles contained in each of the point masses determines the number of gravitons exchanged. Furthermore, since gravitons move at the speed of light, there is a delay in transmitting the force of gravity that was overlooked by Newton, which will be closely examined later. The graviton particles do not interact with each other, just like the photons do not interact with photons (through some force exchange process).

If we now let $M_1$ be a test particle of unit mass, then dividing the force of gravity by $M_1$ provides the *gravitational attraction* **g** produced by a mass $M_1$ at the location of the test particle P(x,y,z) along the vector **r**:

$$\mathbf{g}(x,y,z) = -\frac{GM}{r^2}\mathbf{r} \tag{19.2}$$

where **r** is a unit vector directed from the mass M to the test mass at point P(x,y,z). Because **g** has units of force divided by mass (or acceleration), it is sometimes called the *gravitational acceleration*. Any particle of any mass value at point P(x,y,z) will have the same acceleration due to the principle of equivalence. Therefore, **g** represents the average acceleration vector of the virtual particles in the quantum vacuum at that point. It is known that the gravitational attraction g is an irrotational classic field because $\nabla \times \mathbf{g} = 0$. From the Helmholtz theorem (ref. 33), gravitational attraction is a conservative field and can be represented as the gradient of the Newtonian scalar potential field $\phi(\mathbf{x},t)$:

$$\mathbf{g} = \nabla \phi, \text{ where } \phi = GM/r \tag{19.3}$$

Newton's law treats gravitational forces between particles as vectors. When this law is generalized to a large number of particles interacting, the concept of a mass distribution as a collection of a large number of particles emerges. But for this concept to work properly, the gravitational potential must obey the *principle of superposition*:

**The gravitational potential of a collection of masses is the sum of the gravitational attractions of the individual masses. The net force on a test particle is the vector sum of the forces due to all the mass particles in space.**

The principle of superposition gives us one of the most important properties of the graviton particles (postulate #2): *graviton particles do not exhibit force interactions with other graviton particles*. Thus the total graviton interaction is the vector sum of the individual graviton interactions. This works the same way as the superposition principle works for photons in QED (ref. 40). The principle of superposition can be applied to find the resultant gravitational attraction as the limit is taken towards a continuous distribution of matter. A continuous distribution of matter with mass m is defined as a collection of a great many very small masses dm = $\rho(x,y,z)$ dv, where $\rho(x,y,z)$ is defined as the mass-density of the distribution, and dv is the change in volume.

We follow the work of Bernard F. Schutz (ref. 34) for the Newtonian gravitational potential. The force on a unit test mass at the coordinate (x,y,z) is the vector sum of an infinite and continuous distribution of particles in the mass. The concept of a vector force field can be thought of as the force that will be applied to a unit mass at point (x,y,z). This force is usually written in the form of the gravitational potential $\phi(\mathbf{x},t)$ and this can calculated by solving Poisson's equation for the mass distribution:

**Poisson's Equation:** $\nabla^2 \phi = 4\pi G\, \rho(x,y,z,t)$ (19.4)

where $\phi$ is the Newtonian gravitational potential field, $\rho$ is the mass density function of the source mass which is a function of position (x,y,z) and of time t (the mass distribution can be in motion). In Newtonian physics, the gravitational potential $\phi$ follows the variation of the mass distribution $\rho$ instantaneously. For example, when the earth orbits the sun, the gravitational potential of the earth follows the earth exactly, no matter what the speed of the orbit.

Poisson's equation has the solution for the Newtonian potential $\phi_N(\mathbf{x},t)$ given by:

**Newtonian Potential:** $\phi_N(\mathbf{x},t) = - G \int \rho(\mathbf{y},t)\, r^{-1}\, d^3y$ , where $r \equiv |\mathbf{x} - \mathbf{y}|$ (19.5)

This can be thought of as a superposition of the 1/r potential fields of each of the mass elements given by mass $\rho\, d^3y$ at position $(\mathbf{y},t)$. The vector $\mathbf{r}$ represents the distance between the unit test mass at (x,y,z) and the mass element $\rho(\mathbf{y},t)\, d^3y$. These equations are formulated with Newton's version of absolute space and time, which we will discuss later.

19.3   THE EMQG FIELD EQUATIONS

Here we will determine the equations for the acceleration of the virtual particles and their direction at each point in space, as determined by a high speed (v approaches c) mass distribution. Here we assume that the velocity of the mass distribution can be comparable to the speed of the graviton particles, so that there will be a delay or retardation between the variations of the mass density $\rho(\mathbf{y},t)$ with time, and thus the corresponding Newtonian potential $\phi_N(\mathbf{x},t)$. The delay is due to the velocity of the graviton particles, which moves at the speed of light. The gravitons propagate from the mass distribution to a unit test mass at point (x,y,z) which are also occupied by a dense collection of virtual particles. The virtual particles in turn, are responsible for the subsequent force of gravity on that unit test mass by means of the electromagnetic force on a test mass, as discussed previously. This retardation can be easily introduced in the Newtonian potential function $\phi_N(\mathbf{x},t)$ as follows:

$\phi_R(\mathbf{x},t) = - G \int \rho(\mathbf{y}, t - r/c)\, r^{-1}\, d^3y$ (19.6)

Here, a change in $\rho$ at $\mathbf{y}$ ought to be felt at $\mathbf{x}$ only after a time $|\mathbf{x} - \mathbf{y}| / c$, the propagation delay of the graviton particles. This leads to a modified high speed potential field $\phi_R(\mathbf{x},t)$.

Again, this can be thought of as a superposition of the 1/r potential fields of each mass element given by mass $\rho \, d^3y$ at position $(\mathbf{y},t)$. This superposition is justified because the graviton flux satisfies linear superposition. As before, the vector $\mathbf{r}$ represents the distance between the unit test mass at $(x,y,z)$ and the mass element $\rho(\mathbf{y},t) \, d^3y$.

It can be shown that $\phi_R$ satisfies the following equation:

$$\nabla^2 \phi_R - \frac{1}{c^2} \frac{\partial^2 \phi_R}{\partial t^2} = 4\pi G \, \rho(x,y,z,t) \tag{19.7}$$

The acceleration vector $\mathbf{a}$ from the gravitational potential $\phi_R(\mathbf{x},t)$ at a point $(x,y,z)$ is:

**THE EMQG EQUATIONS FOR THE VIRTUAL PARTICLE ACCELERATION FIELD AT A POINT (X,Y,Z) FOR A HIGH SPEED MASS DISTRIBUTION:**

We have, $\mathbf{a} = \nabla \phi_R$, or $\tag{19.8}$

$A_x = \dfrac{\partial \phi}{\partial x}$

$A_y = \dfrac{\partial \phi}{\partial y}$

$A_z = \dfrac{\partial \phi}{\partial z}$

where $\phi_R(\mathbf{x},t) = - G \int \rho(\mathbf{y}, t - r/c) \, r^{-1} \, d^3y \tag{19.9}$

Again, we must emphasize that these equations are <u>not</u> formulated in the Minkowski four-dimensional curved space-time formalism of general relativity, which is derived for an arbitrary observer with his space-time measuring instruments and chosen coordinate system. Instead, these equations are based on *absolute* cellular automata space and time units. The x, y, and z distances are measured as a count of the number of cells occupying a given length in space, and the time t as a count of the number of 'clock' cycles that has elapsed between two events on the CA. The origin (0,0,0) is usually the center of the mass distribution. These equations approximately represent the inner workings of the gravity on the cellular automata as seen by the cells themselves, and are independent of a physical observer and his instruments. Because of this, these equations cannot be verified directly by experiment.

Therefore, these equations are not generally covariant because they are not formulated for an arbitrary observer in any reference frame with measuring instruments made from matter.. They are formulated in the specific coordinate system of the CA cell space, and are all written in vector form (not as tensors). The mass-density $\rho$ is also treated as the absolute mass-density, which is independent of the observer. The source of the Newtonian gravitational field is the mass-density which is given by $\rho$ = mass/volume. According to relativity, mass and volume are observer dependent, i.e. they vary with the state of motion of the observer. For example, if an observer is moving at relativistic speeds with respect to the mass distribution, the mass varies according to $m = m_0 (1-v^2/c^2)^{-1/2}$, where v is the relative velocity. Similarly, the volume is Lorentz contracted, which is also an observer dependent entity. Thus, the mass-density varies from observer to observer. In EMQG, there exists absolute space, absolute time, and an absolute mass distribution that occupies a definite number of cells. An event takes a definite number of 'clock' cycles. EMQG is not formulated for an arbitrary observer. How would one formulate these same laws of gravity from the perspective of an arbitrary observer in an arbitrary state of motion, who chooses an arbitrary coordinate system for his measurements? It turns out that Einstein has already accomplished this task beautifully in his gravitational field equations.

### 19.4   EINSTEIN'S GRAVITATIONAL FIELD EQUATION REVISITED

Einstein's goal was to propose a theory of gravity while retaining the basic postulates of special relativity in regards to the speed of light being a universal constant. Simple considerations of the motion of light inside an accelerated rocket and the principle of equivalence led him to *postulate* the curvature of four-dimensional Minkowski space-time. This postulate allowed him to retain light as an absolute constant (which means light still moves 'straight', somehow) while having the space-time that light moves in 'curve'. This 4D space-time curvature guides light to move along the curved geodesic paths of the background space-time, which are the 'straight lines' of Riemann geometry. This idea, along with the principle of general covariance, led Einstein to formulate his theory of gravitation in the form of quasi-Riemannian geometry. Einstein started with Poisson's equation, just as we have done in EMQG. Through the use of tensors and the principle of equivalence, Einstein was able to formulate 4D space-time curvature in a form that was generally covariant. This allowed observers to *switch* between accelerated and gravitational frames at will, or to switch between different coordinate systems, and yet have the same general form for the gravitational equations.

The postulates of general relativity, along with the special theory of relativity and the mathematics of Riemann Geometry lead to the famous Einstein gravitational field equation:

$$G_{\alpha\beta} = \frac{8\pi G}{c^2} T_{\alpha\beta} \quad \text{.... Einstein's Tensor Gravitational Field Equations} \quad (19.10)$$

$G_{\alpha\beta}$ is the Einstein tensor, which is the mathematical statement of space-time curvature, that is reference frame independent, and generally covariant. $T_{\alpha\beta}$ is the stress-energy tensor, which is the mathematical statement of the special relativistic mass-energy density (and observer dependent), and G is Newton's gravitational constant, and c the velocity of light. The constant G here still reflects the Newtonian aspects of general relativity. The constant $8\pi G/c^2$ is chosen to adjust the strength of coupling between matter and 4D space-time, so that it corresponds to the correct amount of Newtonian gravitational force.

Thus, Einstein has accomplished an *observer* dependent formulation of gravity. In contrast, EMQG formulates the law of gravity that is **not** observer dependent. Instead, it is based on the only truly important units of measurement: the cellular automata absolute units of space and time, which is not directly accessible to measurement. Einstein concluded that space-time is four-dimensional, and curved in order to be compatible with a constant light velocity in **all** frames. Einstein observed that when light moves parallel to the surface of the earth, light will curve (also for an accelerated observer). Either he abandons the postulate of constancy of light velocity, and allows light move in curved paths, or he must have space and time curved somehow, by the unknown action of the nearby mass distribution, in which case light velocity can be constant and follow a geodesic path. Einstein chose the later approach, but could never find the physical action that causes matter to curve 4D space-time.

A full quantum field theory of the graviton exchange process must be developed in order to complete EMQG. This theory is expected to closely resemble QED theory for the electron and photon, because of the close similarities discussed in section 17. Therefore, a fully renormalizable quantum field theory of the masseon-graviton particles is possible. The Feynman diagrams and rules must also be developed. The properties of the masseon given must be fully developed. In particular, how does the masseon particle fit in with the other forces of nature (the strong and weak nuclear forces) in the standard model? What is the exact nature of the force that binds masseons together to form the particles of the standard model?

How does this fit in with the various unification schemes like super-symmetry, super-gravity, etc? Why are there large gaps in the allowed mass of the particles of the standard model? We call this the mass hierarchy problem. In other words, there is a large jump in the mass of particles as you go from the lepton family (an electron, for example) to the baryon (an up quark, for example) family of particles, with no other particle types in between. One solution to the hierarchy problem is to postulate that leptons and quarks are made up of masseons in some kind of orbital arrangements, where only certain orbital arrangements are allowed. Thus, the quark has tightly bound orbits with highly relativistic masseon orbital speeds, and the electron with lower speeds. The mass would look higher from our frame of reference, when the particle orbit has higher speeds.

## 20. COSMOLOGY AND CELLULAR AUTOMATA

*"Nothing can be created out of nothing"* - *Lucretius*

Is our universe expanding from an initial singularity at the time t=0 in accordance with the Big Bang theory? In other words, is curved 4D space-time expanding from a point of infinite curvature, infinite density, and infinitesimal size to it's present size? Was there nothing before the initial big bang? Using our CA model, we believe that this is simply ***not*** possible. It is quite clear from the basic ideas of CA that at time t=0 there was *no* singularity, and that the basic structure of the CA must have been there before the simulation started evolving. It is also clear that *all* the particles that exist *today*, were definitely *not* present exactly at t=0 on the CA. We believe that in order to pre-program all the matter particles and their respective states of motion that exist in the universe today at the start of the CA simulation is ***totally unthinkable***.

Instead we believe there must have been an era where matter (information patterns) was created from a some simple, initial numeric state. Therefore we conclude that there must have been a *'matter creation'* era in the very early universe, which must have halted shortly afterward since no matter creation process like this is observed by astronomers today. The 'matter creation era' is a new idea that is not a part of standard big bang physics, nor is it part of the inflationary cosmology. Nevertheless we believe that this phase is an absolute necessity in the very early universe, if our universe is a CA.

Figure 12 shows a highly schematic view of the CA version of the big bang. In conventional big bang theory, space and time does not exist before the big bang. Instead, space and time are created as the universe expands. All matter is assumed to be created during the initial big bang. In CA big bang theory, cell space existed before the big bang.

Can matter particles be created out of nothing in today's universe? Not really. For starters, this would violate the existing conservation laws. However, matter particles can be created from energy (i.e., a sufficiently energetic boson) according to the standard model. For example, an energetic photon can transform to an electron-positron pair which possess mass and move at very high velocities. In this way, energy can convert to matter. Therefore, if a matter creation era existed in the early universe, there must have been a large amount energy available to start the process.

Did the volume of our 4D space-time really increase greatly (in accordance with the general relativistic picture) since the initial big bang? Was the entire universe ever the size of a pea or even an atom? This question depends on the answer of another very important question; *what is 4D space-time*? Recall that according to EMQG, space and time at a fundamental level consists of quantized cells that are capable of holding numeric data, and also of CA clock cycles which cause the information to evolve (section 2). From this basic structure, we have seen that *two different* measurement systems can be applied to the measure of space and time. These two measurement systems are summarized below:

(1) At the lowest distance scales, we find that the absolute, quantized 3D CA space can be measured in units of *'number of cells'* between two absolute cell address locations on the CA. Similarly, at the lowest time scales we find that time can be measured as the *'number of elapsed clock cycles'* between two events on the CA. These units of measure are only accessible to hypothetical observers that are capable of determining the numeric contents and absolute locations of CA cells. These units are totally inaccessible to ordinary matter-based observers such as ourselves (section 18.3).
(2) At the large scale there is the familiar relativistic 4D space-time, which can be

measured in meters and seconds (SI system) by matter-based observers using instruments composed of real matter, like clocks and rulers for example. Occasionally when the virtual particles of the quantum vacuum are disturbed such that there exists a relative state of acceleration with respect to the observer and his instruments, these measurements can be modified through the Fizeau-like, quantum vacuum electrical scattering process described in detail in section 18.3.

*So what has actually happened to our space-time in the universe for the past 10 billion years or so*? According to EMQG theory, the cosmological 4D space-time curvature must be strictly a quantum vacuum process, where the state of relative acceleration of the electrically charged virtual particles of the quantum vacuum determines the path that light and real matter take (section 18). **A*ll*** the matter contributed by the real particles in the entire universe contributes to the state of relative acceleration of ***all*** the virtual matter particles of the quantum vacuum through the long-range graviton exchanges. In other words it is the state of relative acceleration of the virtual particles of the quantum vacuum (with respect to real matter) which varies from place to place in the universe that determines the variation of 4D space-time curvature from place to place (section 18.3).

**Note**: *The virtual particles of the quantum vacuum themselves contribute very little, if anything, to the overall curvature of the universe in spite of the fact that many virtual particles possess mass-charge and thus ought to contribute to the cosmological graviton flux. This is because the virtual fermion particles are created and destroyed in virtual matter and anti-matter particle pairs of opposite mass-charge, and thus the total vacuum mass-charge at any time roughly cancels out (in the same way that electrical charge cancels out). In other words, the vacuum contains equal numbers of virtual particles that have gravitational attraction and gravitational <u>repulsion</u> (ref. 36).*

The existing popular big bang model of an expanding 4D space-time from an initial singularity is obviously not compatible with CA and EMQG theory. The very first problem is the initial singularity. The initial singularity supposedly contains *no* space-time and infinite matter-energy density. According to the CA model, the cells are storage locations for numbers, and form the integral part of the CA construct, and cannot be expanded, created or destroyed.

Therefore the curvature of 4D space-time must be the result of the activities of the virtual particles of the quantum vacuum, which act on real matter and light through the 'Fizeau-like' scattering process (described in section 18.3). In EMQG, Einstein's curved 4D space-time does ***not*** represent the actual geometry of the universe on the lowest possible scales. The geometry of the universe at the CA distance scales is utter simplicity, it is a collection of a vast number of cells interconnected in a 3D geometric CA arrangement (section 2).

We have seen that the motion of light is deeply related to the issue of 4D space-time curvature (section 18.3). Imagine that a sufficiently powerful laser light source could be constructed and pointed in a fixed direction in the sky. If the universe turns out to be closed in the general relativistic sense, then this light beam would return to the original source from the opposite direction of the sky, after traveling around the entire universe! This is not the only possibility for the behavior of light. According to EMQG, the laser light beam photons scatters with the accelerated, electrically charged virtual particles of the quantum vacuum (where the acceleration was due to graviton exchanges with all the matter in the universe) such that the path is very slightly curved. After a sufficiently long distance of travel, it is quite conceivable in this model that the light beam can *loop back* on itself and return to the original location. This does not necessarily imply that space-time has expanded. In other words, the path that light take does not necessarily follow the true geometry of the universe!

Now if this same laser beam experiment were performed 5 billion years earlier when the solar system had just formed, the curvature of space-time would have to have been greater than it is now. This is because all the mass in the universe would have been more concentrated as compared to now. The graviton flux is larger for a denser universe. Therefore, the overall virtual particle acceleration would be greater, and the corresponding 4D space-time curvature would have been more tightly curled 5 billion years ago. Therefore the laser beam would return to the starting point quicker.

Does this imply that low level CA 3D space and time has changed within the last 5 billion years? Not at all! At the cellular level, everything is absolute and fixed. ***The only thing that physically has changed in 5 billion years is the overall cosmological virtual particle acceleration*** as measured in absolute CA units. Just because the force free path for matter and light in the cosmos is curved, this does not necessarily imply that the fundamental geometry of the universe is curved and expanding. It is easy to see that all the low level CA cell space existed before the virtual particle phenomena, and the matter expansion took root and dominated the dynamics of the universe. Another definite consequence of the CA model is that the ***universe cannot expand forever***. This is because the CA must be finite, and the number of cells must be definitely limited.

Therefore we propose that it is simply the general outward motion of matter from the initial big bang, and the subsequent variation of cosmological virtual particle acceleration with time that is responsible for the observed Hubble law for the expansion of the universe. What caused matter to move outward at the initial big bang so violently? The answer can be found from basic quantum field theory. During the 'matter creation era' discussed above, vast numbers of particle pairs (virtual and real particle and anti-particle pairs, in order to satisfy the conservation laws) must have been created. In general, particle pairs were born with very high velocities. Therefore, after the 'matter creation era', particles were in a general state of very high velocity, thermal motion. Countless scattering events would have distributed the velocities of the particles into a very broad spectrum. Somehow, this must have led to a distribution of velocities given in the Hubble law. What mechanism could have caused the velocities of galaxies to distribute themselves in this fashion?

Has there been any previous proposals for the big bang based that are based solely on the motion of matter outlined above, rather than on the concept of 4D space-time expansion?

The answer to this question is *yes*. Currently we believe that the best existing model that most closely fits our CA paradigm for cosmology is the Milne Kinematic cosmology model (ref. 43, page 198-199). This model was introduced by E.A. Milne in 1934 to account for the expansion of the universe as a very simple outward flow of matter (possessing a random velocity distribution) in an already existing flat 4D space-time. He was able to explain the observed characteristics of the red shift by the dynamics of the outward motion of matter alone.

Milne reasoned that the initial velocity of matter might have taken on a broad and isotropic distribution of velocities from some violent event akin to an explosion event. Then as the particles move apart freely, the ones with the greatest velocities move the furthest away, and eventually particles with similar velocities sort themselves into zones of similar outward motion. The cosmological redshift, and isotropy are explained by this model without recourse to an expanding 4D space-time. This process naturally leads to a linear relationship between the recession velocity 'v' and the distance 'r' from the explosion. Milne showed that Hubble's law follows from these following basic assumptions:

(1) The universe is homogenous or expanding in the sense that the proper distances between neighboring moving observers are increasing.
(2) We can apply the usual law of vector addition of relative velocities.

However, a drawback of Milne Kinematic Cosmology is the violation of the cosmological principle. The cosmological principle states that any typical observer located anywhere in the universe at the present will see the general isotropy and expansion of the universe in the same way as any other observer. In Milne cosmology, the cosmological expansion does not follow the cosmological principle in that there will be reference frames for observers in the Milne universe where the distribution of galaxies looks different then for a 'typical observer' in other frames. For example, an observer that evolved on a galaxy born out of very high velocity matter (with respect to the general average) would see one part of the sky containing a denser population of galaxies then the another side. It is for this reason, i.e. the violation of the cosmological principle (and also the rise of general relativity) that Milne Cosmology has been long abandoned. However, there is no known reason why the universe should follow this principle, other than for reasons of symmetry.

## 21.   CONCLUSIONS

We have presented a new paradigm for physical reality, which restores a great unity to all physics. We have concluded that our universe is a vast Cellular Automaton (CA) simulation, the most massively parallel computer model known. This CA structure is a simple 3D geometric CA. All physical phenomena, including space, time, matter, and forces are the result of the interactions of numeric *information* patterns, governed by the mathematical laws and the connectivity of the CA. Because of the way the CA functions, all the known global laws of the physics must result from the local mathematical law that governs each cell, and each cell contains the same mathematical law.

We have seen that the CA structure automatically presents our universe with a speed limit for motion. Quantum field theory requires that all forces are the result of the boson particle exchange process. This particle exchange paradigm fits naturally within CA theory, where the boson exchange represents the transfer of boson information patterns between (fermion) matter particles. All forces (gravity is no exception) originate from exchange processes dictated by quantum field theory.

We have concluded that Einstein's special relativity is already manifestly compatible with Cellular Automata theory. We showed that the Lorentz Transformations of Special Relativity can be derived by assuming that light moves on the Cellular Automata by simply shifting from cell to adjacent cell at every 'clock cycle', the maximum possible speed on a CA.

We modified a new theory of inertia first introduced in ref. 5, which we now call Quantum Inertia. Quantum Inertia is based on the idea that inertial force is due to the tiny electromagnetic force interactions originating from each charged particle that consists of real matter undergoing relative acceleration with the virtual electrically charged particles of the quantum vacuum. These tiny forces is the source of the total resistance force to accelerated motion in Newton's law 'F = MA', where the sum of each of the tiny particle forces equals the total inertial force.

This new approach to classical inertia automatically resolves the problems and paradoxes of accelerated motion introduced in Mach's principle, by suggesting that the virtual particles of the quantum vacuum serve as Newton's universal reference frame (which he called absolute space) for accelerated motion only. Thus, Newton was correct that it is the relative accelerated motion with respect to absolute space that somehow determines the inertia of a mass, but absolute space is totally useless in determining the *absolute* velocity of a mass.

A new theory of quantum gravity called ElectroMagnetic Gravity (or EMQG) was developed, which involves *two* pure force exchange processes. EMQG is manifestly compatible with Cellular Automata theory. **Both** the photon and graviton exchanges occur simultaneously inside a large gravitational field. Both particle exchange processes follow the particle exchange paradigm originated in QED, where the photon and the graviton have very similar quantum numbers and characteristics. .

We found that gravity also involves the same 'inertial' electromagnetic force component that exists in an accelerated mass, which reveals the deep connection between inertia and gravity. Inside large gravitational fields there exists a similar quantum vacuum process that occurs for inertia, where the roles of the real charged particles of the mass and the virtual electrically charged particles of the quantum vacuum are reversed. Now it is the charged virtual particles of the quantum vacuum that are accelerating, and the mass particles are at relative rest.

The general relativistic Weak Equivalence Principle (WEP) results from this common physical process existing at the quantum level in both gravitational mass and inertial mass. Gravity involves *both* the electromagnetic force (photon exchanges) and the pure gravitational force (graviton exchanges) that are occurring simultaneously. However, for a gravitational test mass, the graviton exchange process (only found in minute amounts in inertial reference frames) occurring between a large mass, the test mass, and the surrounding vacuum particles upsets perfect equivalence of inertial and gravitational mass, with the gravitational mass being slightly larger. One of the consequences of this is that if a very large, and a tiny mass are dropped simultaneously on the earth, the larger mass would arrive slightly sooner. Since this is in violation of the WEP, the strong equivalence principle is no longer applicable.

We have found that graviton exchanges occur between a large mass and the surrounding virtual particles of the quantum vacuum, and they also directly occur between the large mass and a test mass. The electromagnetic force (photon exchanges) between the virtual particles and the test mass (occurring in inertial frames and in gravitational frames) is responsible for the equivalence of inertial and gravitational mass. The pure gravitational force (graviton exchanges) is responsible for the distortion of the (net statistical average) acceleration vectors of the virtual particles of the quantum vacuum near the earth (with respect to the earth).

The state of accelerated motion of the electrically charged virtual particles of the quantum vacuum with respect to a test mass is very important in considerations of inertia and gravitation, and is the root cause of the equivalence principle. The state of accelerated motion of the quantum vacuum is also extremely important in consideration of the origin of 4D space-time curvature. We introduced a new concept for the origin of 4D, curved Minkowski space-time near a large mass which is compatible with Cellular Automata theory. We found that 4D space-time is simply a consequence of the behavior of matter (fermions) and energy (photons) under the influence of the (net statistical average) downward accelerated 'flow' of the charged virtual particles of the quantum vacuum.

This accelerated flow of the vacuum can be thought of as a special 'Fizeau-like fluid' that was unknown to Einstein at that time. Like in the Fizeau experiment which was performed with constant velocity water, the behavior of light, clocks, and rulers are now affected by the accelerated 'flow' of the virtual particles of the quantum vacuum with respect to a large mass (and in accelerated frames). This accelerated flow can now **act** on motion of matter and light, to distort space and time. This conclusion was based on the concept that photons scatter off the virtual particles of the quantum vacuum, thus maintaining the same acceleration as the downward 'flow' of virtual particles (in absolute CA units). Photons, however, still move at an absolute constant speed between the virtual particle scattering.

On quantum distance scales Minkowski 4D space-time gives way to the secondary (quantized) absolute 3D space and separate absolute (quantized) time required by CA theory. Curved 4D space-time is replaced by a new paradigm where curvature is a result of pure particle interaction processes. Particles occupy definite locations on the CA cells, and particle states are evolved by a universal 'clock'. All interactions are *absolute,* because they depend on absolute space and time units on the CA. However, we cannot probe this scale, because we are unable to access the absolute cell locations, and numeric contents of the cells. In this realm, the photon particle (as well as the graviton) is an information pattern, that moves (shifts) with an absolute constant 'velocity', since it merely shifts from cell to neighboring cell in every 'clock' cycle of the CA.

## 23.   FIGURE CAPTIONS

The captions for the figures are shown below:

Figures 1 to 4: Schematic Diagram of the Principle of equivalence
Figure 5:      Motion of Real Photons in the Presence of Virtual Particles Near Earth
Figure 6:      Motion of Real Photons in Rocket Accelerating at 1g
Figure 7:      Schematic Diagram of What Space Looks Like on the Cellular Automata
Figure 8:      Block Diagram of the Relationship of CA and EMQG with Physics
Figure 9:      Simplified Motion of a Photon Information Pattern on the CA
Figure 10:     Light Velocity Measurement from Two Observers
Figure 11:     Definition of an Inertial Reference Frame
Figure 12:     Simplified Model of the Big Bang on the 3D Geometric CA

# ILLUSTRATIONS

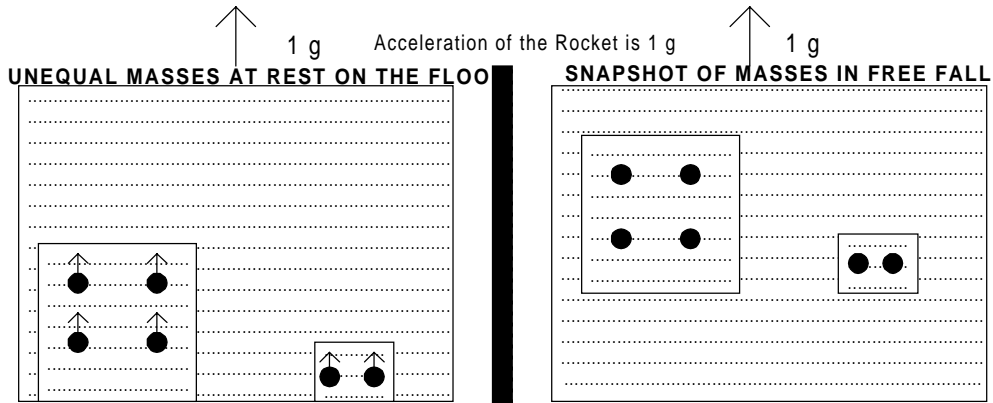

**Figure #1** - Masses '2M' and 'M' at rest on the floor of the rocket

**Figure #2** - Masses '2M' and 'M' in free fall inside of a rocket

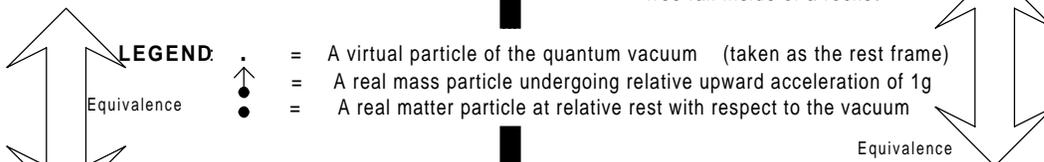

LEGEND:
- **.** = A virtual particle of the quantum vacuum (taken as the rest frame)
- ↑ (large) = A real mass particle undergoing relative upward acceleration of 1g
- • = A real matter particle at relative rest with respect to the vacuum

Equivalence

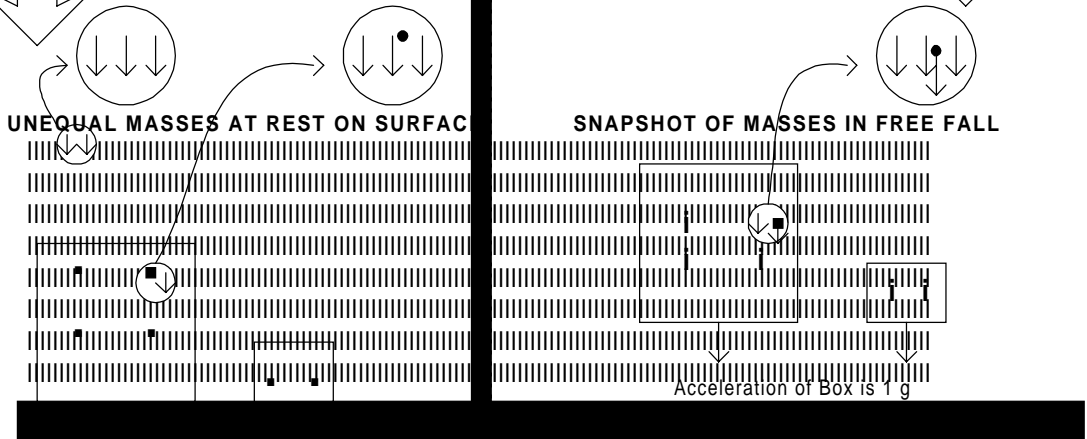

Surface of the Earth where gravity produces a 1 g acceleration

**Figure #3** - Masses '2M' and 'M' at rest on Earth's surface

**Figure #4** - Masses '2M' and 'M' in free fall above the Earth

LEGEND:
- **l** = Relative downward acceleration (1g) of a virtual particle
- **i** = Relative downward acceleration (1g) of a real matter particle
- **.** = A real stationary matter particle (with respect to the earth's center)

## FIGURES 1 TO 4 - THE PRINCIPLE OF EQUIVALENCE FOR A STATIONARY MASS ON THE EARTH AND INSIDE A ROCKET

**Figure #5 - MOTION OF REAL PHOTONS IN THE PRESENCE OF VIRTUAL PARTICLE NEAR EARTH**

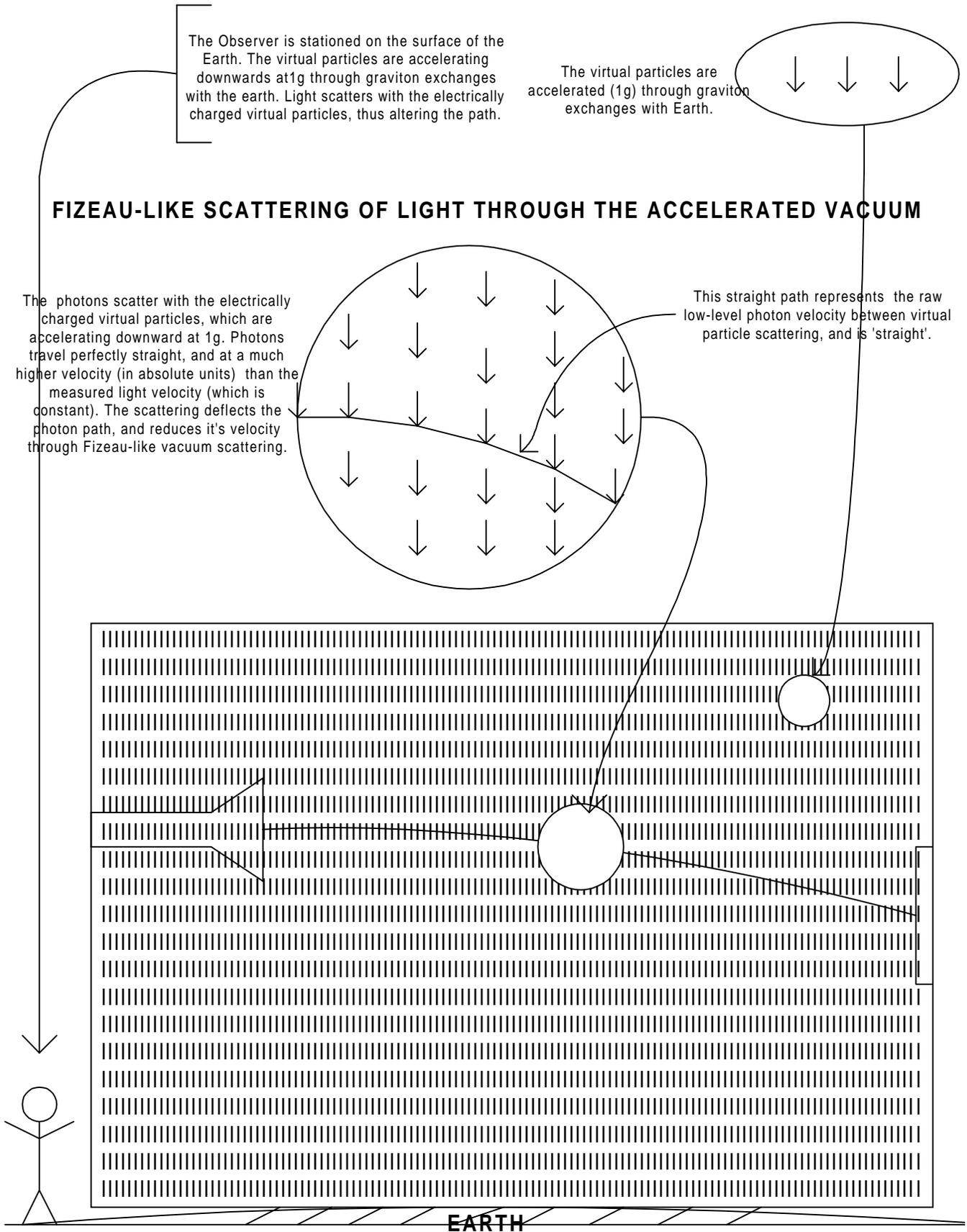

**Figure #6 - MOTION OF REAL PHOTONS IN A ROCKET ACCELERATING AT 1g**

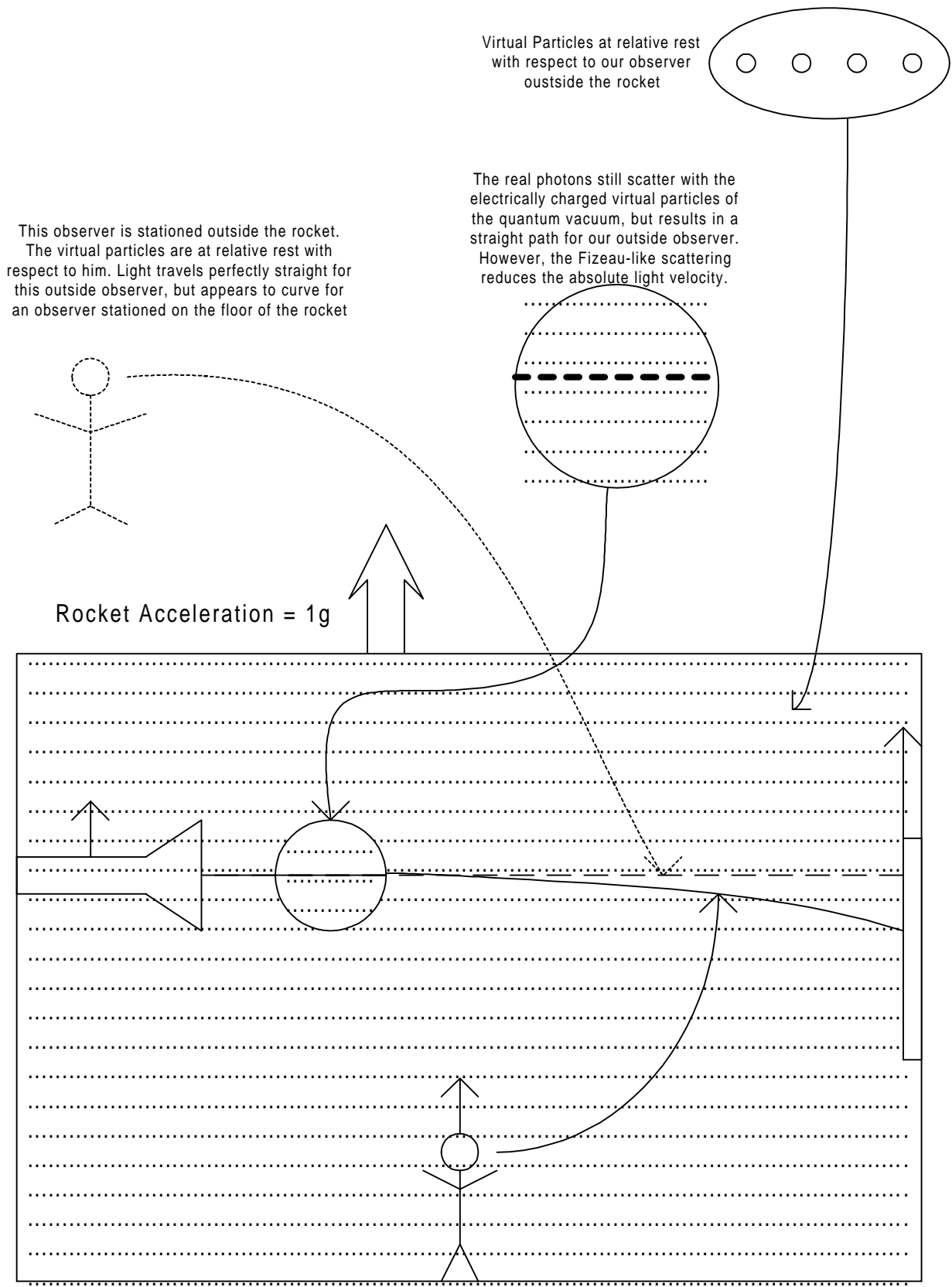

In a 3D Geometric Cellular Automata, the numeric content of Cell $C_{i,j,k}$ is uniquely determined by the numeric contents of each of the surrounding 26 neighbouring cells (and possibly with it's own numeric state). On the next CA 'clock cycle' the contents of cell $C_{i,j,k}$ is determined by a function (or algorithm) $F_{i,j,k}$ such that the contents of the cell $C_{i,j,k} = F(C_{i+x,j+y,k+z})$ where x,y,z take on all the following values: -1,0,1. This same function F is programmed into each and every cell in the entire CA. In the figure below, the binary number system is chosen for illustration purposes (any number system can be used). The dotted lines indicate what cells affect cell $C_{i,j,k}$.

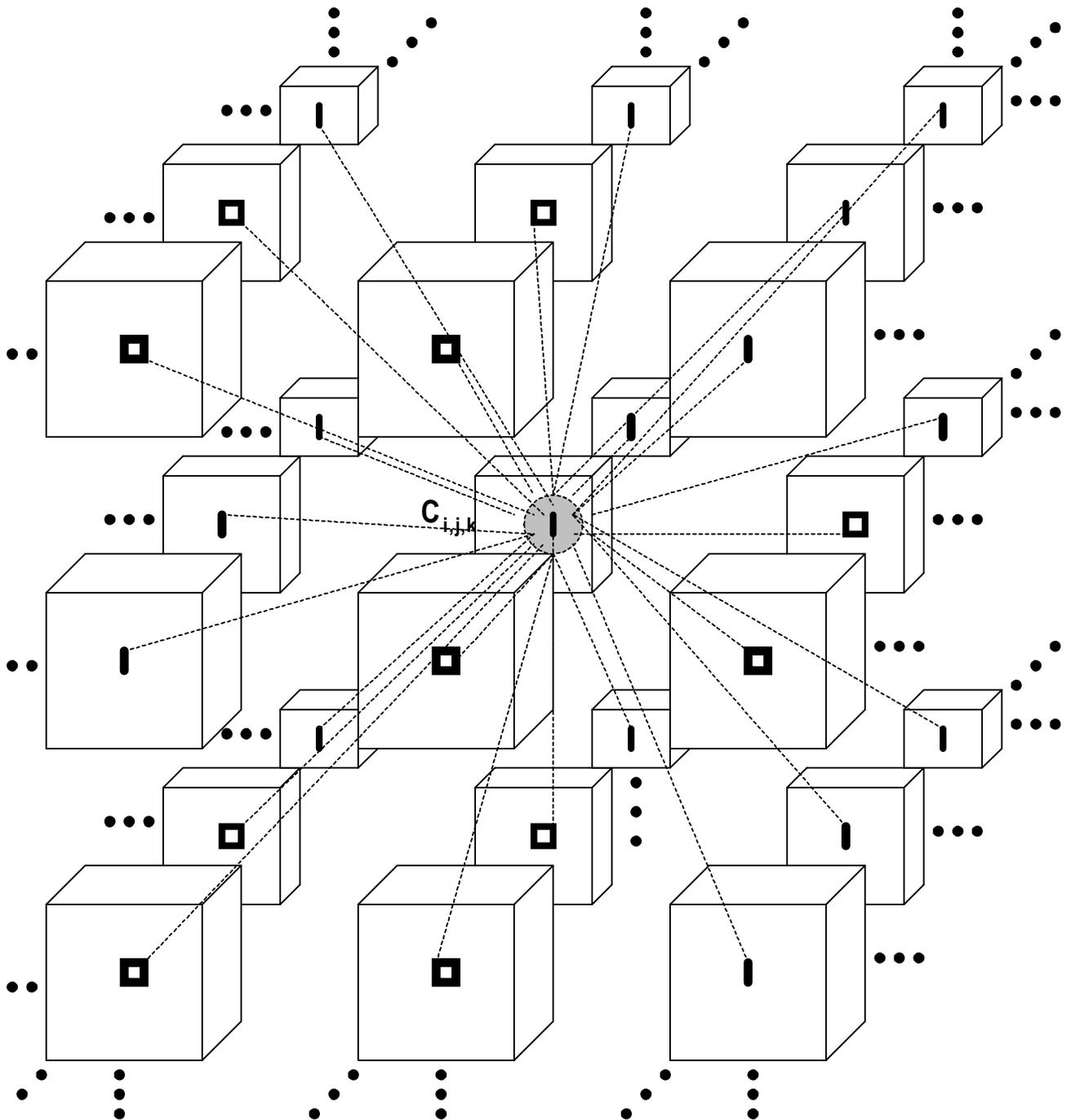

**Figure #7   Schematic Diagram of what space looks look on the Cellular Automata**

**Figure #8 - BLOCK DIAGRAM OF RELATIONSHIP OF CA AND EMQG WITH PHYSICS**

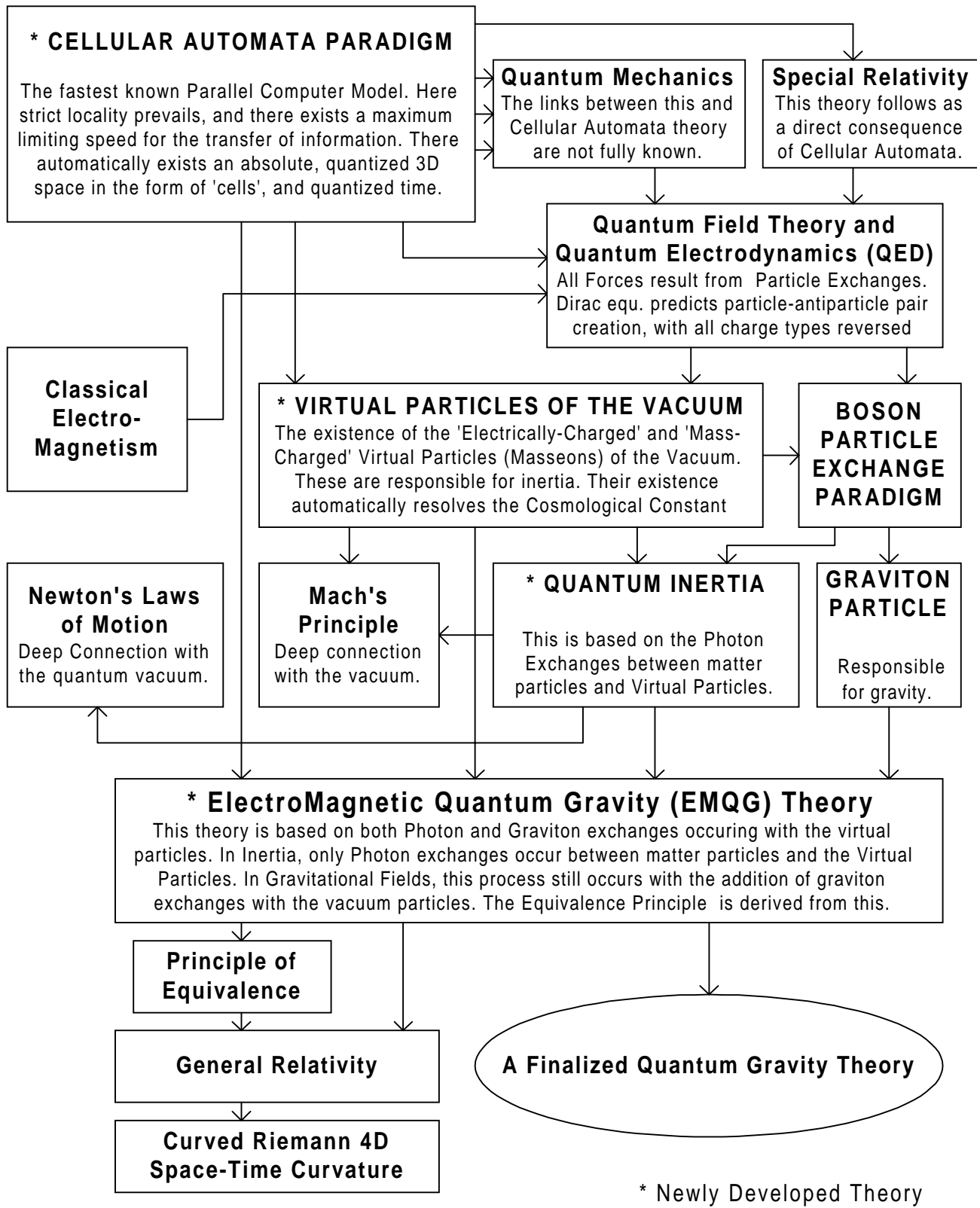

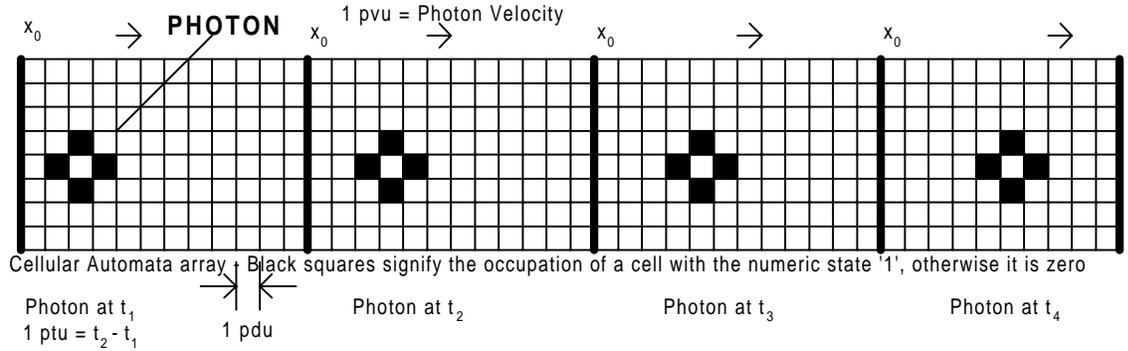

Cellular Automata array - Black squares signify the occupation of a cell with the numeric state '1', otherwise it is zero

Photon at $t_1$          Photon at $t_2$          Photon at $t_3$          Photon at $t_4$
1 ptu = $t_2 - t_1$    1 pdu

**Figure #9** - Simplified model of the motion of the photon information pattern on the CA.
The photon information pattern moves 1 plank unit to the right at every plank 'clock cycle'
(Note: The photon is actually an oscillating wavepattern (the wavefunction not shown in this simplified diagram)

**Absolute CA units:** 1 pdu is the shifting of information by 1 cell; 1 ptu is the time to shift 1 cell; 1 pvu = photon velocity

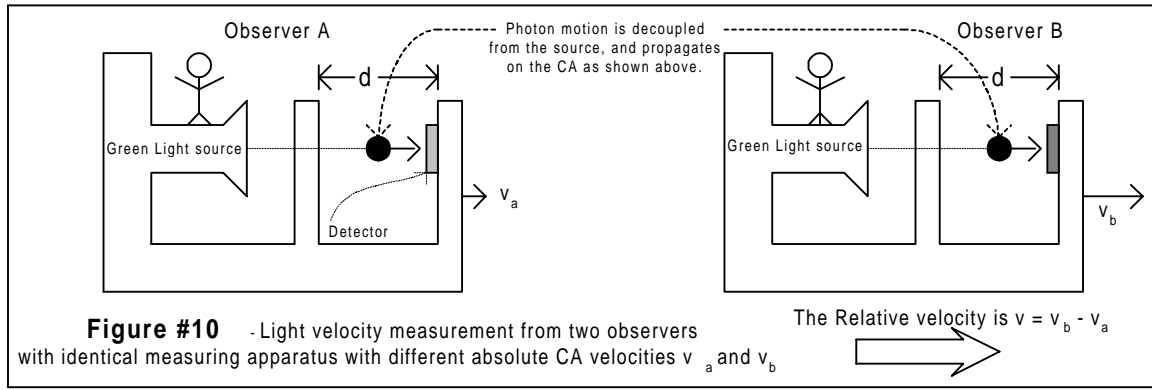

**Figure #10** - Light velocity measurement from two observers with identical measuring apparatus with different absolute CA velocities $v_a$ and $v_b$

The Relative velocity is $v = v_b - v_a$

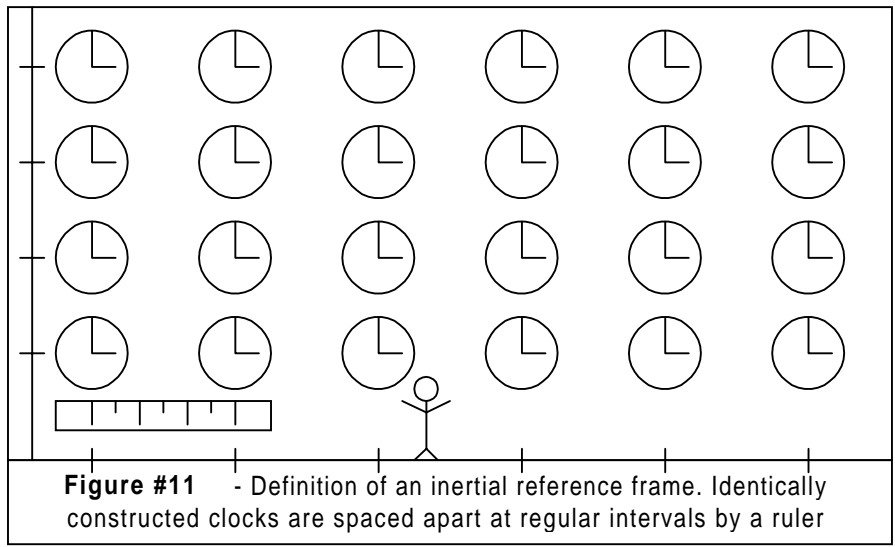

**Figure #11** - Definition of an inertial reference frame. Identically constructed clocks are spaced apart at regular intervals by a ruler

# THE GREAT INFORMATION EXPLOSION

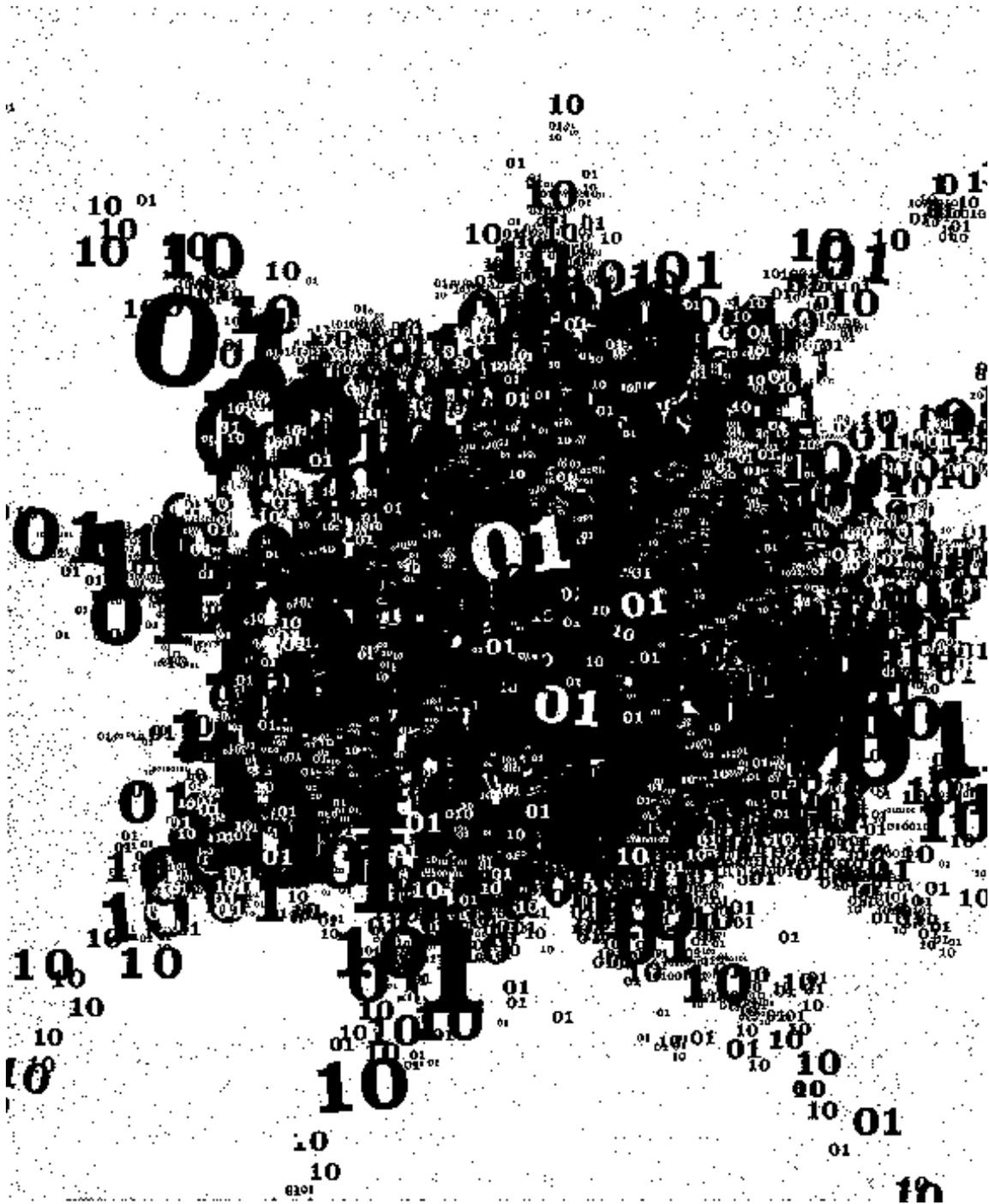

**FIGURE #12**      **Simplified Model of the Big Bang on the 3D Geometric CA**